\pdfoutput=1
\documentclass{JFM-FLM_Au}
\usepackage{xcolor}
\usepackage{mathrsfs}
 \usepackage{subfigure}

\lefttitle{J. Ge, J. Rolland and J.C. Vassilicos}
\righttitle{Journal of Fluid Mechanics}

\title{The prediction of extreme uncertainty-production events in three-dimensional Navier-Stokes turbulence}

\author{Jin GE\aff{1}, Joran ROLLAND\aff{1} \and John Christos VASSILICOS\aff{1}}

\affiliation{\aff{1}Univ. Lille, CNRS, ONERA, Arts et Métiers ParisTech, Centrale Lille, UMR 9014 - LMFL - Laboratoire de Mécanique des Fluides de Lille - Kampé de Feriet, F-59000 Lille, France}

\corresau{Jin GE, \email{jin.ge@cnrs.fr} Joran ROLLAND, \email{joran.rolland@centralelille.fr} John Christos VASSILICOS, \email{john-christos.vassilicos@cnrs.fr}}

\begin{document}
\maketitle

\begin{abstract}
We investigate the exponential growth of uncertainty energy in 3D
Navier-Stokes turbulence, emphasising the intermittent and highly
localized amplification/production of uncertainty, a critical factor
in understanding the predictability of turbulent systems. From the
Navier-Stokes equations one can identify some key fields contributing
to the growth/decay of uncertainty-production term $P_{\Delta}$: strain rate, vorticity,
and vortex deformation. The dynamics of these fields are examined in
the $Q-R$ plane, where $Q$ and $R$ are the second and third invariants
of the velocity gradient tensor, to understand their role in
the evolution of uncertainty-production term $P_{\Delta}$. We proceed by estimating
committor functions across the entire spatiotemporal domain of direct
numerical simulations (DNS) of turbulence in a periodic domain at different Reynolds numbers. Our estimates of
the probability of rare extreme events of local uncertainty-production term
as a function of uncertainty energy, strain rate, vorticity, and
vortex deformation confirm the role of strain rate in driving
uncertainty. Where strain rate and vorticity are too close to their
space-average values, stable probabilistic forecasts appear impossible
solely on the basis of the fields considered here.
\end{abstract}



\section{Introduction}

Since the pioneering work of \citet{lorenz1963deterministic, lorenz1969predictability}, the extreme sensitivity of non-linear
dynamical systems with multiple degrees of freedom to any non-zero
initial perturbation has been widely recognized. Turbulence, as one of
the most emblematic non-linear systems with a vast number of degrees
of freedom, is no exception. \citet{deissler1986navier}
demonstrated that small initial perturbations in the Navier-Stokes
equations could eventually lead to significantly different outcomes in
fully developed turbulent solutions. This sensitivity to initial
perturbations, which are inevitable in practical scenarios, is
fundamentally linked to the uncertainty that results in the long-term
unpredictability of atmospheric and oceanic circulations
\citep{kraichnan1970instability, leith1971atmospheric,
	nonaka2016potentially}, as well as in magnetohydrodynamics
\citep{ho2020fluctuations} and plasma physics
\citep{cheung1987chaotic}. It may not be enough, however, to only check
the global tendency of perturbation evolution if one wants to
understand uncertainty. This paper aims to locate and predict the
``most uncertain" points in turbulence.

An incompressible turbulent flow is entirely characterised by its
velocity field, which is the solution of the Navier--Stokes equations
complemented by a divergence free condition. The uncertainty of a
time-dependent velocity field, $\boldsymbol{u}^{(1)}(\boldsymbol{x},t)$, is determined
by comparing it to another velocity field, $\boldsymbol{u}^{(2)}(\boldsymbol{x},t)$,
that starts with almost the same initial conditions,
i.e. $\boldsymbol{w}(\boldsymbol{x},t)=\boldsymbol{u}^{(2)}(\boldsymbol{x},t)-\boldsymbol{u}^{(1)}(\boldsymbol{x},t)$. Since the uncertainty considered here is defined as an Eulerian point-to-point velocity difference at a given instant, the corresponding uncertainty-related quantities and the event definition adopted in the present work are also naturally formulated in the Eulerian framework. Using
the Eulerian velocity-difference field, $\boldsymbol{w}$, the overall separation
between two turbulent systems, $\boldsymbol{u}^{(1)}$ and $\boldsymbol{u}^{(2)}$, can
be quantified by the mean uncertainty energy $\left\langle
E_{\Delta}\right\rangle\equiv\left\langle
\left|\boldsymbol{w}\right|^{2}/2\right\rangle$, where
$\left\langle\cdot\right\rangle$ signifies the spatial average. The
evolution of $\left\langle E_{\Delta}\right\rangle$ has attracted most
attention in the study of uncertainty in turbulence. As \citet{lorenz1963deterministic, lorenz1969predictability} articulated,
the growth of $\left\langle E_{\Delta}\right\rangle$ is sequentially
driven by chaoticity followed by stochasticity. At early times when
$\boldsymbol{u}^{(1)}$ and $\boldsymbol{u}^{(2)}$ are very close, the chaotic nature
leads to exponential growth of $\left\langle E_{\Delta}\right\rangle$
\citep{ruelle1981small}, i.e. $\left\langle
E_{\Delta}\right\rangle\sim\exp\left(\lambda t\right)$, where
$\lambda$ is the maximal Lyapunov exponent. Using the incompressible Navier-Stokes equations, \citet{ge2023production} derived the time-evolution equation for $\left\langle E_{\Delta}\right\rangle$ in the case of homogeneous/periodic turbulence:\par
\begin{equation}
	\label{eq:eq of uncertainty energy}
	\frac{\mathrm{d}}{\mathrm{d}t}\left\langle E_{\Delta}\right\rangle = \left\langle P_{\Delta} \right\rangle + \left\langle \varepsilon_{\Delta} \right\rangle + \left\langle F_{\Delta} \right\rangle.
\end{equation}
Equation (\ref{eq:eq of uncertainty energy}) shows that the global separation between two flow realizations is determined by three mechanisms: dissipation of uncertainty $\varepsilon_{\Delta}=-\nu\partial_{j}w_{i}\partial_{j}w_{i}$, external input rate of uncertainty $F_{\Delta}=g_{i}w_{i}$ and internal production of uncertainty $P_{\Delta}=-w_{i}S_{ij}w_{j}$, where $\boldsymbol{g}$ is the point-to-point forcing (power input) difference and $\boldsymbol{\mathsf{S}}$ denotes the strain rate of the fluid (a detailed explanation can be found in the following sections). It follows from (\ref{eq:eq of uncertainty energy}) that $P_{\Delta}$ is the only internal term that can contribute positively to the growth of the global separation between the two flows. This is why we refer to $P_{\Delta}$ as the uncertainty-production term. Analysis of this term revealed the mechanism through which uncertainty energy is amplified: stretching events decrease uncertainty ($P_{\Delta}<0$) while the compression events increase uncertainty ($P_{\Delta}>0$). At the global level, the preferential alignment of the uncertainty field with the compressive direction \citep{ge2023production} leads to a positive average value of $P_{\Delta}$ ($\left\langle P_{\Delta}\right\rangle>0$), i.e. uncertainty is produced.

In this paper, we are concerned with this uncertainty production mechanism during the exponential growth regime of $\left\langle E_{\Delta}\right\rangle$. Furthermore, we are actually interested in $P_{\Delta}$ rather than its spatial average,
	$\left\langle P_{\Delta}\right\rangle$, and we investigate the
uncertainty-production term within the turbulence locally in space. This is
motivated by the observation that uncertainty is generated by intense
local events \citep{ge2023production}. During the exponential growth
of $\left\langle E_{\Delta}\right\rangle$, the probability
distribution functions (PDF) of $P_{\Delta}$ at different times were
found to collapse after normalisation by their maximum probability
density and the spatial standard deviation of $P_{\Delta}$, suggesting
self-similarity of uncertainty production during the exponential
growth regime of $\left\langle E_{\Delta}\right\rangle$
\citep{ge2023production}. Furthermore, the mode of these PDFs are
always located at $P_{\Delta}=0$, and their kurtosis and skewness are
extremely high \citep{ge2023production}.  These observations suggest a
high intermittency of $P_{\Delta}$, implying that rare extreme events
dominate the total chaotic production of uncertainty in
turbulence. The main aim of the present work is to forecast local
extreme production events in the reference flow. The localisation of
evolutions towards extreme uncertainty-production events is crucial
for understanding the decorrelation process between two initially
similar flows, and enhancing the accuracy of predictions of turbulent
system at the lowest possible cost.

To achieve this goal, the most natural first step is to directly
derive the evolution equation for $P_{\Delta}$ from the Navier-Stokes
equations, as presented in section \ref{sec:Theoretical analysis of
	the uncertainty production}. As a result, some quantities, such as
the strain rate and vorticity, are found to potentially affect the
evolution of uncertainty production. Therefore, we further investigate
the relationship between extreme uncertainty production events and the
topology of the flow field defined in terms of strain rate and
vorticity \citep{hunt1988streams} at different Reynolds numbers
(section \ref{sec:The relation of extreme uncertainty production
	events and flow topology}). However, the evolution equation for
$P_{\Delta}$ is too complex to be used in this investigation.
Indeed, as shown in section \ref{sec:Theoretical analysis of the
	uncertainty production}, many different physical effects contribute
to the variability and evolution of uncertainty production. Moreover
the equation driving $P_\Delta$ is not closed and these physical
effects also influence one another in non-trivial ways.


Given the complexity and absence of medium range deterministic
predictability of the systems considered, probabilistic forecasts are
considered instead. We are therefore interested in the probability of
an extreme event occurring, given the value of certain scalars
characterizing the system, for example the strain rate and vorticity,
included in the analytical equation mentioned above. This probability,
known as the committor function, can theoretically be estimated from
data in any type of system either deterministic or stochastic
\citep{lucente2022committor}.  Once this function is computed, it can
not only be used for forecast in practice, but it can also be used to
discriminate between the events that will lead to an extreme
(committor close to $1$) or not ($0$), study their physical
characteristics separately and understand what flow organization can
lead to an extreme of production or not.  The committor function was
originally introduced by \citet{onsager1938initial} to estimate
the probability of chemical reactions and was later developed in the
field of chemical kinetics before extending to other branches of
physics, like climate projections
\citep{miloshevich2023probabilistic,miloshevich2024extreme}. In systems
governed by a stochastic differential equation, \emph{posing} the
problem yielding the committor as a solution is formally
straightforward.  Indeed its state probability evolves via a Master or
Fokker-Planck equation (for continuous time) or a Chapman-Kolmogorov
equation (for discrete time).  The committor function is then the
solution of the backward version of these equations once the future
condition is given
\citep{weinan2005transition,vanden2006towards}. However, it is seldom
computed \emph{via} this direct approach. Firstly, solving these
equations in systems with multiple degrees of freedom is impossible in
practice. Secondly, we would like to perform committor computations in
dynamics which are not strictly speaking stochastic processes. To
overcome this, \citet{lucente2022coupling} proposed
using effective random dynamics through the analogue method.
Deterministic analogues were originally proposed by \citet{lorenz1969atmospheric,lorenz1969three} for
fast weather forecast.
They consist in sampling a time series of states, termed analogues,
and perform a forecast given an initial condition by finding its
nearest neighbor among the analogues and propose the same time
evolution (figure \ref{fig:schematic}, top). They have evolved into
stochastic weather generators in the recent years
\citep{yiou2014anawege,yiou2019stochastic}, where transitions are then
selected at random among $K-$nearest neighbors
(figure \ref{fig:schematic}, bottom).  The analogue Markov chain used
for committor estimation is built upon these principles
\citep{lucente2022coupling}. On top of that, the committor is computed
\emph{via} a spectral approach for smoother and more efficient
computation, as designed by \citet{prinz2011efficient}. This approach
is doubly interesting for us as we use it to create the probabilistic
dynamics modeling our turbulent flow on top of simplifying the
committor computation in our complex system. In this paper we also
address another challenge. Previous computations concerned global
properties of systems
\citep{weinan2005transition,vanden2006towards,lucente2022coupling}, or
an extreme occurring locally but at a fixed position and forecast
using fixed localized or global features of the system
\citep{miloshevich2023probabilistic,miloshevich2024extreme}.  In our
case the extreme can occur anywhere in the flow.  We aim to apply this
method to predict probabilities of extreme events in the system we
study, as introduced in section \ref{sec:Statistical analysis for the
	probability evolving to the extreme uncertainty production events}.

In the next section, we present the evolution equation of uncertainty-production term $P_{\Delta}$ in the flow field, which is directly derived from the
Navier-Stokes equations. Although complex, the equation exhibits
potential relationships between uncertainty-production term $P_{\Delta}$ and local
strain rate/vorticity/vortex-deformation. In section
\ref{sec:Statistical analysis for the probability evolving to the
	extreme uncertainty production events}, we introduce the
probabilistic forecast method proposed by \citet{lucente2022coupling}, and apply it to the prediction of
extreme uncertainty-production event. The setups of our DNS and of the
resulting time series used for committor computations are detailed in
section \ref{sec:DNS configurations and data sampling}. The numerical
results are presented in the following sections. In section
\ref{sec:The relation of extreme uncertainty production events and flow topology}, we investigate the relationship between
uncertainty-production term $P_{\Delta}$ and different flow structures defined by strain
rate and vorticity. The results on the committor function calculated
from different combinations of predictors are shown in section
\ref{sec:result}. Finally, we conclude in section
\ref{sec:Conclusion}.

\section{Theoretical analysis of the uncertainty production\label{sec:Theoretical analysis of the uncertainty production}}

In the ﬁrst part of this section we derive the evolution equation for the uncertainty-production term $P_{\Delta}$, then in the second part we discuss how the quantities, derived from the local inertial term, affect the evolution of $P_{\Delta}$.

\subsection{Time evolution of uncertainty production using Navier-Stokes equations\label{sec:Time evolution of uncertainty production using Navier-Stokes equations}}

 \citet{ge2023production} had derived the governing equation of the velocity field difference $\boldsymbol{w}$ from the incompressible Navier-Stokes equations:\par
\begin{equation}
	\label{eq:difference NS equation}	
	\begin{aligned}	
		\partial_{t}w_{i}+\left(u_{j}+w_{j}\right)\partial_{j}w_{i}+w_{j}\partial_{j}u_{i}=&-\partial_{i}h+\nu\partial_{j}\partial_{j}w_{i}+g_{i},		\\
		\partial_{i}w_{i}=&0,
	\end{aligned}
\end{equation}
where $\boldsymbol{u}=\boldsymbol{u}^{(1)}$ is the reference flow velocity. In the
following discussion, we omit the superscripts indicating the flow
field to which quantities belong, and fluid quantities are assumed to
be taken from the reference flow field unless otherwise stated. $\partial_{t}$ denotes the partial derivative with respect to time, and $\partial_{i}$ denotes the partial derivative with respect to the $i$-th spatial coordinate. In
equation~(\ref{eq:difference NS equation}), $h=p_{r}^{(2)}-p_{r}^{(1)}$
denotes the pressure-difference field (implicitly divided by the
fluid density), and $\boldsymbol{g}=\boldsymbol{f}^{(2)}-\boldsymbol{f}^{(1)}$ represents the
external forcing difference. In the present work, we assume that there
is no external input from the forcing, i.e. $\boldsymbol{g}=0$.

We first derive the time evolution of the stress $\boldsymbol{\mathsf{R}}$, with each component defined as $R_{ij}\equiv w_{i}w_{j}$. By calculating the outer product of equation (\ref{eq:difference NS equation}) with $\boldsymbol{w}$ and then add the transpose of this outer product, we obtain\par
\begin{equation}
	\label{eq:Stress NS equation}
	\begin{aligned}
		\partial_{t}R_{ij}
		&+\left(u_{k}+w_{k}\right)\partial_{k}R_{ij}
		+R_{jk}\partial_{k}u_{i}
		+R_{ik}\partial_{k}u_{j} \\
		&= -\left(w_{j}\partial_{i}+w_{i}\partial_{j}\right)h
		+\nu\partial_{k}\partial_{k}R_{ij}
		-2\nu\left(\partial_{k}w_{i}\right)\left(\partial_{k}w_{j}\right).
	\end{aligned}
\end{equation}
Considering that the local uncertainty-production term according to (\ref{eq:eq of uncertainty energy}) is $P_{\Delta}=-w_{i}S_{ij}w_{j}=-S_{ij}R_{ij}$ \citep{ge2023production}, where the strain-rate tensor $\boldsymbol{\mathsf{S}}$ is symmetrical with each component equal to $S_{ij}=\left(\partial_{j}u_{i}+\partial_{i}u_{j}\right)/2$.  We know the time evolution of the strain rate tensor is\par
\begin{equation}
	\label{eq:Strain equation}
	\partial_{t}S_{ij}+u_{k}\partial_{k}S_{ij}+S_{ik}S_{kj}+\Omega_{ik}\Omega_{kj}=-P_{ij}+\nu\partial_{k}\partial_{k}S_{ij}+F_{ij},
\end{equation}
where $\boldsymbol{\Omega}$ is the anti-symmetrical rotation-rate tensor with each component defined as $\Omega_{ij}=\left(\partial_{j}u_{i}-\partial_{i}u_{j}\right)/2$. $P_{ij} \equiv \partial_{i}\partial_{j}p_{r}$ is the pressure Hessian tensor and $F_{ij} \equiv \left(\partial_{i}f_{j}+\partial_{j}f_{i}\right)/2$ is the effect of
	the forcing on the strain rate variation. Equations (\ref{eq:Stress NS equation}) and (\ref{eq:Strain equation}) give the tempo-spatial evolution of local uncertainty-production term :\par
\begin{eqnarray}
	\label{eq:Production NS equation}	
	\partial_{t}P_{\Delta}+\left(u_{k}+w_{k}\right)\partial_{k}P_{\Delta}=&-&R_{ij}w_{k}\partial_{k}S_{ij}+\underbrace{R_{ij}\left[\left(S_{ik}\Omega_{kj}-\Omega_{ik}S_{kj}\right)+\Omega_{ik}\Omega_{kj}+3S_{ik}S_{kj}\right]}_{\text{(i)}}\nonumber\\
&+&\underbrace{S_{ij}\left(w_{j}\partial_{i}+w_{i}\partial_{j}\right)h+R_{ij}P_{ij}}_{\text{(ii)}}\nonumber\\
&+&\underbrace{\nu\left[\partial_{k}\partial_{k}P_{\Delta}+2S_{ij}\left(\partial_{k}w_{i}\right)\left(\partial_{k}w_{j}\right)+2\left(\partial_{k}S_{ij}\right)\left(\partial_{k}R_{ij}\right)\right]}_{\text{(iii)}}\nonumber\\
&-&\underbrace{R_{ij}F_{ij}}_{\text{(iv)}}.
\end{eqnarray}
While the uncertainty energy $E_{\Delta}$ quantifies the similarity or correlation between two flow fields at a given instant, the uncertainty-production term $P_{\Delta}$ instead characterizes the temporal ``stability'' of points in the flow field. Equation (\ref{eq:Production NS equation}), which
describes the space and time evolution of uncertainty-production term, is
complex and provides limited intuitive insight.  In Navier-Stokes
turbulence, $P_{\Delta}$ evolves due to four factors: (i) inertial
effects caused locally by strain and vorticity in the flow, as shown
in the first line on the right-hand-side of
(\ref{eq:Production NS equation}); (ii) pressure effects, including
the global difference between two flows (characterized by $h$) and the
global influence of the reference flow (characterized by
$P_{ij}$), as shown in the second line on the
right-hand-side of (\ref{eq:Production NS equation}); (iii)
viscosity effects; and (iv) external forcing. Here, we focus on the exponential growth regime of the uncertainty energy $\left\langle E_{\Delta}\right\rangle$. This regime is observed only when the perturbation norm is
	infinitesimal, i.e. $\left|\boldsymbol{w}\right|\ll\left|\boldsymbol{u}\right|$ \citep{ruelle1981small}.
In this limit, the contribution $-R_{ij}w_k\partial_k S_{ij}$ on the right-hand side of (\ref{eq:Production NS equation}) is much smaller than the retained production terms, and is therefore neglected in the following analysis.

Given the complexity of (\ref{eq:Production NS equation}), in the present work we focus primarily on the local inertial contribution to $P_{\Delta}$. This contribution originates from the coupling between the perturbation field $\boldsymbol{w}$ and the local velocity-gradient field. It therefore provides a direct connection between $P_{\Delta}$ and local kinematic quantities, such as strain, rotation and the associated flow topology. This connection offers an intuitive way to relate uncertainty production to the local structure of the flow, and provides a basis for identifying and visualizing the regions that are most susceptible to subsequent uncertainty growth.

However, the pressure-induced and viscosity-induced contributions involve different mechanisms and are not analysed in detail here. The pressure is intrinsically nonlocal in incompressible turbulence, since it is determined by the global velocity gradient through the Poisson equation. As a result, its contribution at a spatial point cannot be interpreted from the local velocity-gradient tensor or from the local sampling used in the following sections. The viscosity-induced contribution is associated with small-scale gradients and dissipative regularization of the perturbation field. A detailed interpretation of this term would require a separate analysis of dissipative and scale-dependent effects. For these reasons, the present work concentrates mainly on the inertial mechanism, with the aim of clarifying how local flow structures contribute to the growth of uncertainty production.

\subsection{Inertial effect on the uncertainty production\label{sec:Inertial effect on the uncertainty production}}
According to (\ref{eq:Production NS equation}), the inertial
effect on uncertainty production can be decomposed into the
interaction of velocity-difference stress with three different
symmetrical tensors. The tensor
$[\boldsymbol{\mathsf{S}},\boldsymbol{\Omega}]$ (with components $[\boldsymbol{\mathsf{S}},\boldsymbol{\Omega}]_{ij}=S_{ik}\Omega_{kj}-\Omega_{ik}S_{kj}$)
represents the vortex stretching rate with zero trace. Using the
principal axes of
$[\boldsymbol{\mathsf{S}},\boldsymbol{\Omega}]$ as a
local orthonormal reference frame, we can write:\par
\begin{equation}
	\label{eq:Ta in principal axe}
	R_{ij}\left(S_{ik}\Omega_{kj}-\Omega_{ik}S_{kj}\right)=\Lambda^{a}_{1}(w_{1}^{a})^{2}+\Lambda^{a}_{2}(w_{2}^{a})^{2}+\Lambda^{a}_{3}(w_{3}^{a})^{2},
\end{equation}
where $\Lambda^{a}_{1}$, $\Lambda^{a}_{2}$, $\Lambda^{a}_{3}$ are the
eigenvalues of
$[\boldsymbol{\mathsf{S}},\boldsymbol{\Omega}]$ and
$w_{1}^{a}$, $w_{2}^{a}$, $w_{3}^{a}$ are the components of the
velocity-difference vector projected on the corresponding principal
axes. Incompressibility forces $[\boldsymbol{\mathsf{S}},\boldsymbol{\Omega}]$ to
be traceless, i.e.,
$\Lambda^{a}_{1}+\Lambda^{a}_{2}+\Lambda^{a}_{3}=0$. Defining the
order of eigenvalues as
$\Lambda^{a}_{1}\leq\Lambda^{a}_{2}\leq\Lambda^{a}_{3}$, we must have
$\Lambda^{a}_{1}<0$ representing local vortex compression and
$\Lambda^{a}_{3}>0$ representing local vortex stretching, while the
sign of intermediate eigenvalue is uncertain. The important point
which can now be made on the basis of equation (\ref{eq:Ta in
	principal axe}) is that uncertainty-production is increased by
vortex stretching ($\Lambda^{a}_{3}>0$) and is decreased by vortex
compression ($\Lambda^{a}_{1}<0$). The total contribution of
$R_{ij}\left(S_{ik}\Omega_{kj}-\Omega_{ik}S_{kj}\right)$
to $P_{\Delta}$ depends on the alignment of uncertainty energy with
the principal axes of
$[\boldsymbol{\mathsf{S}},\boldsymbol{\Omega}]$.

Vortex stretching increases uncertainty production. At the same time, it also enhances vorticity. The vorticity term $R_{ij}\Omega_{ik}\Omega_{kj}$, however, reduces uncertainty production. Since
$\boldsymbol{\mathsf{\Omega}}$ is an anti-symmetrical tensor,
$\boldsymbol{\Omega}\cdot\boldsymbol{\Omega}$ (with components $\Omega_{ik}\Omega_{kj}$) is symmetrical with three non-positive
eigenvalues, defined to be
$\Lambda^{b}_{1}\leq\Lambda^{b}_{2}\leq\Lambda^{b}_{3}\leq0$. Following
equation (\ref{eq:Ta in principal axe}), we rewrite
$R_{ij}\Omega_{ik}\Omega_{kj}$ on the principal axes
of $\boldsymbol{\Omega}\cdot\boldsymbol{\Omega}$ as\par
\begin{equation}
	\label{eq:Tb in principal axe}
	R_{ij}\Omega_{ik}\Omega_{kj}=\Lambda^{b}_{1}(w_{1}^{b})^{2}+\Lambda^{b}_{2}(w_{2}^{b})^{2}+\Lambda^{b}_{3}(w_{3}^{b})^{2},
\end{equation}
$w_{1}^{b}$, $w_{2}^{b}$, $w_{3}^{b}$ are the components of the
velocity-difference vector projected on the principal axes of
$\boldsymbol{\Omega}\cdot\boldsymbol{\Omega}$. We have
$R_{ij}\Omega_{ik}\Omega_{kj}\leq0$, meaning that the
enstrophy decreases uncertainty production.

We now check the tensor $\boldsymbol{\mathsf{S}}\cdot\boldsymbol{\mathsf{S}}$ (with components $S_{ik}S_{kj}$), representing the magnitude of strain rate. Given that $\boldsymbol{\mathsf{S}}$ is symmetrical, the eigenvalues of $\boldsymbol{\mathsf{S}}\cdot\boldsymbol{\mathsf{S}}$, whose the order is defined as  $0\leq\Lambda^{c}_{1}\leq\Lambda^{c}_{2}\leq\Lambda^{c}_{3}$, are all positive. Following equation (\ref{eq:Ta in principal axe}), we have\par
\begin{equation}
	\label{eq:Tc in principal axe}
	R_{ij}S_{ik}S_{kj}=\Lambda^{c}_{1}(w_{1}^{c})^{2}+\Lambda^{c}_{2}(w_{2}^{c})^{2}+\Lambda^{c}_{3}(w_{3}^{c})^{2},
\end{equation}
where $\boldsymbol{w}^{c}$ are the velocity-difference vector rewritten on the principal axes of $\boldsymbol{\mathsf{S}}\cdot\boldsymbol{\mathsf{S}}$.
We have that $R_{ij}S_{ik}S_{kj}\geq0$, suggesting all the strains of flow, no matter compression or stretching, increase the uncertainty production.

Actually, equation (\ref{eq:Production NS equation}) has such a
complex form that we cannot use it to localize or predict the extreme
events of uncertainty production. However, the connection between
uncertainty production and the strain rate or vorticity of the
reference field, as shown in equation (\ref{eq:Production NS
	equation}), suggests the possibility of relating uncertainty-production term to the flow field's topographical structures, which are 
defined by strain rate and vorticity. This approach is presented in
section \ref{sec:The relation of extreme uncertainty production events
	and flow topology}, and will help us analyze the DNS data to
intuitively understand and localize extreme events of uncertainty
production. To predict the extreme uncertainty-production events,
considering the complexity of the deterministic governing equation
(\ref{eq:Production NS equation}), we will use the committor function
for probabilistic forecasting. The committor function is estimated
using DNS data and the method proposed by \citet{lucente2022coupling}, as shown in section
\ref{sec:Statistical analysis for the probability evolving to the
	extreme uncertainty production events}.

\begin{figure}
	\centering
	\includegraphics[width=0.6\textwidth]{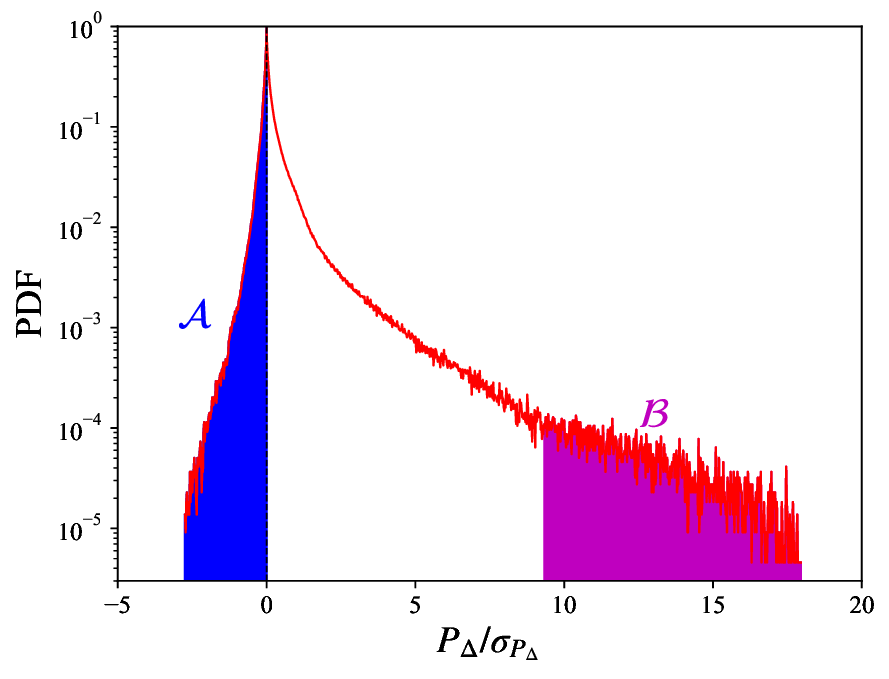}
	\caption{An example of probability density function (PDF) of
		the local uncertainty-production term $P_{\Delta}$ obtained
		as explained in section 4. The sets $\mathcal{B}$ and
		$\mathcal{A}$ are schematically represented in this plot:
		$\mathcal{B}$ contains rare extreme events in the positive
		$P_{\Delta}$ tail of the PDF, and $\mathcal{A}$ contains all
		the negative production events. The red line is the PDF of
		$P_{\Delta}$ normalized by the probability at $P_{\Delta}=0$
		and the standard deviation $\sigma_{P_{\Delta}}$ of
		$P_{\Delta}$ estimated during the exponential growth of
		$\left\langle E_{\Delta}\right\rangle$. The working
		definition of $\sigma_{P_{\Delta}}$ is given in
		figure \ref{fig:pdf}.}
	\label{fig:PDF rare_events}
\end{figure}

\section{Statistical analysis for the probability of evolving to the extreme uncertainty-production events\label{sec:Statistical analysis for the probability evolving to the extreme uncertainty production events}}

We now turn from the dynamical evolution of the uncertainty-production term i.e., equation (\ref{eq:Production NS equation}), to its statistical prediction. Although the local dynamics of $P_{\Delta}$ may naturally evoke an evolution along the flow, the prediction target considered here is Eulerian, namely the occurrence of extreme uncertainty-production events within fixed observation windows. For this reason, the statistical framework adopted below is also Eulerian. Our goal is to assess whether a reduced description of the flow, built from a small number of physically relevant quantities, contains useful information for predicting future extreme uncertainty-production events. Although the underlying Navier-Stokes dynamics is deterministic, its full phase space is extremely high-dimensional and involves too many coupled mechanisms to make a direct predictive analysis practical. For this reason, we adopt reduced phase-space descriptions and formulate the prediction problem in probabilistic terms. Probabilistic reduced descriptions are also familiar in turbulence research, for example in PDF methods and in Markov/Fokker--Planck descriptions inferred from turbulent data \citep{pope2000turbulent,FriedrichPeinke1997,Friedrich2011}.

In this section, we first explain why a probabilistic description is needed for a coarse-grained representation of a deterministic system in \ref{spb}. We then introduce the committor as the quantity used to measure the probability of reaching an extreme-event set in \ref{amc}, describe how it is estimated from DNS data using analogue Markov chains in \ref{affine}, and finally how we can measure the precision of that estimation using the Brier score in ref \ref{brier}.

\subsection{Position of the problem}\label{spb}

The central motivation for our study comes from the observation by \citet{ge2023production} that rare extreme production events contribute to a large proportion of the spatially averaged uncertainty production.  Our goal is therefore not only to isolate these high-production events, but also to diagnose which flow properties are most relevant to their occurrence and to the increase of uncertainty.

For this purpose, one could follow an \emph{a posteriori} approach.
By this, we mean sampling histograms of the uncertainty production
(figures \ref{fig:PDF rare_events},~\ref{fig:pdf}), then determining the
weight of tails where $P_\Delta$ is larger than its average by several
standard deviations $\sigma_{P_\Delta}$.  From that point, it is
sometimes possible to analyze the flow states corresponding to these
tails, and in some cases trace the time evolution backward to determine
what flow configuration, and at what time, led to these events. While conceptually simple, this approach has several limitations
related to the rarity of events and the difficulty to sample
them. Moreover, following this approach it is not straightforward to
forecast the probability of an extreme event in the future given the
current flow state.

To address this predictive problem, we introduce the
\emph{committor}. We write a realization of the dynamics as $z(t)$,
with initial condition $z_0=z(t=0)$. We define $\mathcal{B}$ as the set
of states $z$ corresponding to an extreme event and $\mathcal{A}$ as the
set of states $z$ corresponding to normal conditions,
for instance using thresholds on one PDF tail (shaded in purple in
figure \ref{fig:PDF rare_events}) and around the mode (shaded in blue in
figure \ref{fig:PDF rare_events}), respectively. These two sets should
be disjoint. We then define the first-hitting times of the sets
$\mathcal{A}$ and $\mathcal{B}$ for a realization $z$ evolving in phase
space as\par
\begin{equation}
	\label{defhtimes}
	\begin{aligned}
		T_{\mathcal{A}}
		&\equiv \inf\left\{ t \ge 0 : z(t)\in\mathcal{A} \right\}, \\
		T_{\mathcal{B}}
		&\equiv \inf\left\{ t \ge 0 : z(t)\in\mathcal{B} \right\}.
	\end{aligned}
\end{equation}
These are simply the first times at which the system reaches either of
the two sets.

Although the underlying Navier-Stokes dynamics considered in this paper are deterministic, the phase-space variable $z$ used here is a coarse-grained representation built from only a finite number of physically relevant observables. Consequently, one point $z_0$ in this reduced phase space may correspond to multiple underlying flow states, which can evolve along different trajectories and reach $\mathcal{A}$ or $\mathcal{B}$ at different times. The probability in the committor therefore arises from this coarse-grained description. The committor is then defined as the probability of reaching set $\mathcal{B}$ before set $\mathcal{A}$, given the starting point $z_0$:\par
\begin{equation}
	\label{defcommgen}
	q(z_0)=P\left(T_{\mathcal{B}}<T_{\mathcal{A}} \mid z(t=0)=z_0\right).
\end{equation}
In the context of rare and extreme events, $z_0$ is often
close to $\mathcal{A}$, and the probability to reach the extreme set
$\mathcal{B}$ is small compared to $1$.  This probability has been an
object of study for more than eight decades in chemical kinetics
\citep{onsager1938initial} and the term committor has been coined about
twenty years ago
\citep{weinan2005transition,vanden2006towards,prinz2011efficient}.
This function can be defined in a deterministic (where it is either
$0$ or $1$) or in a probabilistic context (where it can span all
values between $0$ and $1$) \citep{lucente2022committor}.  If the
committor is known, we can then study extreme events \emph{a priori}.
This probability depends on the initial condition $z_0$, it can be
used to perform forecasts, and it is possible to construct new
composites and to perform averages of states ``$z_0$'' conditioned on
their probability to evolve into an extreme
\citep{miloshevich2023probabilistic}.  The features of these flow
states can be studied, to try to highlight precursors of given
extremes.

While the system we study is entirely deterministic, taking a
probabilistic approach by using the committor function is practically
simpler. Firstly, choosing to estimate a committor from a deterministic
point of view requires a fine scanning of all points in phase space
with minimal error bars on event properties; this is in contradiction
with our object of study, where there is by default a perturbation to
velocity fields. Secondly, if we make some additional reasonable
assumptions on our process, the committor (equation
(\ref{defcommgen})) becomes strongly structured because it follows a
given algebraic or differential equation.

Indeed, this is the case if we assume that the time evolution can be
modelled by a \emph{Markov chain}, that is to say a stochastic process
where each random change between $t$ and $t+\Delta t$ is solely
determined by the state of the system at $t$. This requires two first
reasonable approximations: the representation of the time evolution of
the system by a process discrete in time with step $\Delta t$ (larger
or equal to ${\rm d}t$, the time step of our simulations) as well as
rapid enough decorrelations in time. In that case, the starting point
for the definition and computation of the committor is the conjugate
probability $p(z(t)\,|\,z(t=0)=z_0)$ of being in a state $z(t)$ at some
time $t>0$ conditioned on having been in state $z_0$ at time $t=0$,
i.e. $z(t=0)=z_0$. Note that while $z$ is the independent variable for $p$, $z_0$ is
	the independent variable for $q$. This probability $p$
contains the information on all possible realizations of the process
at a given time.

In our discrete-in-time Markov chain, the advance in time of $p$ is
given by the application of a linear operator $G^\dagger$, i.e.\par
\begin{equation}
	p(t+\Delta t)=G^\dagger(p(t))\,,
	\label{CK}
\end{equation}
which is known as a Chapman--Kolmogorov equation
\citep{gardiner2009stochastic,Friedrich2011}. The operator $G^\dagger$ contains the
transition probabilities from any state $z$ to any other state $z'$
during a time $\Delta t$, and its action is expressed as the sum of
transition probabilities times probabilities (an actual sum for
discrete systems or an integral for continuous ones). We chose this
notation because the committor is then a steady solution of the affine
adjoint problem\par
\begin{equation}
	q=G(q)\,,\qquad
	q(z_0)=0\ \text{if}\ z_0 \in \mathcal{A}\,,\qquad
	q(z_0)=1\ \text{if}\ z_0 \in \mathcal{B}\,.
	\label{PbC}
\end{equation}
The two conditions directly follow from the definitions of the hitting
times (equation (\ref{defhtimes})) and of the committor (equation
(\ref{defcommgen})) \citep{prinz2011efficient}: if the process starts
in $\mathcal{A}$ (respectively $\mathcal{B}$), it stops at the first
step, $T_{\mathcal{A}}=0$, $T_{\mathcal{B}}>0$ (respectively
$T_{\mathcal{B}}=0$, $T_{\mathcal{A}}>0$), and
$\mathbb{P}(T_{\mathcal{B}}(z_{0})<T_{\mathcal{A}}(z_{0}))=0$
(respectively
$\mathbb{P}(T_{\mathcal{B}}(z_{0})<T_{\mathcal{A}}(z_{0}))=1$).

The proxies $\tilde{q}$ and $G^\dagger$ are interesting because it is
simple and moderately expensive to estimate them from data with enough
precision given an unconstrained, representative, sample of the
dynamics. Note that this adds the additional assumption that our
system is ergodic: sample means converge to ensemble averages and,
among other conditions, there are no isolated areas in phase space. In
particular, transition probabilities contained in $G$ can be estimated
from unconstrained realizations of the process. Conversely, a direct
sampling of the committor from data would require sampling many
realizations of the dynamics (until they reach $\mathcal{A}$ or
$\mathcal{B}$) for each value of $z_0$ \citep{lucente2022committor}.
Using that method, the estimate is the proportion of realizations that
reach $\mathcal{B}$ before $\mathcal{A}$, and the variance of the
estimated committor is at least the inverse of the number of
realizations started from $z_0$.

The final stage of approximation consists in discretising the phase
space in which our process $z(t)$ evolves. In studies such as the one
by \citet{lucente2022coupling}, the main reason for this discretisation
was numerical cost reduction. Indeed, the purpose of this study was to
estimate the committor for a system described by an explicit,
analytic stochastic differential equation. In that case, the direct
and adjoint of the exact linear operator $G$ were known. In the limit
$\Delta t \rightarrow 0$ the probabilities $\tilde{q}$ and $q$ were
respectively solutions of the direct Fokker--Planck equation and the
steady adjoint Fokker--Planck equation \citep{gardiner2009stochastic,Friedrich2011},
which are the time-continuous equivalents of equation (\ref{CK}) and
equation (\ref{PbC}). However, the random variable had up to six
degrees of freedom: the cost of the numerical resolution of the
Fokker--Planck equation was prohibitive, while the analytical
resolution was intractable. In the present case, the original system,
a flow governed by the incompressible Navier--Stokes equations, does
not follow a Markov chain. This Markov chain has to be created, and
creating this discrete process is one of the possible approaches. This
is our last stage of modelling: describing our system in a
probabilistic manner, assuming that it is Markovian with a time step
$\Delta t$, and finally reducing its dimension to a finite number of
states.

After that stage of discretisation, our process can now visit only a
finite number $N$ of discrete states. As a consequence, the committor
$\mathbf{q}$ and the probability $\mathbf{p}$ are now vectors in
$\mathbb{R}^N$ and $\mathbf{G}$ and $\mathbf{G}^\dagger$ are $N\times N$
matrices. Each component $q_j$ and $p_j$ corresponds to a state $z_j$,
$1\le j \le N$. Meanwhile the entries $G_{nj}$ of $\mathbf{G}^\dagger$
are the transition probabilities from state $n$ to state $j$. The
Chapman--Kolmogorov equation~(\ref{CK}) for advancing $\mathbf{p}$ in
time then reads\par
\begin{equation}
	p_j(t+\Delta t)=\sum_{n=1}^N G_{nj}p_{n}(t)\,,\qquad
	G_{nj}\ge 0\,,\qquad
	\sum_{n=1}^N G_{nj}=1\,,\qquad
	1\le j \le N\,,
	\label{emat}
\end{equation}
where the transition probability follows a normalization condition,
because the system necessarily transits somewhere. Thus, if we are
able to estimate from data the transition probabilities contained in
$G_{nj}$, through one representation of the process or another, we can
then obtain an estimate of the committor on the discrete states by
solving equation~(\ref{PbC}). Using that estimate, it may then be
possible to interpolate the committor on any state, for instance using
an approach derived from the method used to discretise the phase
space.

\subsection{Construction of the analogue Markov chain}\label{amc}

In this section, we explain how the phase space is discretised in practice to construct the $N$ states on which our analogue Markov chain evolves, and how the corresponding transition probabilities are estimated from DNS data. The resulting process is termed an analogue Markov chain, and it provides the basis for the committor computation presented in the next section.

\begin{figure}
	\centering
	\includegraphics[width=0.6\textwidth]{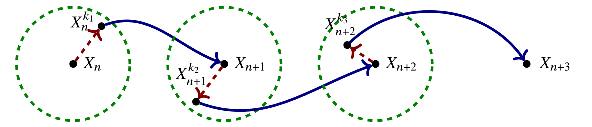}	
	\includegraphics[width=0.6\textwidth]{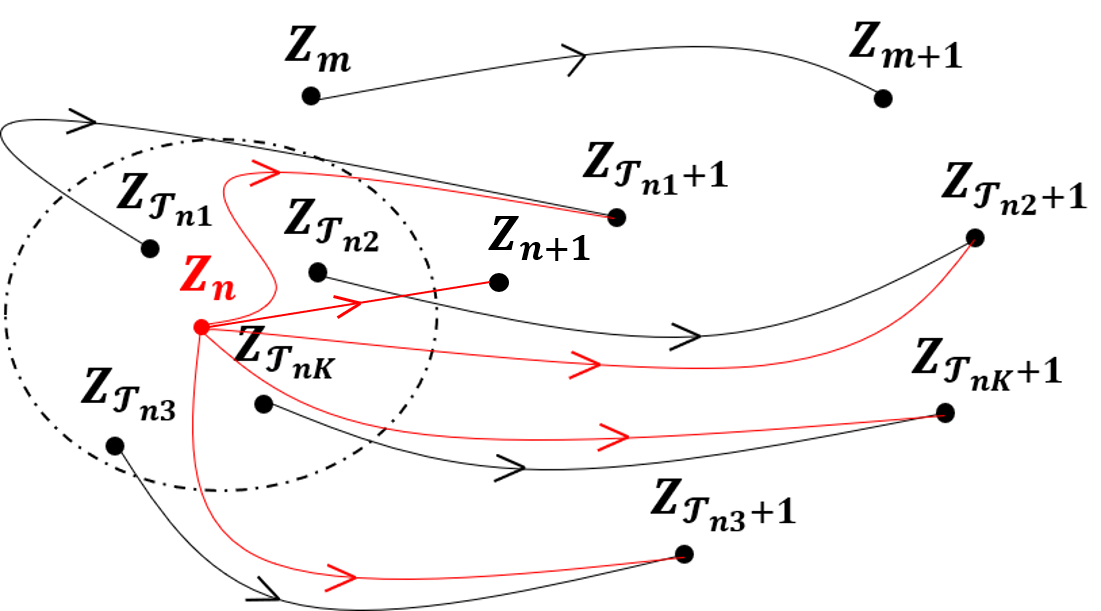}
	\caption{Top: Schematic of the analogue method to predict a deterministic evolution (figure reproduced from \citet{miloshevich2024extreme}).
		Bottom: Schematic of the analogue Markov chain method taking $K=5$ as an
		example. On the left-hand side, the event $\boldsymbol{Z}_{n}$ (shown in red and not corresponding to the event at the final instant of any time series) is surrounded by its $K$ nearest events
		$\left\{\boldsymbol{Z}_{\mathcal{T}_{nk}}\right\}_{1\leq
			\mathcal{T}_{nk} \leq N_{t}-1}, 1\leq k\leq K$ including itself, as indicated within the circle. On the right hand side are their subsequent
		observation events and the transition to these subsequent
		observations is denoted by black lines. The red lines are
		the possible transitions in the analogue Markov chain. The
		schematic is adapted from figure 3 of \citet{lucente2022coupling}. The observations
		$\boldsymbol{Z}_{m}$ and $\boldsymbol{Z}_{m+1}$, are added here to represent
		events outside of the neigbourhood of $\boldsymbol{Z}_{n}$. }
	\label{fig:schematic}
\end{figure}

The method that we follow to construct our effective Markov dynamics
on discrete states is based on the method of analogues proposed by
\citet{lorenz1969atmospheric,lorenz1969three} for simple atmospheric
predictability. The original approach is deterministic and is based on
the observation that, up to a predictability time horizon, similar
initial conditions will remain close as they evolve in time. Thus, if
one has sampled a long enough time series of a process $\{ x_a(t)\}$,
the most basic formulation of the analogue method to forecast the
future $x(t+\Delta t)$ of a new observation $x(t)$ is to search for
the closest state, its nearest neighbor, according to a norm in phase
space, among the so-called analogues $x_a(t)$. The prediction is then
$x_a(t+\Delta t)\simeq x(t+\Delta t)$ (figure \ref{fig:schematic},
top). In order to account for the chaotic nature of the turbulent flow
and to propose a process from which we can estimate a committor, we
use a variant of the analogue method, termed the analogue Markov
chain. In this method, the process does not deterministically evolve
along the same path as its nearest neighbor among the analogues.
Instead, at each time step, it is allowed to jump at random to one out
of $K$ states: the jump is chosen among the $K$ time evolutions of the
$K$ nearest neighbors of our state (figure \ref{fig:schematic},
bottom). This creates a Markov chain in the state space comprising the
observed analogues.

In order to construct this analogue Markov chain, we first need to
define the state space where it takes place, which is constructed from
data observed in our turbulent flow. For this matter we sample
\emph{predictors}: vectors $\boldsymbol{Z}$ in the phase space
$\boldsymbol{\chi}\subset\mathbb{R}^{D}$. Each of the $D$ components of
$\boldsymbol{Z}$ is given by the value of a predictor field at a given
spatial point ${\bf x}$, at a given instant $t$. Actually, the main
challenge in this paper is the choice of predictor fields in the first
place. Thus, in this paper we consider more than one spatial point
${\bf x}$, say a number $N_x$ of such points, where the predictor
fields are calculated. At each one of these points we sample
uninterrupted time series of $N_t$ samples of $\boldsymbol{Z}$ with
time separation $\Delta t$, i.e. for $t=0, \Delta t, 2\Delta t,
3\Delta t, ..., (N_{t}-1)\Delta t$ (where $t=0$ stands for an
arbitrary initial time). Our state space is constituted by the
resulting $\left\{\boldsymbol{Z}_{n}\right\}_{1\leq n \leq N_{t} N_x}$, concatenated as $N_x$ successive time series. For all $1\le m\le N_x$,
$\boldsymbol{Z}_{l+1+(m-1) N_t}$ comprises the $D$ predictors sampled at
points ${\bf x}_m$ and $\boldsymbol{Z}_{n}$ at instants $l\Delta t$ for $0\le l\le
N_{t}-1$.

We now present the construction of the transition matrix
$\mathbf{G}^\dag$ of the analogue Markov chain from the series of
observed events $\boldsymbol{Z}_{n}\in\left\{\boldsymbol{Z}_{n}\right\}_{1\leq n \leq N_{t}
	N_x}$ at the $N_x$ spatial points. In our analogue Markov chain,
each state $\boldsymbol{Z}_{n}$ can follow the transition of each of
its $K$ nearest neighbors with equal probability (see figure
\ref{fig:schematic}, bottom, for $K=5$). The nearest neighbors are
determined according to their closeness in Euclidean norm to
$\boldsymbol{Z}_{n}$ in phase space among all
$\boldsymbol{Z}_{l+1 + (m-1)N_t}$ with $1\le m\le N_x$ and
$0 \le l \le N_t-1$ (note that, in the nearest-neighbor search, we
exclude the $\boldsymbol{Z}$ at the final instant of every time series
corresponding to every point ${\bf x}_m$ so as to ensure that each of
the $K$ neighbors has a recorded evolution in state space). We denote
by $\mathcal{T}_{nk}$ the indices of the $K$ nearest neighbors
($1\le k\le K$) of $\boldsymbol{Z}_{n}$, which means that the
$\boldsymbol{Z}_{n'}$ with $n'=\mathcal{T}_{nk}$ realise the $K$
smallest values of the Euclidean norm
$\parallel \mathbf{Z}_n-\mathbf{Z}_{n'}\parallel$. Since each
$\boldsymbol{Z}_{\mathcal{T}_{nk}}$ evolves into
$\boldsymbol{Z}_{\mathcal{T}_{nk}+1}$ in the recorded time series, in
our analogue Markov chain $\boldsymbol{Z}_{n}$ can transit to any of
the $\boldsymbol{Z}_{\mathcal{T}_{nk}+1}$ with probability $1/K$ and
does not transit to any other state. Given that rule, we can write all
the entries of the $(N_tN_x)\times(N_tN_x)$ transition matrix
$\boldsymbol{\mathsf{G}}^\dag$ as follows\par
\begin{equation}
	\label{eq:Transition}
	G_{nj}^\dag=\left\{
	\begin{array}{ccl}
		1/K       &      &\text{if } \exists\, k\in[1,K] \text{ such that } j=\mathcal{T}_{nk}+1,\\
		0         &      &\text{otherwise.}
	\end{array} \right.
\end{equation}
In this article, we use this transition matrix as a basis to compute
the committor in our state space, as we detail in the next section.

\subsection{Computing the committor function from the analogue Markov chain \label{sec:Computing the committor function from the analogue Markov chain}}\label{affine}

Now that we have presented the construction of the transition matrix
$\tilde{\boldsymbol{\mathsf{G}}}$ of our analogue Markov chain. We first explain how $\tilde{\boldsymbol{\mathsf{G}}}$ is modified compared to its original form $\boldsymbol{\mathsf{G}}^\dag$ in order to get the committor.  We then detail how we use this transition matrix
$\tilde{\boldsymbol{\mathsf{G}}}$ of analogue Markov chain to compute the committor \citep{prinz2011efficient,lucente2022coupling}.

A series of $\mathbf{Z}$ that are representative enough for a
committor computation by our method necessarily contain a number
$N_{\mathcal{A}}>0$ of instants $n$ where $\mathbf{Z}_n$ is in
$\mathcal{A}$ and a number $N_{\mathcal{B}}>0$ of instants where
$\mathbf{Z}_{n'}$ is in $\mathcal{B}$. In the initial Markov chain,
the system can enter these states by following recorded transitions
but can also exit them in the same way. However, if we wish to
compute a committor $\mathbf{q}$, the dynamics must be allowed to
enter either of these two sets but not to exit them: they must be
absorbing. We therefore modify the transition matrix and create an
auxiliary Markov chain where the two sets are strictly absorbing. A
simple way to modify our process consists in changing the rows of
$\mathbf{G}$ corresponding to the states $\mathbf{Z}_n \in
\mathcal{A}\cup\mathcal{B}$ so that each of these states transits only
to itself. We thus create a new transition matrix
$\tilde{\mathbf{G}}$ such that\par
\begin{align}
	\label{eq:transition matrix G}
	\notag \tilde{G}_{nj}=G_{nj}\,,\, 1\le j \le N_{x} N_t
	\,\text{if}\,\mathbf{Z}_n \notin \mathcal{A}\cup
	\mathcal{B}\,,\\
	\tilde{G}_{nn}=1\,\text{and}\, \tilde{G}_{nj}=0 \,,\,
	1 \le j \le N_{x} N_t\,,\, j\ne n \,\text{if}\,\mathbf{Z}_n \in
	\mathcal{A}\cup \mathcal{B}\,.
\end{align}
Note that this modification respects the normalization condition of
equation~(\ref{emat}).

For our Markov chain, we can define the committor as a vector
$\tilde{\mathbf{q}}\in\mathbb{R}^{N_{x} N_t}$, each of its
components $q_n$ being the committor value associated with state
$\mathbf{Z}_n$. For our finite-dimensional analogue Markov chain
process, a time-invariant committor statisfying the backward Chapman--Kolmogorov equation (\ref{emat}) then reads\par
\begin{equation}
	\label{eq:affine problem}
	\sum_{j=1}^{N_tN_x}\tilde{G}_{ij}q_j=q_i\,,\,1\le i\le N_xN_t\,,\,
	q_n=0\,\text{if}\, \mathbf{Z}_n \in{\mathcal{A}}\,,\,
	q_{n'}=1\,\text{if}\, \mathbf{Z}_{n'} \in{\mathcal{B}}\,.
\end{equation}
Following \citet{prinz2011efficient,lucente2022coupling}, we solve (\ref{eq:affine problem}) in two steps: first by computing the eigenvectors of
$\tilde{G}$ for its largest eigenvalue, and secondly by expressing $q$
as a linear combination of these eigenvectors. Perron--Frobenius
theorems indicate that the matrix $G$, which is non-negative (zero or
positive entries) and irreducible (it conserves probability)
(equation~(\ref{emat})), has a single eigenvalue equal to $1$, its spectral
radius. The corresponding eigenspace therefore has dimension one. The
same properties are found for $G^\dag$. Because of the modifications
applied to turn $G$ into $\tilde{G}$ (turning $\mathcal{A}$ and
$\mathcal{B}$ into absorbing states), the spectral radius of
$\tilde{G}$ remains $1$; however, the eigenvalue $1$ is no longer
simple: its corresponding eigenspace has dimension
$N_{\mathcal{A}}+N_{\mathcal{B}}$
\citep{prinz2011efficient,lucente2022coupling}. Since $\tilde{G}q=q$,
the committor vector lies in that eigenspace. If we term
$\{\mathbf{v}_m\}_{1\le m\le N_{\mathcal{A}}+N_{\mathcal{B}}}$ a basis
of this eigenspace, there exists a unique collection of $M$ scalars
$\alpha_m$ such that\par
\begin{equation}
	\mathbf{q}=\sum_{m=1}^M \alpha_m\mathbf{v}_m\,.
	\label{dqa}
\end{equation}
These scalars are in turn the solution of the system of equations
given by the $N_{\mathcal{A}}$ conditions\par
\begin{equation}
	0=q_n=\sum_{m=1}^M \alpha_m v_{n,m}\,,
	\qquad \text{for } \mathbf{Z}_n \in\mathcal{A}\,,
	\label{eqa}
\end{equation}
and the $N_{\mathcal{B}}$ conditions\par
\begin{equation}
	1=q_{n'}=\sum_{m=1}^M \alpha_m v_{n',m}\,,
	\qquad \text{for } \mathbf{Z}_{n'} \in\mathcal{B}\,,
	\label{eqb}
\end{equation}
where $v_{n,m}$ is the component $n$ of eigenvector $\mathbf{v}_m$.
Thus, in our python implementation, we first compute a number of
eigenvectors for the largest eigenvalues, strictly larger than $M$, so
as to be able to span the eigenspace of $\tilde{\mathbf{G}}$ for
eigenvalue $1$. From the eigenvectors associated with eigenvalue $1$,
we construct the system of $N_{\mathcal{A}}+N_{\mathcal{B}}$
equations (equations~(\ref{eqa}) and (\ref{eqb})) and solve it. Using
the computed scalars, we then construct our committor vector
(equation~(\ref{dqa})) defined on the states of the analogue Markov
chain.

Note that, in principle, all states in $\mathcal{A}$ and all states in $\mathcal{B}$ could be further aggregated into two internally mixed macrostates. In that case, a reduced transition matrix compared with (\ref{eq:transition matrix G}) could be constructed. The multiplicity associated with the eigenvalue $1$ would then decrease from $N_{\mathcal{A}}+N_{\mathcal{B}}$ to $2$, which would lower the dimension of the corresponding linear system and simplify the computation. Since the committor only depends on which set is reached first, such an aggregation would preserve the same committor. In the present work, however, we retain the original formulation in order to remain consistent with the current implementation of the transition matrix and the associated event indexing.

Note that this result also provides a method to estimate the value of
the committor for any new state $\mathbf{Y}\in\chi$ sampled in the
flow. According to the principle of the representation of the
dynamics by an auxiliary analogue Markov chain, we can start a new
realization of the chain starting from $\mathbf{Y}$. It can transit
toward the future step of any of its $K$ nearest neighbors among the
$N_{t} N_x$ sampled analogues $\mathbf{Z}_n$, drawn uniformly. The
chain then proceeds until it is absorbed by either $\mathcal{A}$ or
$\mathcal{B}$. Thus, the probability that a trajectory starting from
$\boldsymbol{Y}$ reaches $\mathcal{B}$ before $\mathcal{A}$ is given by
the arithmetic average of the committor values over the $K$ nearest
neighbors of $\boldsymbol{Y}$. Here, the out-of-sample event $\boldsymbol{Y}$ is an isolated query event, so transitions from in-sample events to $\boldsymbol{Y}$ are not admissible, and the original transition structure and committor values remain unchanged. If we denote the indices of
these $K$ states by $\{\mathcal{T}_{k}(\boldsymbol{Y})\}_{1 \leq k \leq K}$,
an estimate for the committor on $\mathbf{Y}$ is thus given by the arithmetic
average\par
\begin{equation}
	q(\boldsymbol{Y}) = \frac{1}{K}
	\sum_{k=1}^{K} q_{\mathcal{T}_{k}(\boldsymbol{Y})+1}\,.
\end{equation}

In the present work, we set $K=100$ but we also tried $K=50$ and
$K=150$ without qualitative change of conclusions. The set
$\left\{\boldsymbol{Z}_{n}\right\}_{1\leq n \leq N_{x}N_{t}}$ is
sampled from the DNS data described in section 4. The sampling
details are presented in section 4.3, where the definitions of the
sets $\mathcal{B}$ and $\mathcal{A}$ are also provided.

\subsection{Evaluation of the quality of an approximate committor function}\label{brier}
Following \citet{lucente2022coupling}, we use the Brier score to evaluate the precision of the obtained approximate committor
function. It was originally proposed by
\citet{brier1950verification} to estimate the precision of weather forecasts.
First, we define $q_{B}$ as a binary
verification function such that $q_{B}(\boldsymbol{Y})=1$ if
$T_{\mathcal{B}}(\boldsymbol{Y})<T_{\mathcal{A}}(\boldsymbol{Y})$, and
$q_{B}(\boldsymbol{Y})=0$ if
$T_{\mathcal{B}}(\boldsymbol{Y})>T_{\mathcal{A}}(\boldsymbol{Y})$. The accurate
committor function is represented by $\hat{q}$. The
outcomes of $q_{B}$ have a Bernoulli distribution, where
$\mathbb{P}(q_{B}(\boldsymbol{Y}) = 1) = \hat{q}(\boldsymbol{Y})$ and
$\mathbb{P}(q_{B}(\boldsymbol{Y}) = 0) = 1 - \hat{q}(\boldsymbol{Y})$.
The Brier score is defined as\par
\begin{equation}
	\label{eq:brier function}
	BT_{N}=\frac{1}{N}\sum_{i=1}^{N}\left(q(\boldsymbol{Y}_{i})-q_{B}(\boldsymbol{Y}_{i})\right)^{2},
\end{equation}
where $N$ is the number of events used for testing the performance of
the approximate committor function. The Brier score therefore takes
values between $0$ and $1$. Assuming that the sampled process in the set
$\{\boldsymbol{Y}_{i}\}$ is ergodic in phase space, we have the equivalence
between the arithmetic mean and the ensemble mean, i.e.\par
\begin{equation}
	\label{eq:eception brier function}
	\lim_{N\to\infty}BT_{N}=\mathbb{E}(BT_{N})=\left\Vert \sqrt{\hat{q}(1-\hat{q})}\right\Vert_{\rho}^{2}+\left\Vert \hat{q}-q\right\Vert_{\rho}^{2},
\end{equation}
where $\left\Vert
f\right\Vert_{\rho}^{2}=\int_{\chi}f^{2}(\boldsymbol{Y})\rho(\boldsymbol{Y})\mathrm{d}\boldsymbol{Y}$
is the norm weighted according to the stationary distribution function
$\rho(\boldsymbol{Y})$. The first term in equation (\ref{eq:eception brier
	function}) depends only on the nature of the system and is
independent of the quality of the approximate committor function; it
is a fixed lower bound resulting from the probabilistic nature of the
prediction. The second term in equation (\ref{eq:eception brier
	function}) represents the quadratic deviation of the approximate
committor function from the accurate one. The closer $q$ is to
$\hat{q}$, the lower the Brier score is.

In the present work, this point is particularly important because each
choice of predictor defines a different coarse-grained phase space.
Therefore, the Brier score does not only evaluate the quality of the
approximate committor itself, but also the amount of relevant
information retained by the chosen predictor. A lower Brier score
means that the selected predictor, or equivalently the reduced phase-space
representation built from it, is more informative about the future
occurrence of extreme uncertainty-production events. In this sense,
comparing the Brier scores obtained from different predictors provides
a quantitative way to assess how closely the corresponding
coarse-grained phase spaces capture the dynamically relevant structure
of the original Navier-Stokes phase space.

\section{DNS configurations and data sampling\label{sec:DNS configurations and data sampling}}

In this section we introduce the DNS and the post-processing
statistical methods, starting in section \ref{sec:DNS numerical setup}
with the DNS setup and its results in section \ref{sec:DNS
	results}. The simulated stationary periodic flow is sampled to
construct the analogue Markov chain and estimate the committor
function. The sampling setup, the definition of the extreme events and
the method of testing the extracted committor function are given in
section \ref{sec:Statistical analysis setup}

\subsection{DNS numerical setup\label{sec:DNS numerical setup}}
To study the growth of average uncertainty energy in three-dimensional
periodic/homogeneous turbulence, we use a fully de-aliased
pseudo-spectral code to perform DNS of forced incompressible
Navier-Stokes turbulence in a periodic box of size $\mathcal{L}=2\pi$
in all three normal directions. Time advancement is achieved with a
second-order Runge-Kutta scheme. We use the same adaptive time step
${\rm d}t$ for both $\mathbf{u}^{(1)}$ and $\mathbf{u}^{(2)}$ as
follows. The Courant-Friedrichs-Lewy (CFL) condition, with CFL number
equal to $0.4$, gives a time step ${\rm d}t^{(1)}$ when applied to
$\mathbf{u}^{(1)}$ and ${\rm d}t^{(2)}$ when applied to
$\mathbf{u}^{(2)}$. We set ${\rm d} t=\min({\rm d}t^{(1)},{\rm
	d}t^{(2)})$. The code strategy is detailed in
\citet{vincent1991spatial} and has been used in our previous study
\citep{ge2023production,ge2025interscale,ge2025eulerian}. We first generate a statistically stationary
reference flow. To generate the reference flow we use a von Kármán
initial energy spectrum with the same coefficients as
\citet{https://doi.org/10.48550/arxiv.1306.3408} and random initial
Fourier phases. We integrate the reference flow until it reaches a
statistically stationary state and then seed it with small
perturbations to create the perturbed flow at a time which we refer to
as $t_{0}=0$. The reference flow and the perturbed flow are simulated
in parallel using the same time step. The key parameters of the
reference flow are given in Table \ref{tab:main parameters}.

	The generation of the perturbed flow is detailed in
\citet{ge2023production}, where\par
\begin{equation}
	\hat{\boldsymbol{u}}^{(2)}(\boldsymbol{k},t_0)=\left\{
	\begin{array}{ccc}
		\hat{\boldsymbol{u}}^{(1)}(\boldsymbol{k},t_0)&       &\text{if } \left|\boldsymbol{k}\right|< k_{0}, \\
		\text{Randomly generated}&       &\text{if } \left|\boldsymbol{k}\right|\geq k_{0}.
	\end{array}\right.\label{const_dec}
\end{equation}
 We denote respectively by $\hat{\boldsymbol{u}}^{(1)}$ and
$\hat{\boldsymbol{u}}^{(2)}$ the velocity fields of the
reference flow and the perturbed flow in Fourier space. In the
present work, we set $k_{0}\eta=1.5$, the same as in our
previous work \citep{ge2023production}. We use this perturbation strategy in order to keep the perturbed flow as close as possible to the reference flow. Specifically, the perturbed field is constructed so that the one-dimensional energy spectrum is preserved at the initial time $\hat{E}^{(1)}(k,t_{0})=\hat{E}^{(2)}(k,t_{0})$ for all resolved wavenumbers. In our previous works \citep{ge2025eulerian}, we have shown that this way of constructing the perturbation does not affect the subsequent uncertainty-growth rate when compared with the more standard approach of directly adding a white-noise perturbation. We have also shown that the uncertainty-growth dynamics is insensitive to both the amplitude of the initial perturbation and the range of perturbed wavenumbers \citep{ge2023production}.

In the present work, we consider two different types of external forcing. For each forcing type, we test two Reynolds-number configurations. The first one corresponds to a Reynolds number chosen to be as low as possible within the present framework. This allows us to maintain a comparatively high spatial resolution and, at the same time, to obtain a relatively long chaotic exponential-growth regime of the uncertainty energy. Such a long growth regime is advantageous for the present statistical analysis because it provides long continuous time series and therefore a large number of temporal samples $N_t$. The second configuration corresponds to a Reynolds number chosen to be as high as possible within our computational limits. The purpose of these higher-Reynolds-number cases is to reduce the direct influence of the forcing on the smaller scales and thereby to better isolate the inertial effects of interest. Because increasing the Reynolds number also shortens the chaotic exponential-growth regime, these higher-Reynolds-number simulations do not provide time series as long as those of the lower-Reynolds-number cases. To compensate for this reduced temporal extent and still obtain sufficient sampling, we increase the number of observation windows $N_x$. The corresponding sampling details are discussed in subsection \ref{sec:Statistical analysis setup}.

One type of forcing, referred to as case F1, is a negative-damping force applied exclusively at low wavenumbers, a method commonly used in previous studies \citep{boffetta2017chaos,berera2018chaotic,ge2023production}. Here, ``negative damping'' refers to the fact that the forcing is aligned with the large-scale velocity modes by injecting energy into them, whereas a damping force is anti-aligned (opposed) to the velocity \citep{linkmann2015sudden,mccomb2015self}. Therefore, the forcing is expressed as\par
\begin{equation}
	\label{eq:negative damping forcing 2}
	\hat{\boldsymbol{f}}^{(2)}\left(\boldsymbol{k},t\right)=\hat{\boldsymbol{f}}^{(1)}\left(\boldsymbol{k},t\right)=\left\{
	\begin{array}{ccc}
		\frac{\varepsilon_{0}}{2E_{f}^{(1)}}\hat{\boldsymbol{u}}^{(1)}\left(\boldsymbol{k},t\right)&       &\text{if } 0<\left|\boldsymbol{k}\right|\leq k_{f},\\
		0&       &\text{otherwise,}
	\end{array}\right.
\end{equation}
where $\varepsilon_{0}=0.1$ and $k_{f}=2.5$.

The other forcing type, referred to as case F2, is a single-mode divergence-free force that remains constant in time:\par
\begin{equation}
	\label{eq:negative damping forcing 3}
	\boldsymbol{f}^{(2)}\left(\boldsymbol{x},t\right)=\boldsymbol{f}^{(1)}\left(\boldsymbol{x},t\right)=f_{0}\left(
	\begin{array}{ccc}
		\cos\left(k_{0}y\right)\sin\left(k_{0}z\right)\\
		\cos\left(k_{0}z\right)\sin\left(k_{0}x\right)\\
		\cos\left(k_{0}x\right)\sin\left(k_{0}y\right)\\
	\end{array}\right),
\end{equation}
where $f_{0}=0.5$ and $k_{0}=2$.

\begin{table}
	\begin{center}
		\def~{\hphantom{0}}
		\begin{tabular}{cccccccccc}
			Case&$N^{3}$&$\nu$&$\varepsilon$&$ U$&$ L$&$ T_{0}$&$\text{Re}$&$\text{Re}_{\lambda}$&$ k_{\max}\eta$\\ [3pt]
			F1$_{128}$&$128^{3}$&0.0378&0.0999&0.5153&1.6806&3.2611&22.9487&16.7970&6.41\\
			F1$_{512}$&$512^{3}$&0.0010&0.0971&0.6038&1.066&1.7646&643.6905&143.7088&1.71\\
		    F2$_{128}$&$128^{3}$&0.0300&0.1466&0.4748&1.087&2.2924&17.1952&13.2071&4.90\\
			F2$_{512}$&$512^{3}$&0.0015&0.2171&0.6501&0.6473&0.9976&280.2203&90.7673&1.90\\
		\end{tabular}
	\caption{Parameters of the reference flows, where all the parameter here is presented after averaged in time when the
	flows are statistically stationary. The large eddy turnover time $T_{0}^{(1)}$ is
	$T_{0}\equiv L/U$ for the reference flow where
	$U=\sqrt{2\left\langle E\right\rangle/3}$ is the rms
	velocity and $L=\left(3\pi/4\left\langle
	E\right\rangle\right)\int k^{-1}\hat{E}(k)\rm{d}$$k$
	is the integral length scale. $N$ is the resolution
	of the simulations, $\nu$ is the kinematic
	viscosity, $\varepsilon$ is the turbulence
	dissipation rate. $\text{Re}= UL/\nu$ is the
	Reynolds number based on $L$ and
	$\text{Re}_{\lambda}\equiv U\lambda/\nu$ is the
	Reynolds number based on the Taylor length
	$\lambda\equiv\sqrt{10\left\langle
		E\right\rangle\nu/\varepsilon}$. $k_{\max}=N/3$
	is the maximum resolvable wavenumber and
	$\eta=\left(\nu^{3}/\varepsilon\right)^{1/4}$
	is the Kolmogorov length.}
\label{tab:main parameters}
	\end{center}
\end{table}
\subsection{DNS results\label{sec:DNS results}}
\begin{figure}
	\centering \subfigure[]{
		\label{fig:growthcertaintyenergy}
		\includegraphics[width=0.48\textwidth]{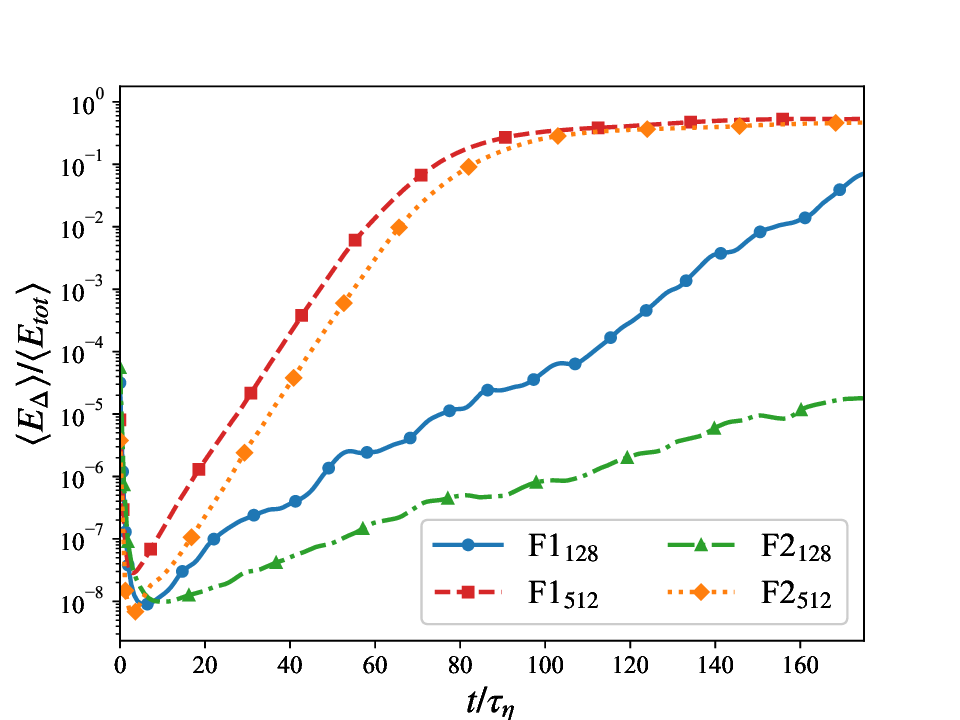}}
	\subfigure[]{
		\label{fig:gamma}
		\includegraphics[width=0.48\textwidth]{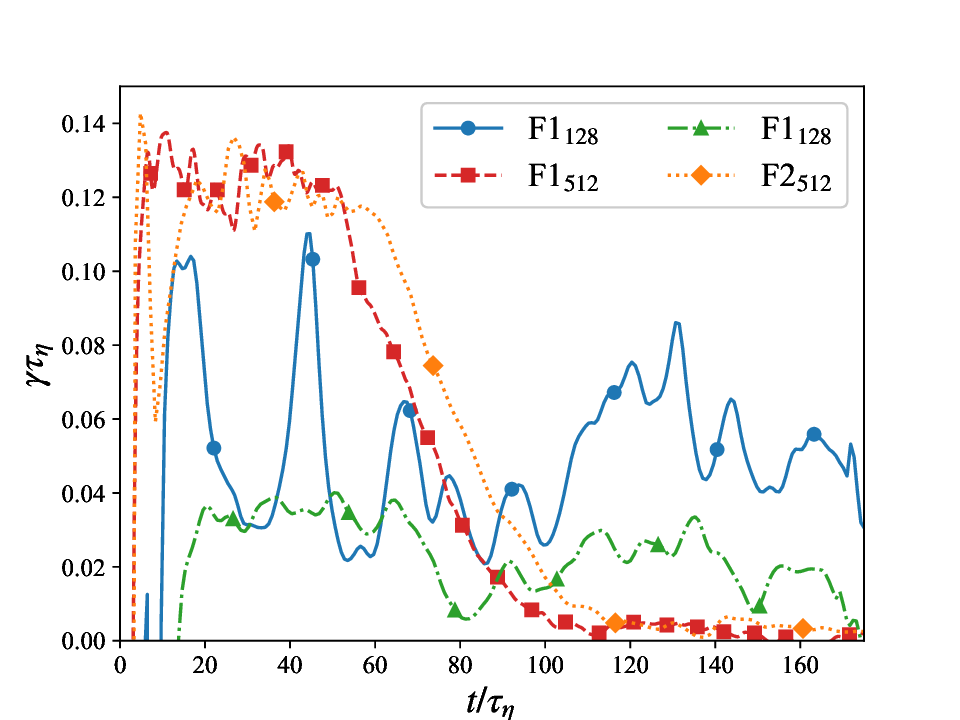}}
	\caption{Time evolution of (a) the average uncertainty energy in a semilogarithmic plot and (b) the corresponding growth exponent in a linear plot. $\gamma \equiv \frac{1}{2}\,\mathrm{d}\ln\langle E_{\Delta}\rangle/\mathrm{d}t$, $\tau_{\eta}$ denotes the Kolmogorov time scale, and $\left\langle E_{tot}\right\rangle=\left\langle \left|\boldsymbol{u}^{(1)}\right|^{2}/2+\left|\boldsymbol{u}^{(2)}\right|^{2}/2 \right\rangle$ is the total mean kinetic energy.}
	\label{fig:time evolution of uncertinty}
\end{figure}

\begin{figure}
	\centering \subfigure[]{
		\label{fig:pdf f1 128}
		\includegraphics[width=0.48\textwidth]{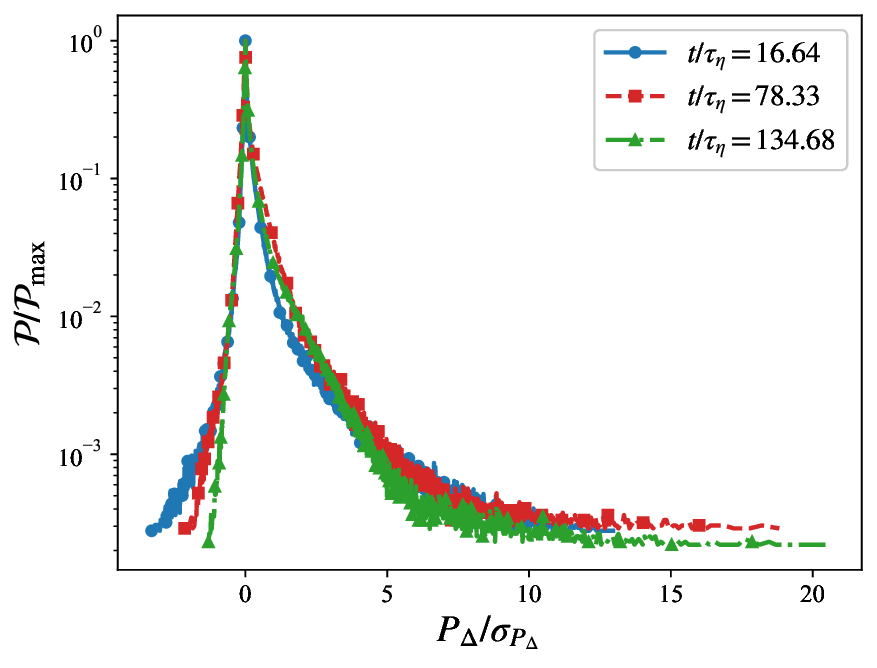}}
 \subfigure[]{
	\label{fig:pdf f1 512}
	\includegraphics[width=0.48\textwidth]{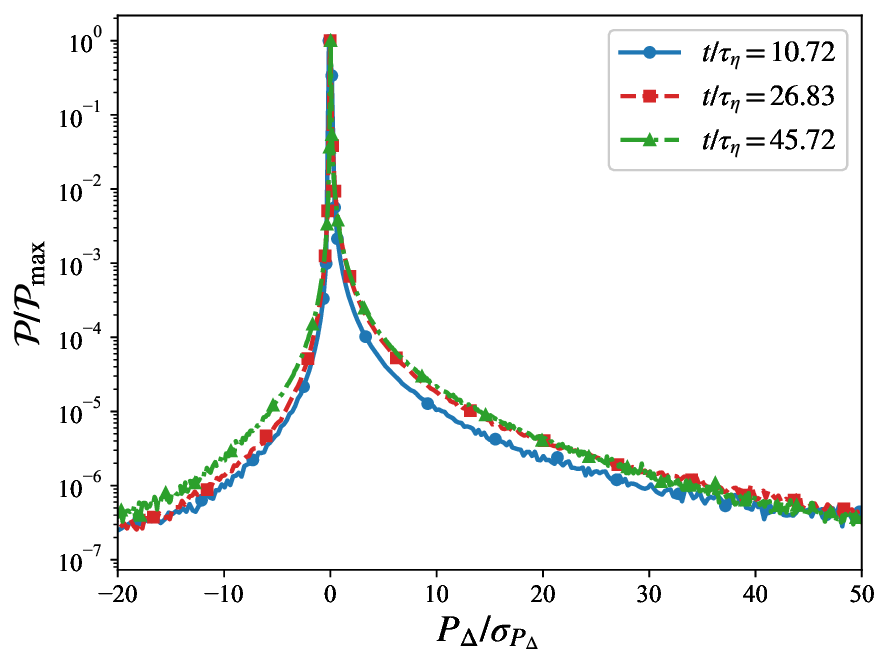}}
	\subfigure[]{
		\label{fig:pdf f2 128}
		\includegraphics[width=0.48\textwidth]{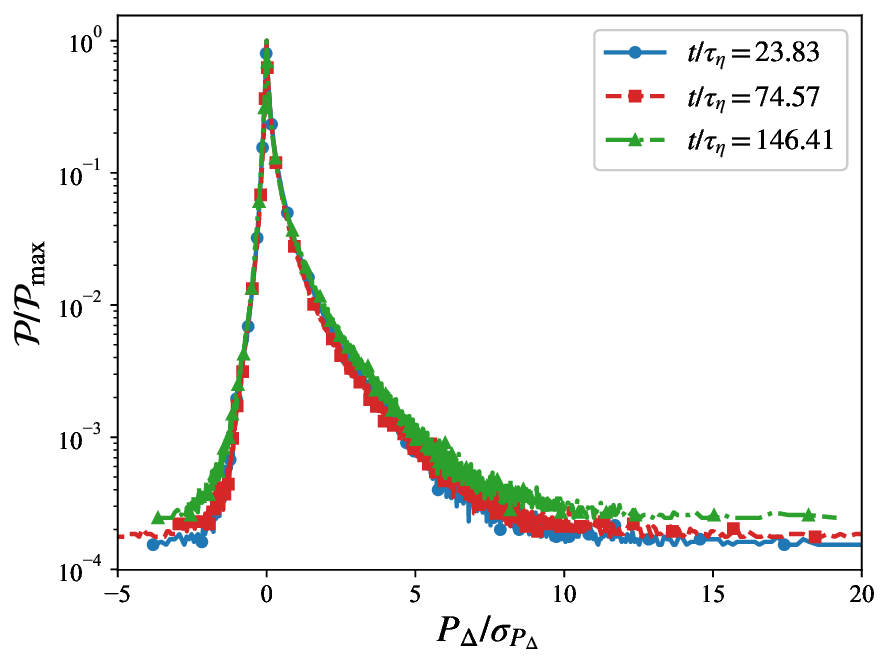}}
	\subfigure[]{
	\label{fig:pdf f2 512}
	\includegraphics[width=0.48\textwidth]{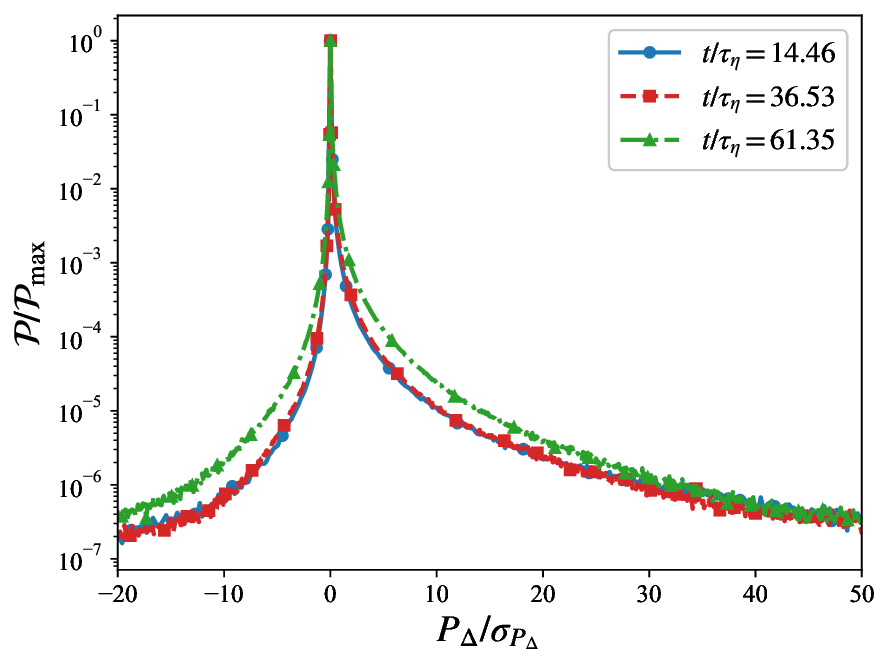}}
	\caption{Time evolution of PDFs of $P_{\Delta}$ for cases (a)
		F1$_{128}$, (b) F1$_{512}$ and (c) F2$_{128}$ and (d) F2$_{512}$. Similarly to \citet{ge2023production}, PDFs
		are normalized by their maximum $\mathcal{P}_{\max}$ and are
		plotted versus $P_{\Delta}/\sigma_{P_{\Delta}}$, where
		$\sigma_{P_{\Delta}}$ is the standard deviation of
		$P_{\Delta}$ with
		$\sigma^{2}_{P_{\Delta}}=\int_{P_{\Delta\text{min}}}^{P_{\Delta\text{max}}}(P_{\Delta}-\left\langle
		P_{\Delta}\right\rangle)^{2}\mathcal{P}(P_{\Delta}){\rm
			d}P_{\Delta}$.}
	\label{fig:pdf}
\end{figure}

\begin{figure}
	\centering
	\includegraphics[width=0.6\textwidth]{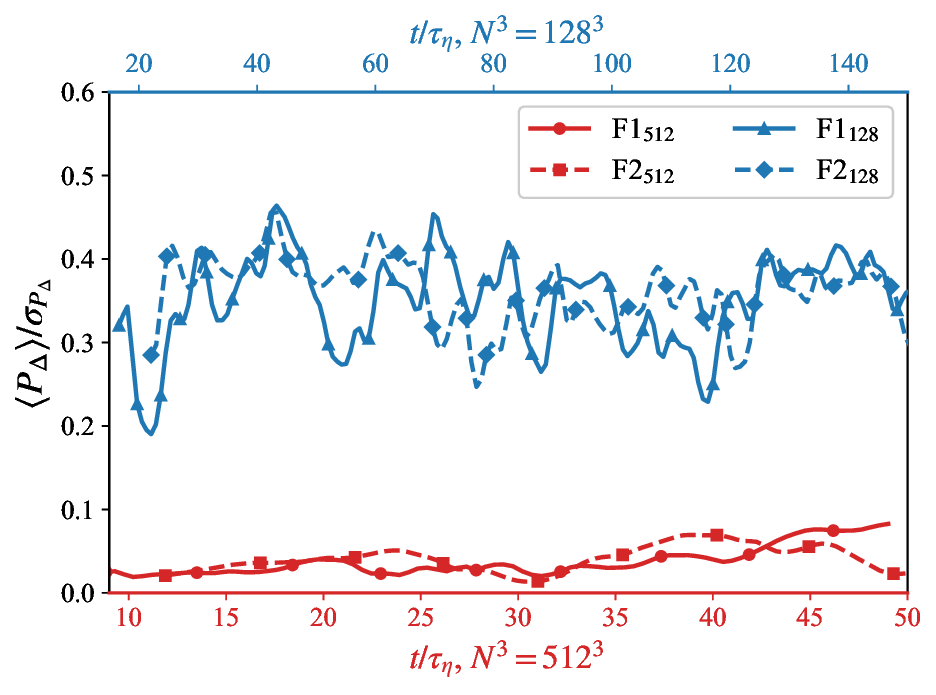}
	\caption{Time evolution of the average uncertainty production, normalized by its standard deviation. The top blue $x$-axis corresponds to cases F1$_{128}$ and F2$_{128}$, whereas the bottom red $x$-axis corresponds to cases F1$_{512}$ and F2$_{512}$.}
	\label{fig:NormalizedmeanProduction}
\end{figure}

Figure \ref{fig:time evolution of uncertinty} presents the time
evolution of $\langle E_{\Delta} \rangle$, where $\tau_{\eta}\equiv\left(\nu/\varepsilon\right)^{1/2}$ denotes the Kolmogorov time scale. It is observed that the average uncertainty energy $\langle E_{\Delta} \rangle$ enters an exponential-growth regime after a sharp initial decrease, as already noted by \citet{ge2023production}. As mentioned in section \ref{sec:DNS numerical setup}, the lower-Reynolds-number cases exhibit a much longer exponential-growth interval. To characterize this interval more clearly, figure \ref{fig:gamma} also shows the time evolution of the instantaneous growth exponent, defined as $\gamma \equiv \frac{1}{2}\mathrm{d}\ln\langle E_{\Delta}\rangle/\mathrm{d}t,$ whose nearly constant value indicates exponential growth. In the present work, we focus
exclusively on the regime of exponential growth of $\langle E_{\Delta}
\rangle$, namely the plateau interval identified in figure \ref{fig:gamma}. The time-sampling intervals used in the following analysis are all chosen within this range, as listed in Table~\ref{tab:main parameters of statistics}.
Figure \ref{fig:pdf} shows the PDF of the uncertainty production at
different times during this regime. The PDFs of $P_{\Delta}$ at
different times during this time range approximately collapse
if normalized by the PDF's maximum value and standard deviation. The
average uncertainty production rate normalized by its standard
deviation is therefore an approximately statistically stationary
process as figure \ref{fig:NormalizedmeanProduction} in fact
demonstrates.  It is therefore reasonable to disregard the chronological order of the observations, after normalizing them by their standard deviation, when constructing the analogue Markov chain.

\subsection{Statistical analysis setup\label{sec:Statistical analysis setup}}

\subsubsection{Data Sampling\label{sec:Data Sampling}}

The data used to establish the analog Markov chain are sampled from
the DNS using a number of observation boxes. These boxes, denoted as
$\mathbb{D}$, are small boxes\par
\begin{equation}
	\mathbb{D} = \left\{ \left( x_1, x_2,
	x_3 \right) \mid \left| x_i - x_i^{\mathbb{D}} \right| <
	L_{\mathbb{D}}, i = 1, 2, 3 \right\}\,,
\end{equation}
where $\boldsymbol{x}^{\mathbb{D}}$ represents the geometrical center of the
sampling box, which is randomly determined within the simulation domain.

The size of the sampling box, $2L_{\mathbb{D}}$, is chosen to be close to the average distance between two velocity stagnation points, which is equal to $\mathbb{P}\left(\boldsymbol{u}=0\right)\lambda$ \citep{goto2009dissipation}, where $\lambda$ is the Taylor microscale and $\mathbb{P}\left(\boldsymbol{u}=0\right)$ is the probability density for the instantaneous fluid velocity to be zero in the reference flow. This characteristic length scale may be interpreted as being close to the average size of a turbulent eddy. In practice, the half-size of each observation box, denoted by $L_{\mathbb{D}}$, is chosen in the form $L_{\mathbb{D}}=2^{n}\Delta x$, where $\Delta x=2\pi/N$ is the DNS spatial resolution. The integer exponent $n$ is selected such that the full box size $2L_{\mathbb{D}}=2^{n+1}\Delta x$ is the smallest grid-compatible length scale larger than $\mathbb{P}\left(\boldsymbol{u}=0\right)\lambda$. This choice ensures that the observation boxes can exactly tile the computational domain.

To reduce spatial correlations between different observation boxes, the distance between their geometric centres is required to be larger than all relevant correlation lengths, including those of the candidate predictors (see section \ref{sec:Predictor fields}) and of the uncertainty-production term. Here, each correlation length is denoted as $L_{P_{\Delta}}$, $L_{\boldsymbol{\mathsf{S}}}$, $L_{\boldsymbol{\Omega}}$ and $L_{[\boldsymbol{\mathsf{S}},\boldsymbol{\Omega}]}$, defined as the integral of the corresponding auto-correlation function \citep{Batchelor1953}. Therefore, the distance between any two geometric centres is larger than $L_{\mathrm{sep}}=\max\{L_{P_{\Delta}}, L_{\boldsymbol{\mathsf{S}}}, L_{\boldsymbol{\Omega}}, L_{[\boldsymbol{\mathsf{S}},\boldsymbol{\Omega}]},L_{\Delta}\}$, where $L_{\Delta}$ is the uncertainty length scale, defined as the integral length scale according to the uncertainty energy spectrum, see \citet{ge2023production} for details. Although $L_{\mathrm{sep}}$ is chosen as the maximum among the relevant correlation lengths, it remains small compared with the size of the periodic domain, $\mathcal{L}=2\pi$. From table~\ref{tab:main parameters of statistics}, one obtains $L_{\mathrm{sep}}/\mathcal{L}\in [0.02,0.2]$ for the all the cases. Thus, even the largest separation length used to decorrelate the observation boxes is much smaller than the domain size. Under this constraint, different observation boxes do not overlap. To further reduce spatiotemporal correlations between samples collected from different boxes, the sampling interval in time is required to satisfy $\Delta t = L_{\mathrm{sep}}/U_{\max}$, where $U_{\max}$ is the maximum flow velocity.

The time sampling is restricted strictly to the exponential-growth regime of the uncertainty energy, namely the plateau region identified in figure~\ref{fig:time evolution of uncertinty}. Together, this temporal restriction and the minimum sampling interval determine the number of temporal samples $N_t$. The number of sampled observation boxes, denoted by $N_x$, is then adjusted accordingly so that the total number of samples satisfies $N_xN_t \geq \left(\pi/L_{\mathrm{sep}}\right)^3.$
All of these statistical parameters are listed in Table~\ref{tab:main parameters of statistics}.

\begin{figure}
	\centering
	\subfigure[]{
		\label{fig:Observation boxes}
		\includegraphics[width=0.48\textwidth]{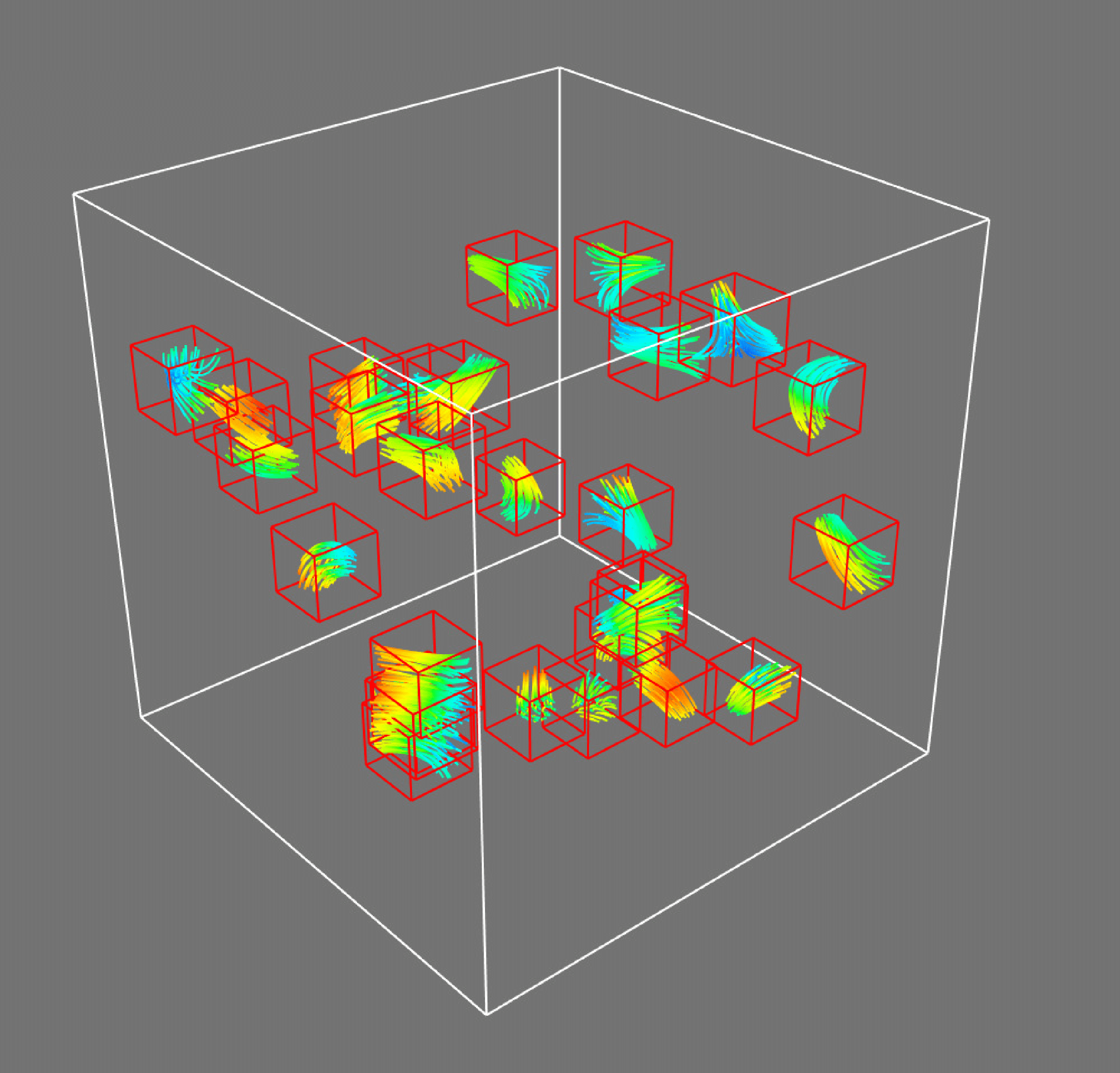}}
	\subfigure[]{
		\label{fig:Testing boxes}
		\includegraphics[width=0.48\textwidth]{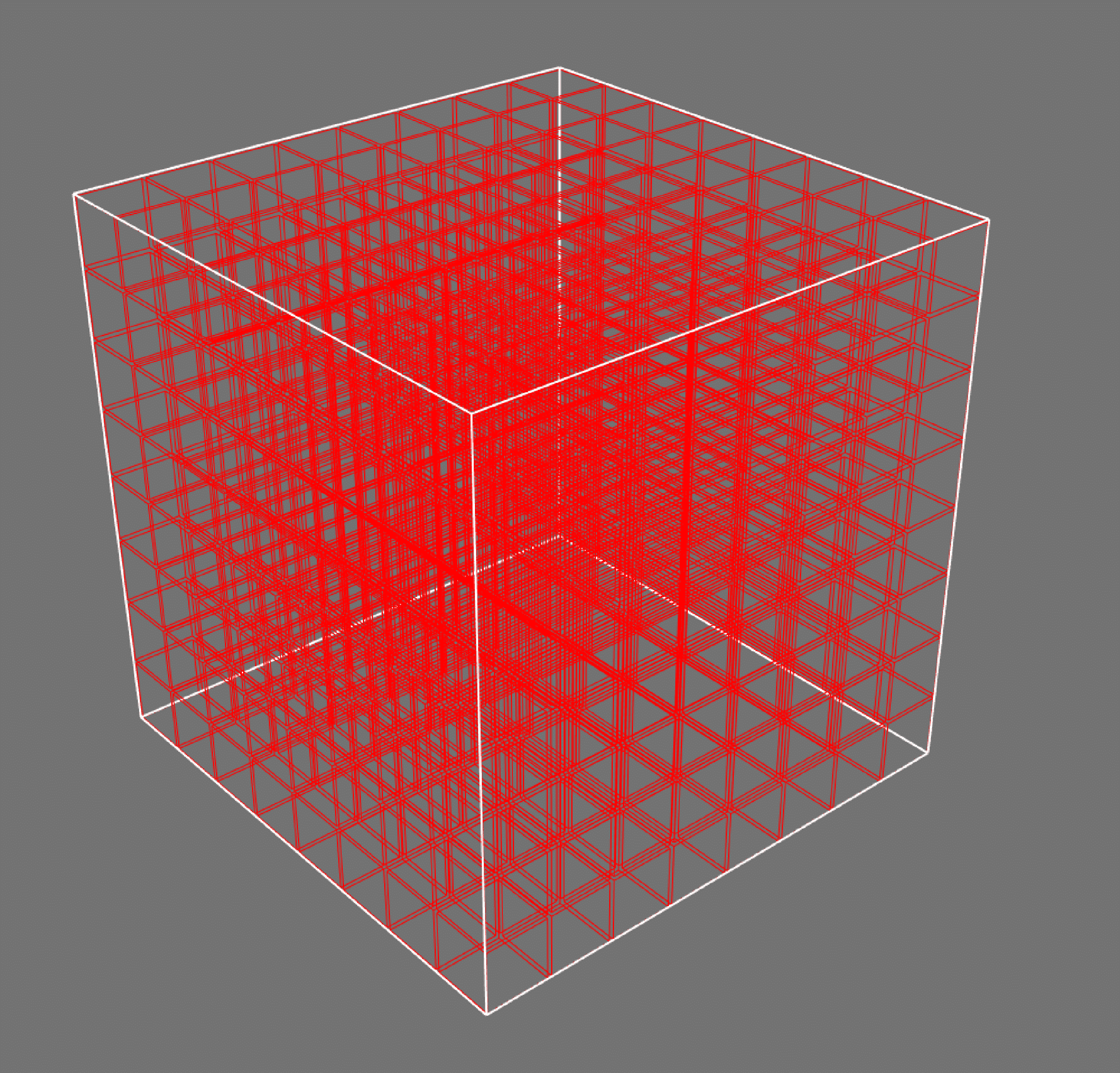}}
	\caption{(a) One example of spatial distribution of
		observation boxes ($N_{x}=28$) within the simulation
		domain. Inside the boxes the streamlines of the velocity of
		the reference flow are plotted. (b) Spatial distribution of
		all potential testing data boxes}
\end{figure}

Rather than using all the spatial points within each observation box to estimate the committor function, we use only the box-averaged observed events, namely $\left\langle \overline{\boldsymbol{Z}}_n \right\rangle_{\mathbb{D}}$, where $\left\langle \cdot \right\rangle_{\mathbb{D}}$ denotes the spatial average over the observation box and the overbar indicates normalization by the standard deviation of the corresponding global field. As shown in section \ref{sec:DNS results}, this normalization makes the statistics of these quantities approximately statistically steady, which is advantageous for the present data-driven construction.
The normalisation here is intended to
help appreciate how the input data deviate from their mean. The key
parameters of the data sampling are gathered in table \ref{tab:main
	parameters of statistics}. Figure \ref{fig:Observation boxes} illustrates the spatial
distribution of observation boxes within the simulation domain in the
case $N_{x}=28$.

\begin{table}
	\begin{center}
		\def~{\hphantom{0}}
		\begin{tabular}{cccccccccccccc}
			Case&$[t_{\min}/\tau_{\eta},t_{\max}/\tau_{\eta}]$&$N_{\mathrm{case}}$&$N_{x}$&$N_{t}$&$2L_{\mathbb{D}}$&$L_{\Delta}$&$L_{P_{\Delta}}$&$L_{\boldsymbol{\mathsf{S}}}$&$L_{\boldsymbol{\Omega}}$&$L_{[\boldsymbol{\mathsf{S}},\boldsymbol{\Omega}]}$&$\Delta t/\tau_{\eta}$&$\mathbb{P}(\mathcal{A})$&$\mathbb{P}(\mathcal{B})$\\ [3pt]
			F1$_{128}$&$[16.64,161.86]$&$15$&45&186&0.39&1.23&0.63&0.54&0.52&0.61&0.78&0.333&0.106\\
			F1$_{512}$&$[8.84,49.05]$&$5$&500&88&0.098&0.075&0.049&0.20&0.12&0.13&0.46&0.417&0.008\\
			F2$_{128}$&$[22.07,153.65]$&$15$&60&160&0.39&0.91&0.38&0.55&0.42&0.49&0.83&0.308&0.105\\
			F2$_{512}$&$[11.88,69.31]$&$5$&250&156&0.049&0.096&0.055&0.15&0.10&0.10&0.37&0.365&0.012\\
		\end{tabular}
		\caption{Parameters of the data sampling. The interval $[t_{\min}/\tau_{\eta},\, t_{\max}/\tau_{\eta}]$ gives the starting and ending times of the sampling, where $\tau_{\eta}$ is the Kolmogorov time scale. $N_{\mathrm{case}}$ is the number of DNS realizations performed from the same reference flow with independently generated random perturbations. $N_{x}$ is the number of observation boxes used in the training dataset, and $N_{t}$ is the length of the time sequence sampled in each observation box. $2L_{\mathbb{D}}$ is the size of the observation box, and $L_{\Delta}$ is the uncertainty-energy length scale. $L_{P_{\Delta}}$, $L_{\boldsymbol{\mathsf{S}}}$, $L_{\boldsymbol{\Omega}}$, and $L_{[\boldsymbol{\mathsf{S}},\boldsymbol{\Omega}]}$ are the correlation lengths, defined from the integrals of the corresponding auto-correlation functions, of $P_{\Delta}$, $\left\Vert\boldsymbol{\mathsf{S}}\right\Vert_{F}$, $\left\Vert\boldsymbol{\Omega}\right\Vert_{F}$, and $\left\Vert[\boldsymbol{\mathsf{S}},\boldsymbol{\Omega}]\right\Vert_{F}$, respectively. $\Delta t$ is the sampling time interval. $\mathbb{P}(\mathcal{A})$ and $\mathbb{P}(\mathcal{B})$ are the probabilities of the event sets $\mathcal{A}$ and $\mathcal{B}$, respectively.}
		\label{tab:main parameters of statistics}
	\end{center}
\end{table}

\subsubsection{Definition of extreme uncertainty-production event\label{sec:Definition of extreme uncertainty-production event}}

In a previous work \citep{ge2023production}, we have shown that the
production of uncertainty in the flow field is dominated by rare but
extreme $P_{\Delta}$ events during the chaotic exponential growth of
uncertainty regime. A set $\mathcal{B}$ of such rare extreme events
can be defined as
$\mathcal{B}\equiv\left\{\overline{P}_{\Delta}>1\right\}$, where the normalized uncertainty-production term, as defined in the previous subsection, is $\overline{P}_{\Delta}\equiv P_{\Delta}/\sigma_{P_\Delta}$. The reference set corresponding to normal conditions is defined as $\mathcal{A}\equiv\left\{\overline{P}_{\Delta}<0\right\}$. Table~\ref{tab:main parameters of statistics} reports, for each case according to figure \ref{fig:pdf}, the probabilities of the events belonging to sets $\mathcal{A}$ and $\mathcal{B}$.
Our objective is to calculate the probability
of hitting set $\mathcal{B}$ before set $\mathcal{A}$ from a given
initial state of weak/moderate uncertainty production
($0<\overline{P}_{\Delta}<1$)
belonging to neither $\mathcal{B}$ nor $\mathcal{A}$ and accounting
for approximately 60\% of all events.

\subsubsection{Predictor fields\label{sec:Predictor fields}}

The phase space (also known as predictor space) $\boldsymbol{\chi}$ of events
is characterized by values of $D$ scalar fields at particular spatial
positions and times. The distance between these scalar fields, also
referred to as predictor fields, is evaluated in this phase space
$\boldsymbol{\chi}\subset\mathbb{R}^{D}$.
As shown in section \ref{sec:Theoretical analysis of the uncertainty
	production}, vortex stretching, vorticity and strain rate in the
reference flow as well as the uncertainty field have a clear role in
the evolution of uncertainty production as described by equation
(\ref{eq:Production NS equation}). Therefore, the scalar fields
$\left\{E_{\Delta}, \left\Vert [\boldsymbol{\mathsf{S}},\boldsymbol{\Omega}] \right\Vert_{F}, \left\Vert
\boldsymbol{\Omega} \right\Vert_{F}, \left\Vert \boldsymbol{\mathsf{S}}
\right\Vert_{F}\right\}$, along with their combinations, are selected
to constitute predictor spaces. $\left\Vert \cdot \right\Vert_{F}$ is
the Frobenius norm of a tensor and is therefore the root of the sum of
the squares of all the tensor's elements. $\left\Vert \boldsymbol{\Omega}
\right\Vert_{F}^{2}=\left|\boldsymbol{\omega}\right|^{2}/2=-tr(\boldsymbol{\Omega}\cdot\boldsymbol{\Omega})$
is the enstrophy ($\boldsymbol{\omega}=\boldsymbol{\nabla}\times\boldsymbol{u}$ being the
vorticity), whereas $\left\Vert \boldsymbol{\mathsf{S}}
\right\Vert_{F}^{2}=tr(\boldsymbol{\mathsf{S}}\cdot\boldsymbol{\mathsf{S}})$. By
assessing the predictive performance of the committor function derived
from different predictor spaces, we aim to identify the most effective
combination that relates to the production of uncertainty.

\subsubsection{Data testing \label{sec:Data testing}}

The committor function is estimated on the basis of a number $N_x$ of
observation boxes of size $L_{\mathbb{D}}$. Its predictions are then
assessed by using as initial state the field values at the data
sampling starting time. This assessment is made on the basis of
boxes with the same size $L_{\mathbb{D}}$ as the observation boxes but
regularly placed throughout the initial field, as shown in figure
\ref{fig:Testing boxes}. As a consequence, the initial field is
divided into $(\mathcal{L}/2L_{\mathbb{D}})^{3}$ boxes.
Within these boxes, we calculate averages of the normalized fields,
as mentioned in section \ref{sec:Data Sampling}. All of these
boxes are then introduced as input to the committor function to
determine the probability of each box to evolve into an extreme
positive production event in the future before reaching a negative
production event. Subsequently, the actual evolution of the DNS field
following its initial state determines whether a testing box actually
evolves first into an extreme positive production event or a negative
production event, and this information is then used to calculate the
Brier score based on equation (\ref{eq:brier function}).

For each case, we perform $N_{\mathrm{case}}$ (shown in table \ref{tab:main parameters of statistics}) simulations starting from the same reference field but with different randomly generated initial perturbations. For each of these simulations, the committor is trained separately following the procedure described above. The corresponding verification datasets are then combined, and the Brier scores of the committors trained from the different perturbation realizations are evaluated on this aggregated testing set. In this way, we also assess the sensitivity of the committor prediction to the particular choice of the initial perturbation.

\section{The relation of extreme uncertainty production events and flow topology\label{sec:The relation of extreme uncertainty production events and flow topology}}

We first use our DNS data to localize the instantaneous extreme
production events based on the flow topology defined by
$\boldsymbol{\Omega}$ and $\boldsymbol{\mathsf{S}}$. Starting from the work of  \citet{hunt1988streams}, the flow field can be divided
into three types of region according to the magnitude of
$Q\equiv\frac{1}{2}\left(\left\Vert \boldsymbol{\Omega}
\right\Vert_{F}^{2}-\left\Vert \boldsymbol{\mathsf{S}}
\right\Vert_{F}^{2}\right)$, the second invariant of the velocity
gradient tensor $\boldsymbol{\nabla}\boldsymbol{u}$: eddy ($Q>2\sigma_{Q}$), convergence
($Q<-\sigma_{Q}$) and streaming ($Q\approx0$) regions, where $\sigma_{Q}$ is the standard deviation of $Q$. The third invariant of
the velocity gradient tensor is
$R=-\det\left(\boldsymbol{\nabla}\boldsymbol{u}\right)$ and it has been repeatedly
observed \citep{tsinober2019essence} that the $Q-R$ diagram has
the same/similar tear drop shape in many turbulent flows. The joint
probability density function of $Q$ and $R$ forms an asymmetrical
teardrop shape along the $(27/4)R^{2}+Q^{3}=0$, $R\ge 0$ line \citep{tsinober2019essence}, showing, in particular, the
importance of strain self-amplification in turbulence.

\begin{figure}
	\centering
	\subfigure[]{
		\label{fig:QFf1128}
		\includegraphics[width=0.48\textwidth]{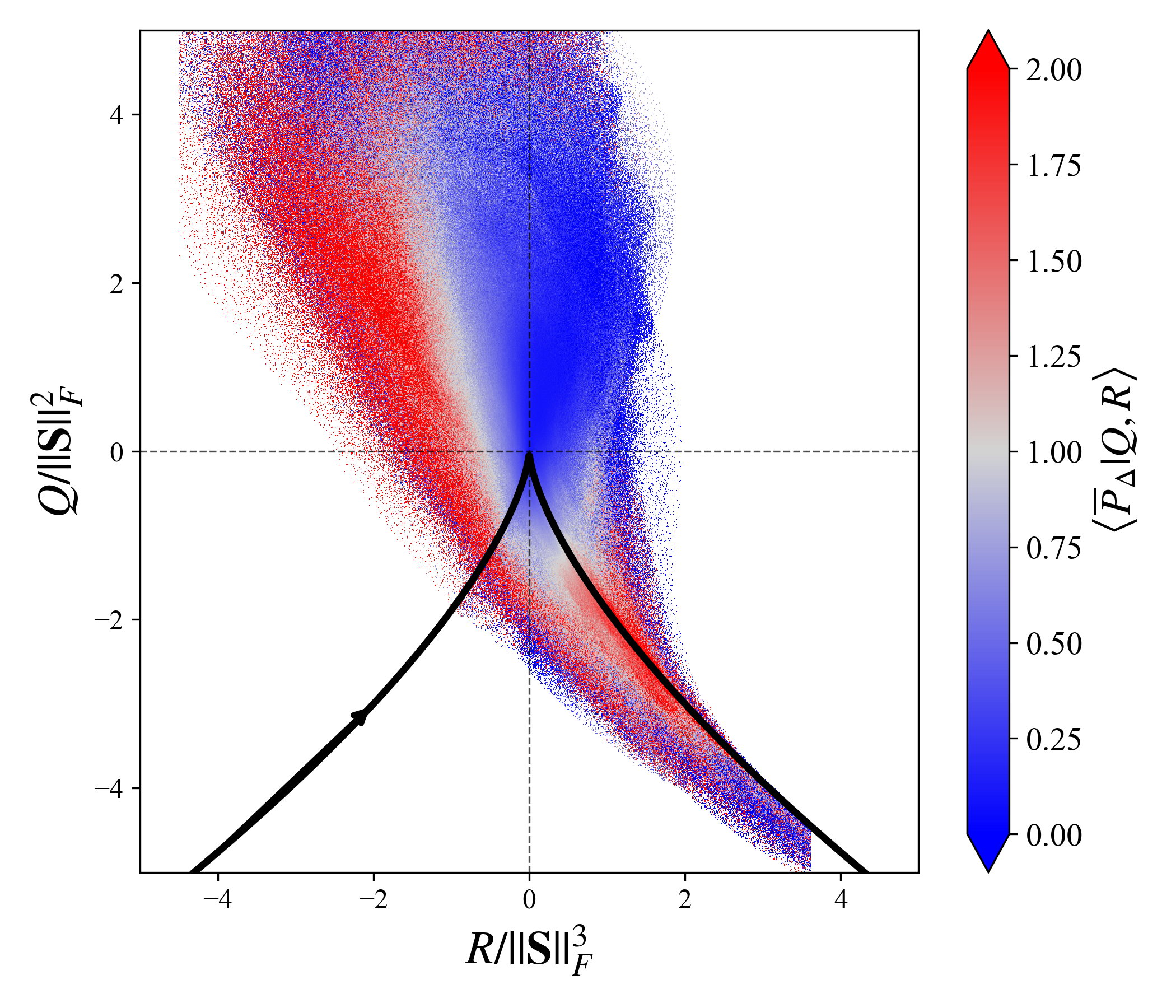}}
	\subfigure[]{
	\label{fig:QFf1512}
	\includegraphics[width=0.48\textwidth]{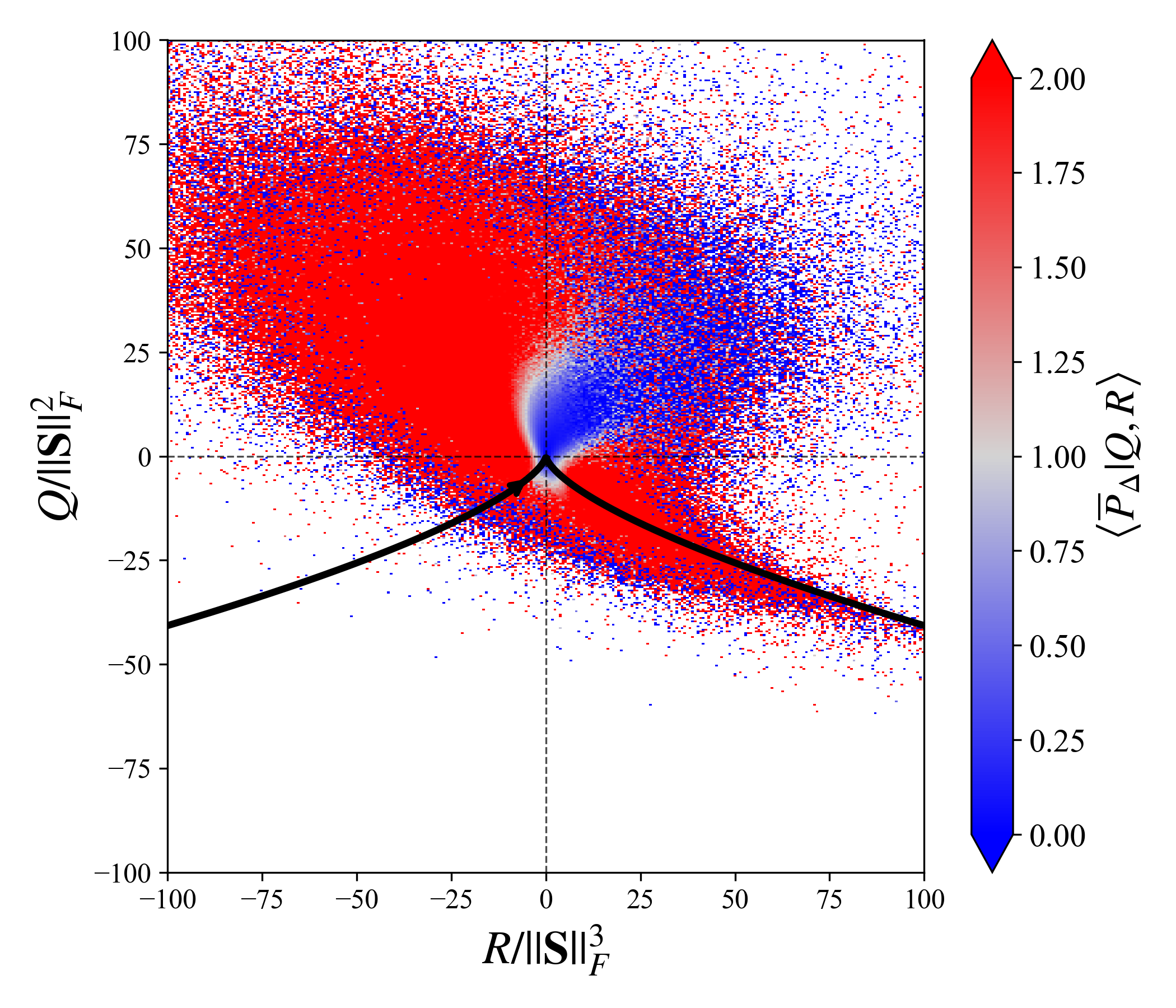}}
	\subfigure[]{
		\label{fig:QFf2128}
		\includegraphics[width=0.48\textwidth]{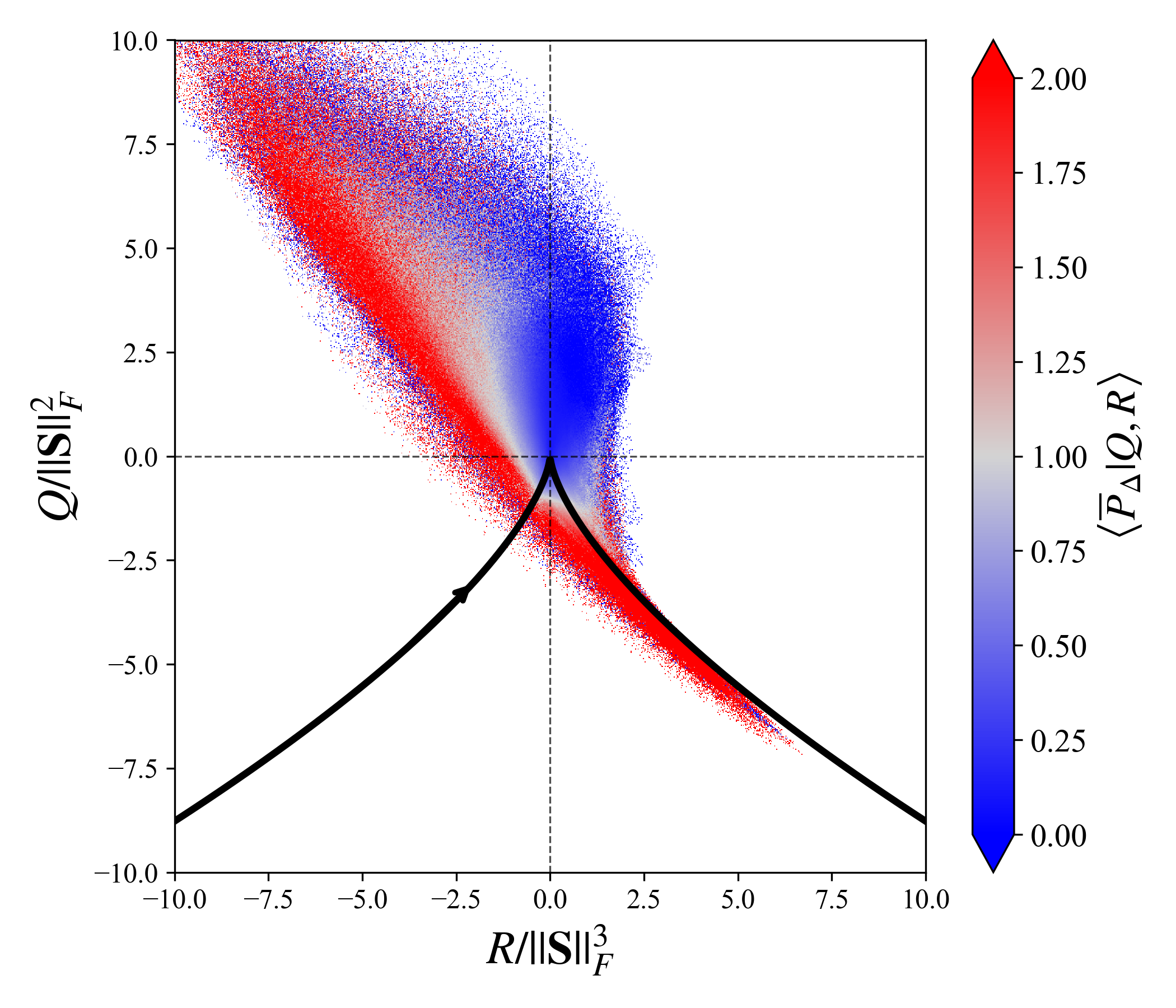}}
	\subfigure[]{
	\label{fig:QFf2512}
	\includegraphics[width=0.48\textwidth]{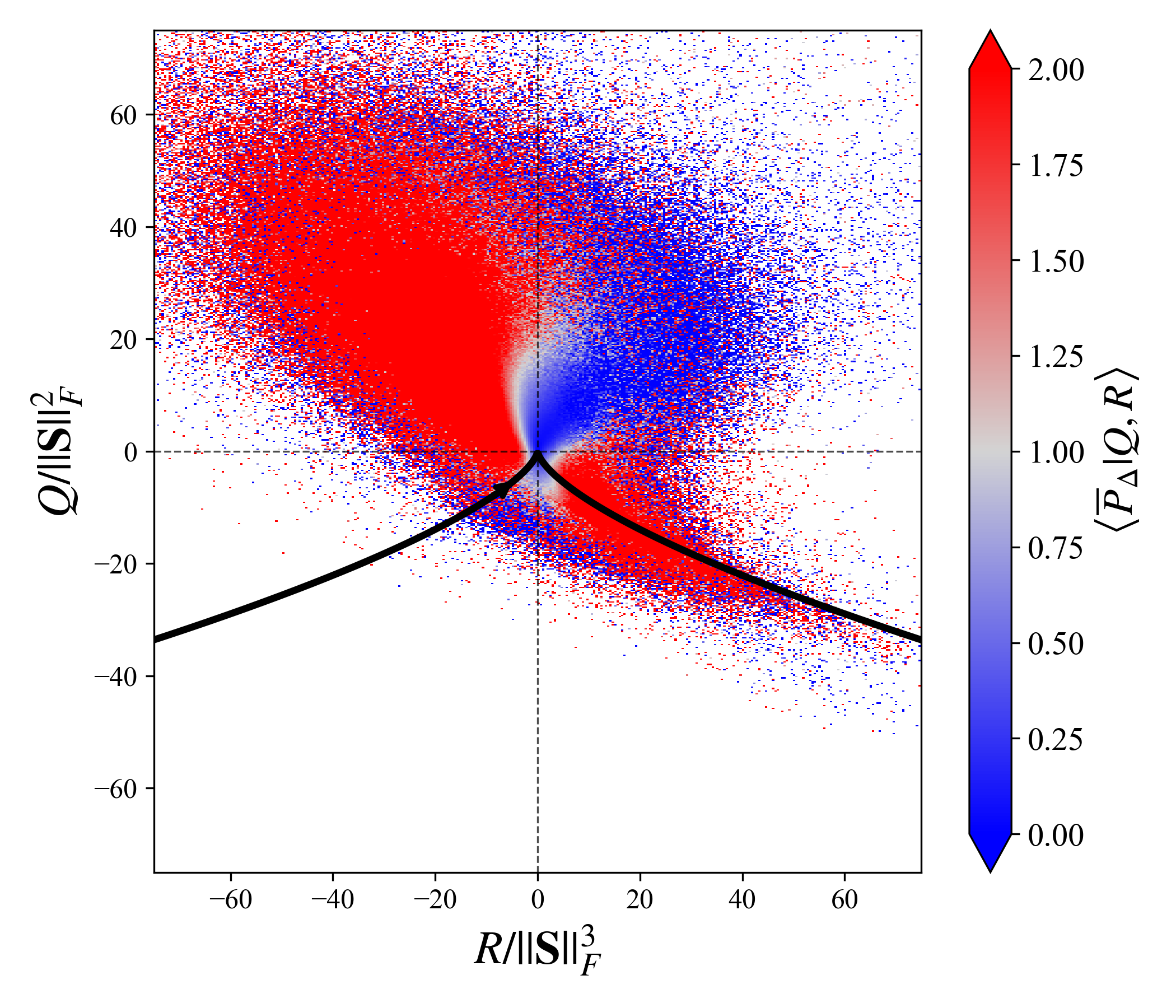}}
	\subfigure[]{
	\label{fig:Q-R_overlay_F1}
	\includegraphics[width=0.48\textwidth]{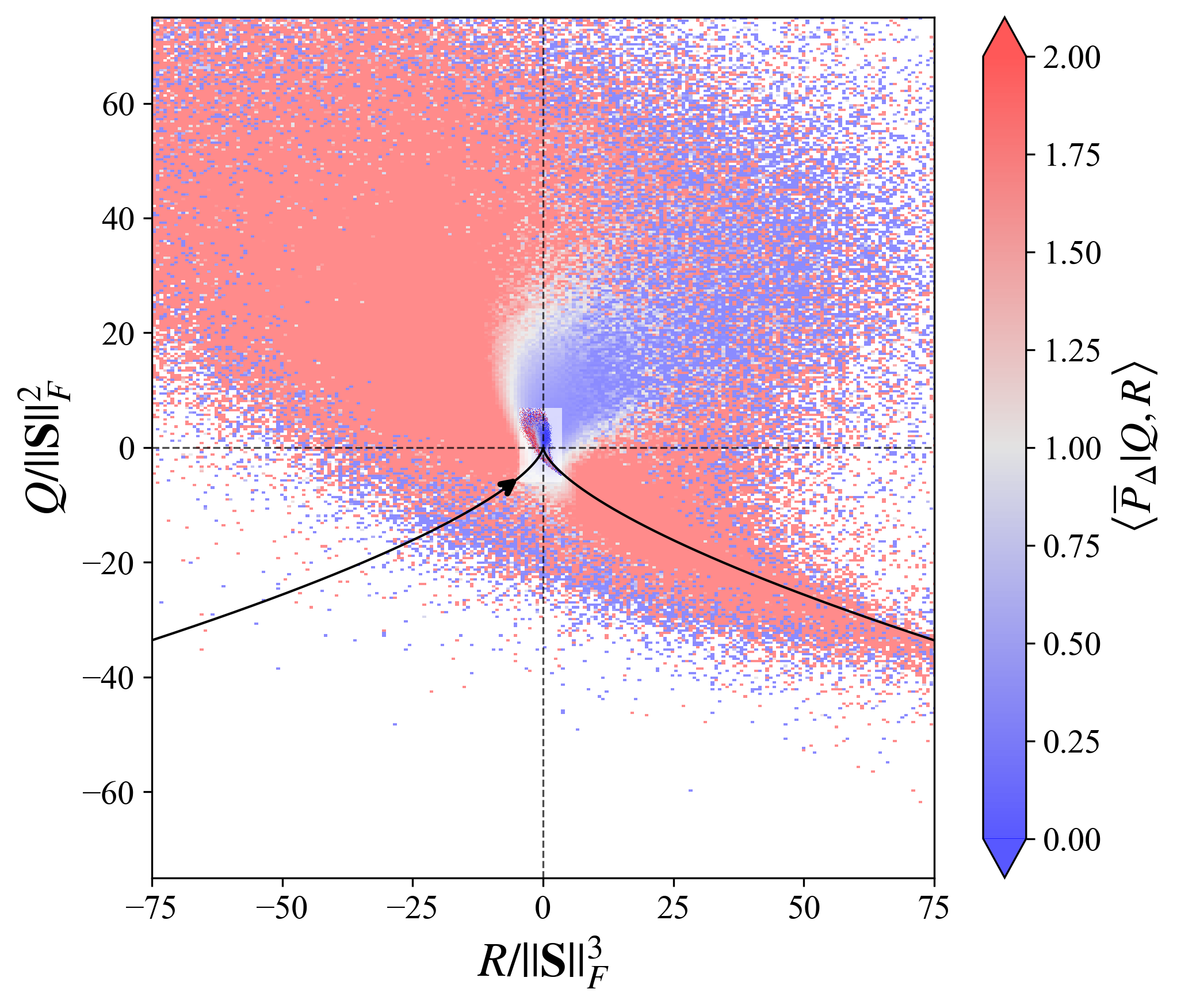}}
	\subfigure[]{
	\label{fig:Q-R_overlay_F2}
	\includegraphics[width=0.48\textwidth]{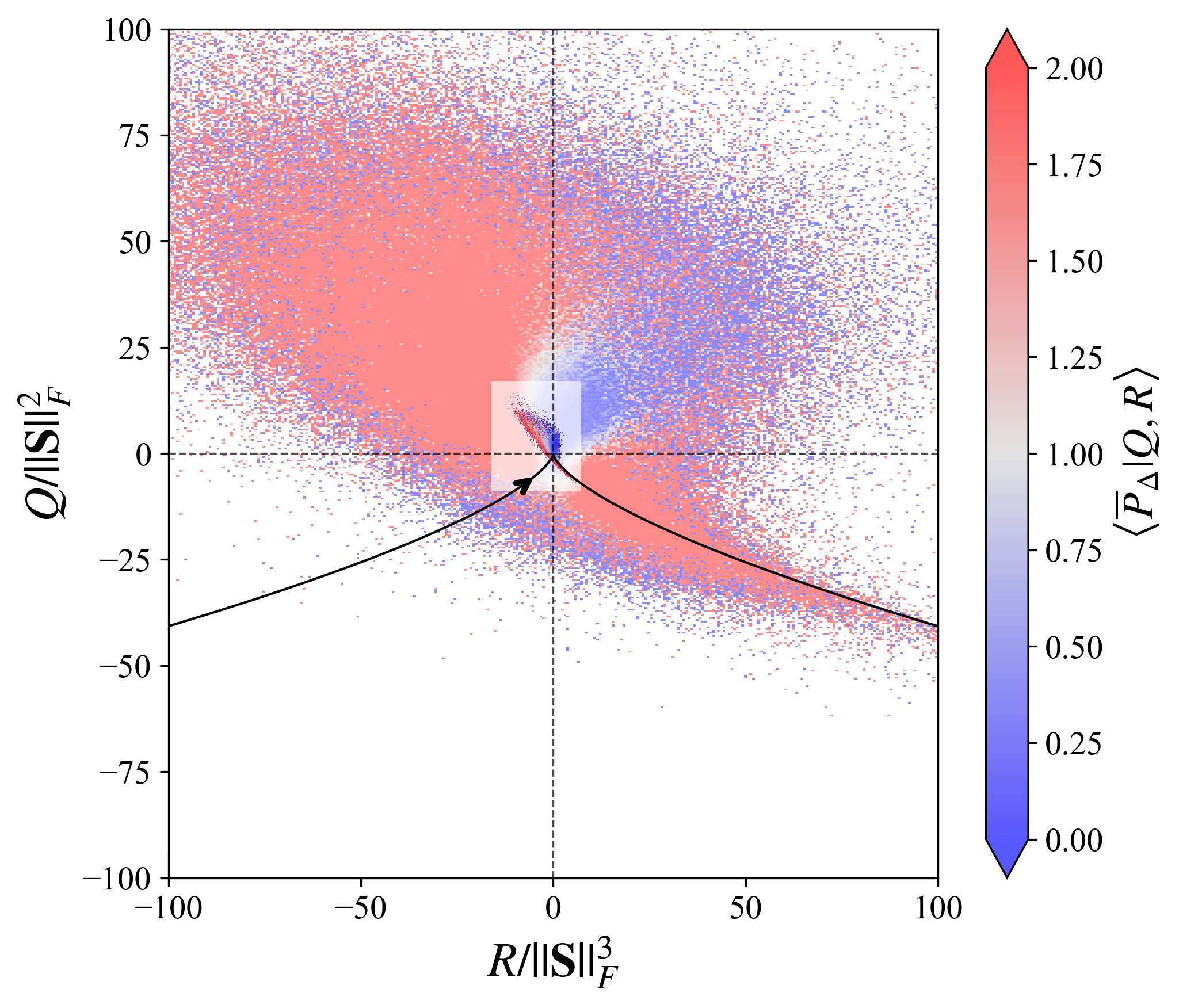}}
	\caption{Time averaged distribution of $\overline{P}_{\Delta}$
		in the $Q-R$ plane for cases (a) F1$_{128}$, (b) F1$_{512}$, (c) F2$_{128}$ and (d) F2$_{512}$,, The
		continuous black line corresponds to $(27/4)R^{2}+Q^{3}=0$
		and is dominated by strain self-amplification in the
		lower-right quadrant. (e) shows panels (a) and (b) overlaid on the same scale, with panel (a) on top and panel (b) shown with a transparent background.
		(f) shows panels (c) and (d) overlaid on the same scale, with panel (c) on top and panel (d) shown with a transparent background.}
	\label{fig:Q-F}
\end{figure}

In figure \ref{fig:Q-F} we plot the conditional average of
	$\overline{P}_{\Delta}$ on the $(Q,R)$ plane using the DNS data, noted as $\left\langle\overline{P}_{\Delta}\vert Q,R\right\rangle $, where
	$\overline{P}_{\Delta}=P_{\Delta}/\sigma_{P_{\Delta}}$ and
	$\sigma_{P_{\Delta}}$ is the standard deviation. More specifically, for
	each bin in the $(Q,R)$ plane, we collect all spacetime points for which the local values of $Q$
	and $R$ fall into that bin, and then compute the average value of
	$\overline{P}_{\Delta}$ over those samples. Figures \ref{fig:pdf} and
\ref{fig:NormalizedmeanProduction} show that $\overline{P}_{\Delta}$ maintains a
statistically steady distribution over time though $P_{\Delta}$
increases with time.  For both F1 and F2 cases, it is observed that
$\left\langle\overline{P}_{\Delta}\vert Q,R\right\rangle $ also forms an asymmetrical
teardrop shape along the $(27/4)R^{2}+Q^{3}=0$, $R\ge 0$
curve. Moreover, for all setups of forcing and Reynolds number, the extreme uncertainty production is mainly aligned
with the curve $(27/4)R^{2}+Q^{3}=0$ in the quadrant $Q<0$ and
$R>0$. This quadrant represents strain self-amplification  \citep{cantwell2013symmetry,tsinober2019essence,vieillefosse1982local,vieillefosse1984internal,johnson2024multiscale}.
This alignment suggests a potential relation between the strain
self-amplification of turbulence
and extreme uncertainty production. There are also extreme uncertainty
production events in the $Q>0$, $R<0$ quadrant which is the
vortex-stretching quadrant \citep{cantwell2013symmetry,tsinober2019essence,vieillefosse1982local,vieillefosse1984internal,johnson2024multiscale}. As the Reynolds number increases, the red region associated with strong uncertainty-production events progressively extends over most of the vortex-stretching quadrant, indicating that the contribution of vortex-stretching-related topologies becomes increasingly important at higher Reynolds numbers. In addition, as shown more clearly by the overlaid plots in figures \ref{fig:Q-R_overlay_F1} and \ref{fig:Q-R_overlay_F2}, the higher-Reynolds-number distribution appears to grow outward from the lower-Reynolds-number one. That is, the low-Reynolds-number pattern is largely retained as an inner core, while the increase in Reynolds number mainly extends this structure towards larger $\left|Q\right|$ and $\left|R\right|$. Furthermore, few extreme positive uncertainty-production events are observed in the vortex-compression and strain self-attenuation quadrants where $QR>0$; instead, these quadrants are mainly associated with negative or weak uncertainty-production events \citep{cantwell2013symmetry,tsinober2019essence,vieillefosse1982local,vieillefosse1984internal,johnson2024multiscale}. At the same time, for all Reynolds numbers considered here, a narrow transition contour can be identified in the weak-$(Q,R)$ region near the origin. This contour is approximately upward-opening and parabola-like, and acts as an empirical boundary separating the red and blue regions of $\langle \overline{P}_{\Delta}\mid Q,R\rangle$. Its persistence across all cases suggests that this local separation between two distinct uncertainty-production regimes is a robust feature of the dynamics.
The distribution of extreme uncertainty events shown in figure
\ref{fig:Q-F} suggests that the extreme uncertainty production events
probably occur when the flow enters into the vortex-stretching zones
far away from the origin and along the line $(27/4)R^{2}+Q^{3}=0$,
representing strain self-amplification.

\section{Statistical results\label{sec:result}}

In this section we present and discuss the results of the committor
computations. We ﬁrstly present in section \ref{sec:Prediction quality
	of the committor functions} how the trained committor improves with
increasing amounts of data, and how the rate of improvement
discriminates between predictors. In section \ref{sec:Evolution of the
	committor functions with the values of predictor} we investigate how
the probabilities given by the committor function evolve in terms of
the normalized predictor $\overline{F}\equiv F/\sigma_{F}$ used as input.

\subsection{Prediction quality of the committor functions  \label{sec:Prediction quality of the committor functions}}

\begin{figure}
	\centering
	\subfigure[]{
		\label{fig:prediction skill singleF1_128}
		\includegraphics[width=0.48\textwidth]{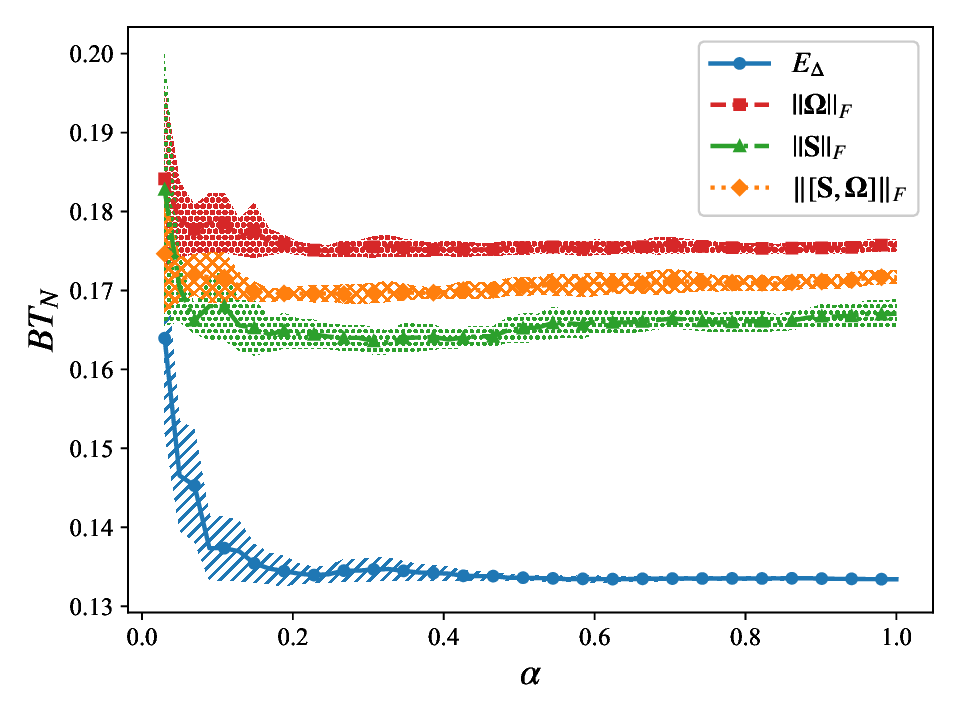}}
	\subfigure[]{
	\label{fig:prediction skill singleF1_512}
	\includegraphics[width=0.48\textwidth]{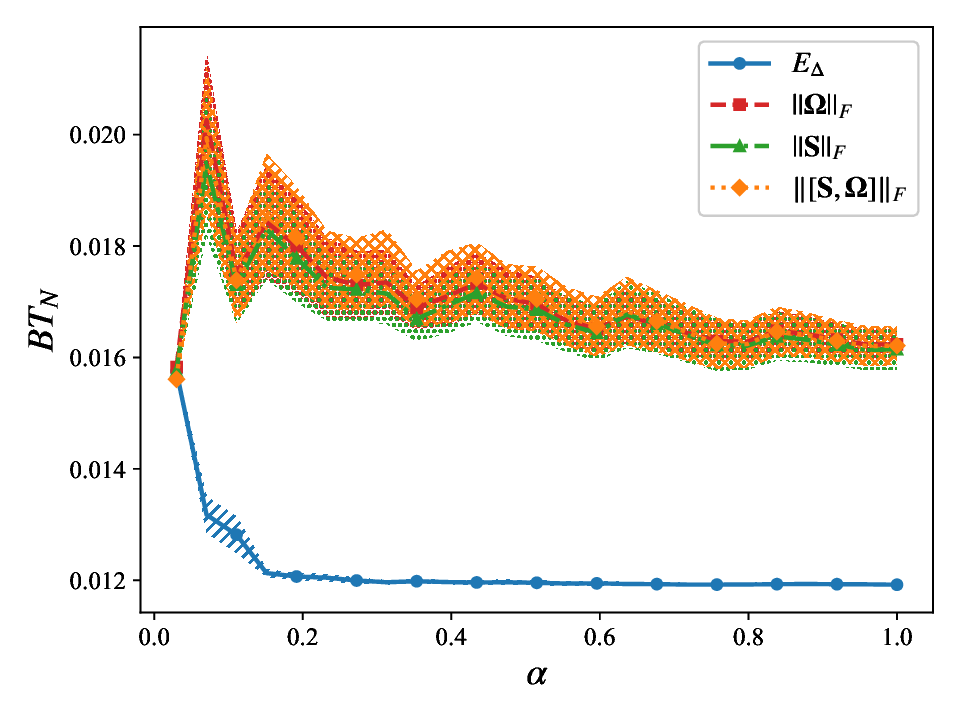}}
	\subfigure[]{
		\label{fig:prediction skill doubleF1_128}
		\includegraphics[width=0.48\textwidth]{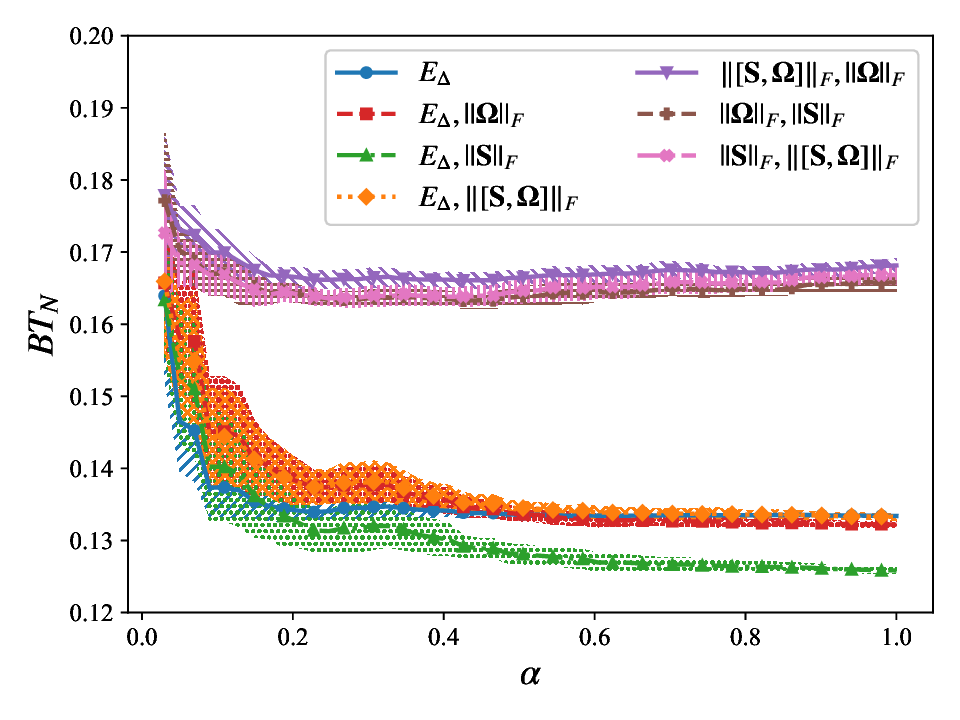}}
	\subfigure[]{
	\label{fig:prediction skill doubleF1_512}
	\includegraphics[width=0.48\textwidth]{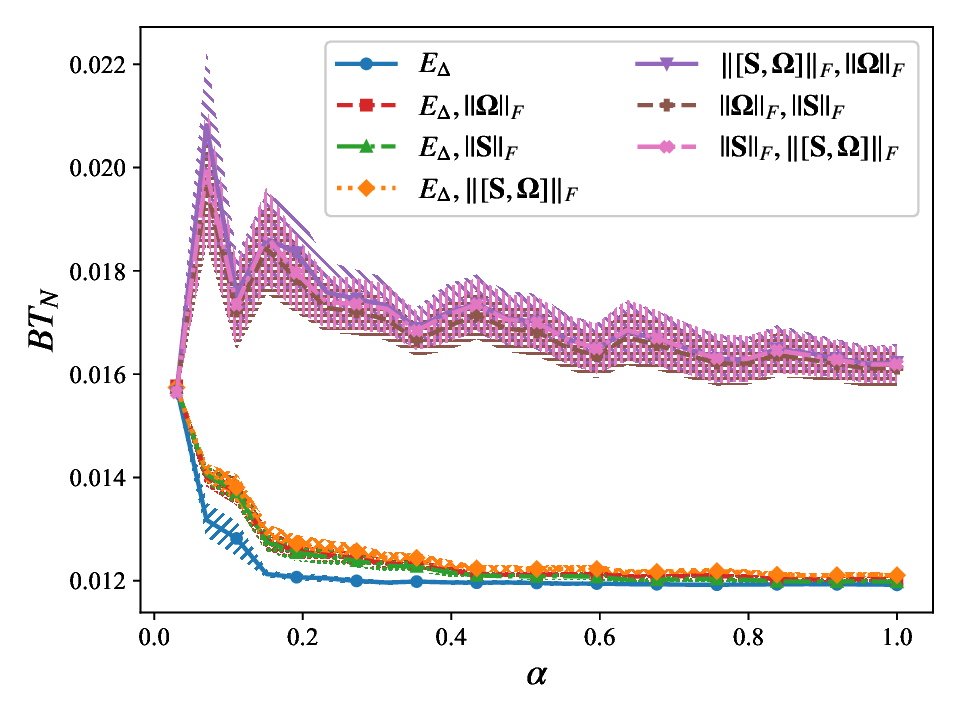}}
	\subfigure[]{
		\label{fig:prediction skill tribleF1_128}
		\includegraphics[width=0.48\textwidth]{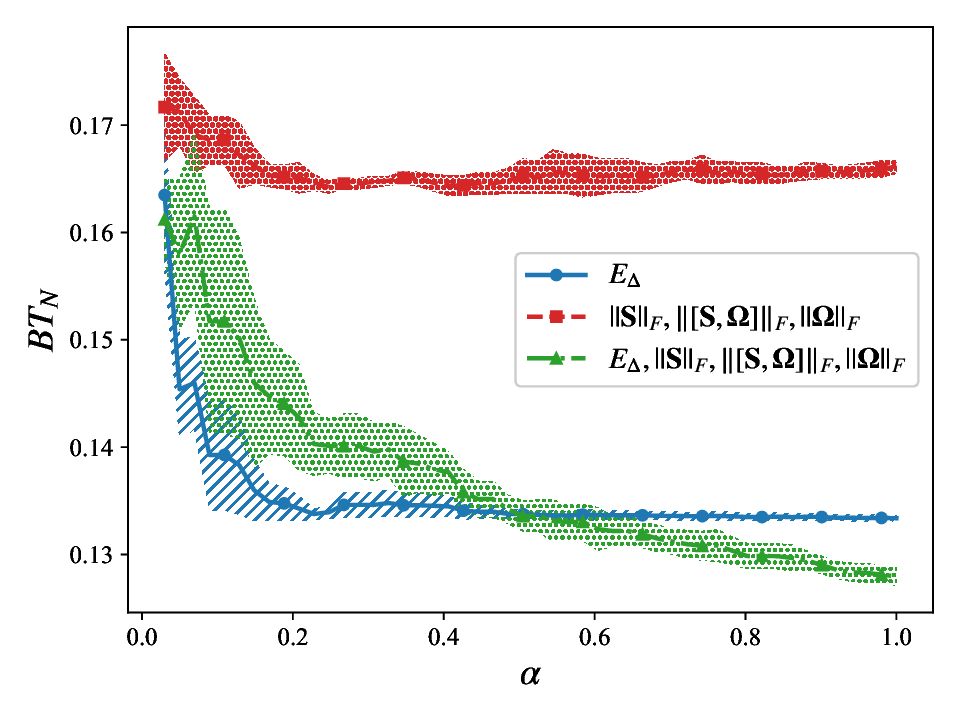}}
	\subfigure[]{
	\label{fig:prediction skill tribleF1_512}
	\includegraphics[width=0.48\textwidth]{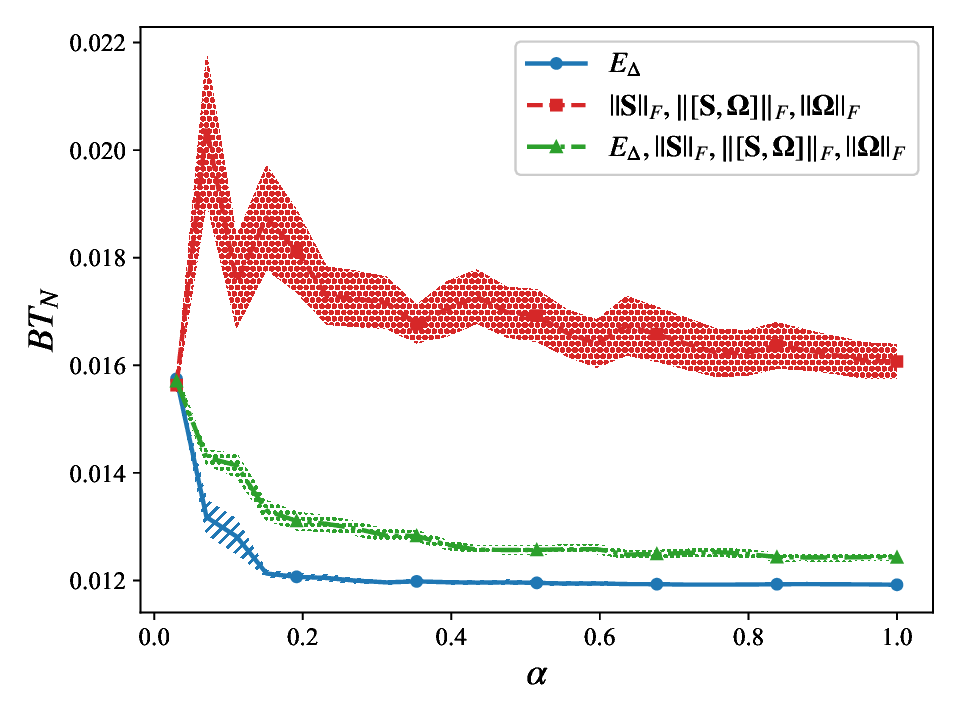}}
	\caption{For (a)(c)(e) case F1$_{128}$ in left panel and case F1$_{512}$ in right panel, the Brier score $BT_{N}$ of the
		approximated committor function versus the number of
		observation boxes $\alpha N_{x}$, obtained from (a)(b) a single scalar
		field, (c)(d) two scalar fields and (e)(f) more than two scalar
		fields. For comparison, the Brier score of the approximated
		committor function obtained from the uncertainty energy is
		included in all plots. The data points and the 
		shadowed zones around them show the average value and the standard
		deviation of the Brier scores obtained from different
		sampled data sets.}
	\label{fig:prediction skill F1}
\end{figure}

\begin{figure}
	\centering
	\subfigure[]{
	\label{fig:prediction skill singleF2_128}
	\includegraphics[width=0.48\textwidth]{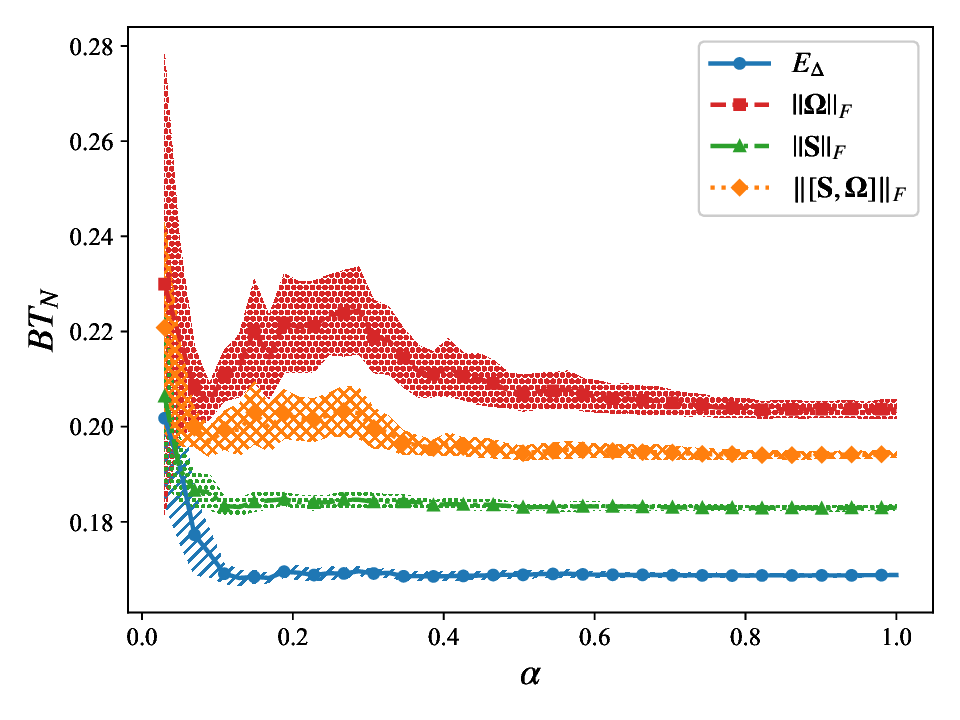}}
\subfigure[]{
	\label{fig:prediction skill singleF2_512}
	\includegraphics[width=0.48\textwidth]{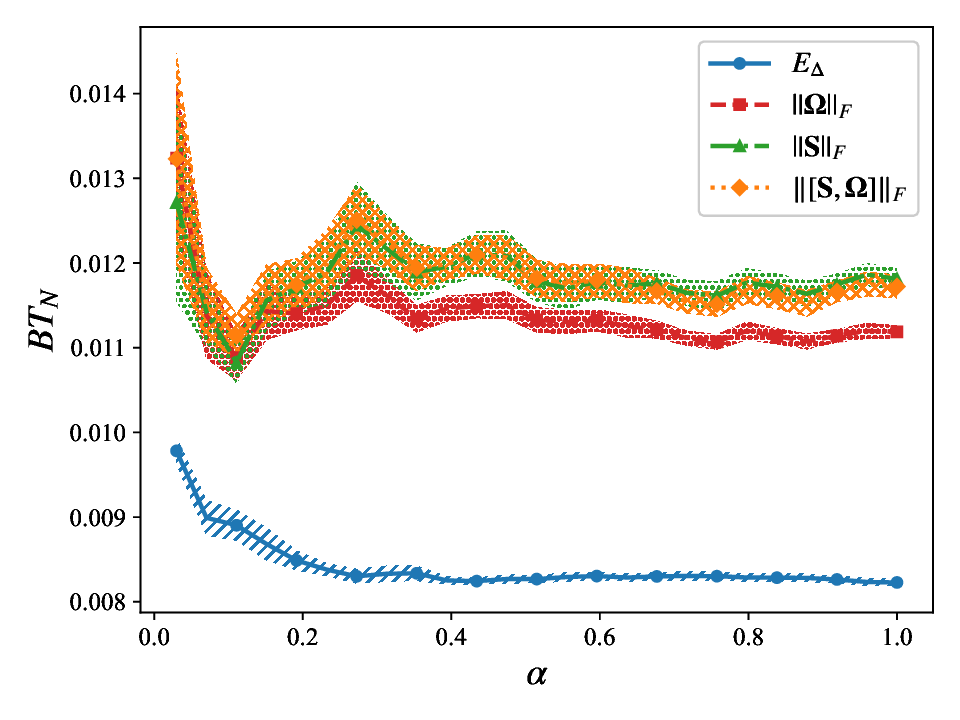}}
\subfigure[]{
	\label{fig:prediction skill doubleF2_128}
	\includegraphics[width=0.48\textwidth]{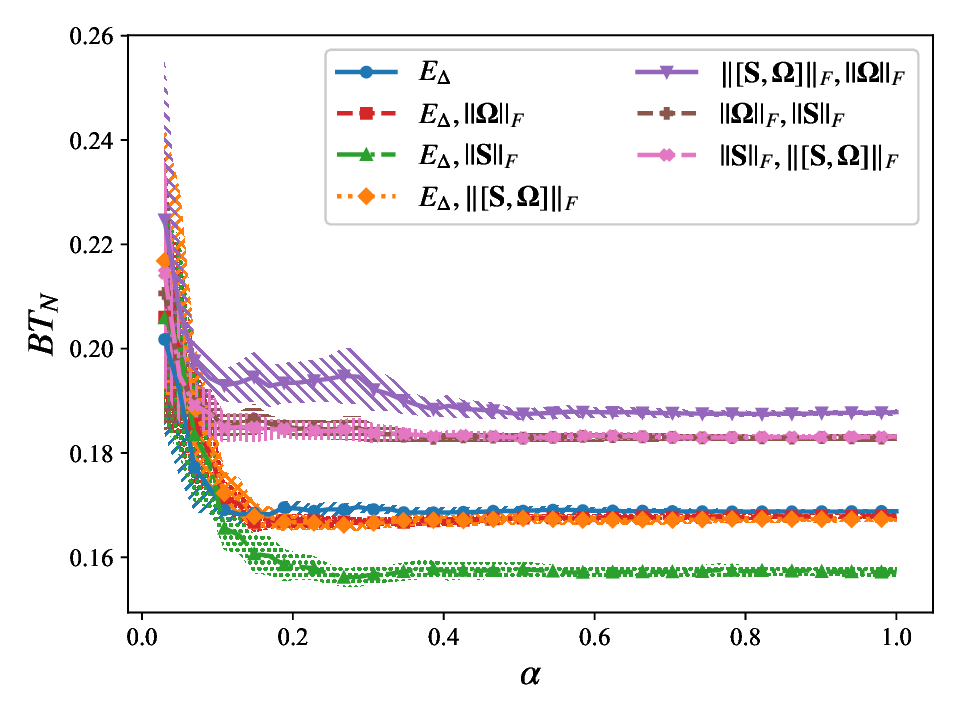}}
\subfigure[]{
	\label{fig:prediction skill doubleF2_512}
	\includegraphics[width=0.48\textwidth]{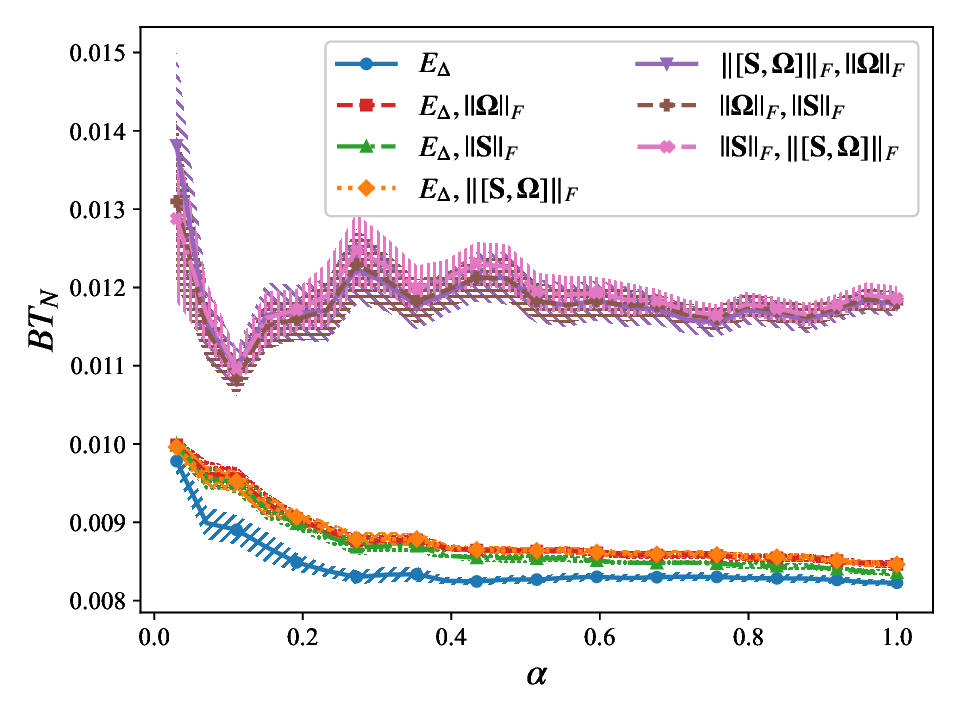}}
\subfigure[]{
	\label{fig:prediction skill tribleF2_128}
	\includegraphics[width=0.48\textwidth]{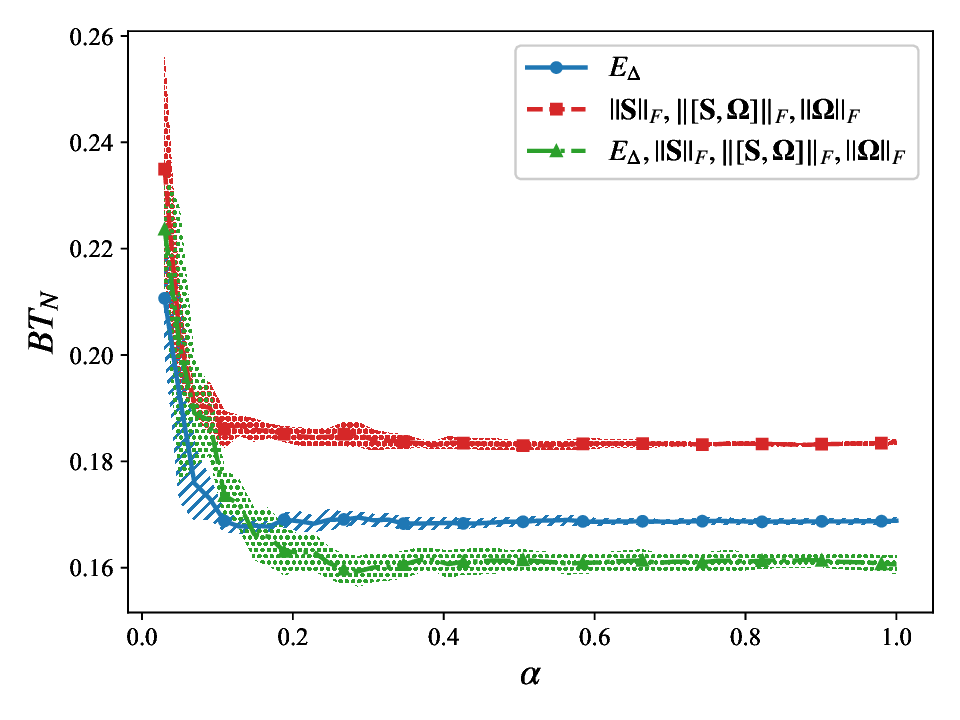}}
\subfigure[]{
	\label{fig:prediction skill tribleF2_512}
	\includegraphics[width=0.48\textwidth]{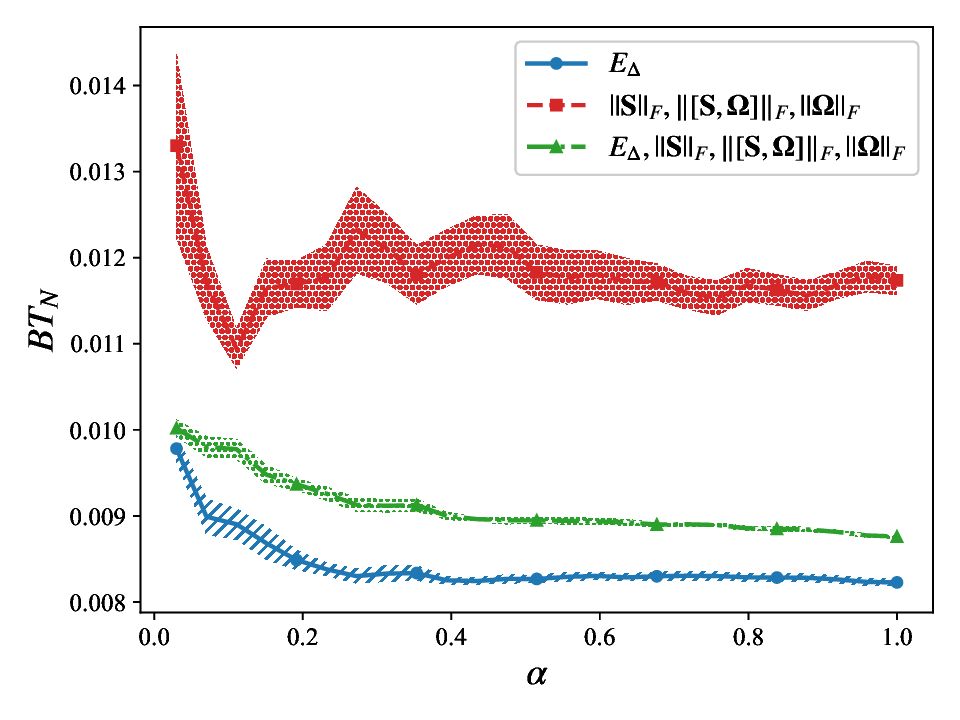}}
	\caption{For (a)(c)(e) case F2$_{128}$ in left panel and case F2$_{512}$ in right panel, the Brier score $BT_{N}$ of the
		approximated committor function versus the number of
		observation boxes $\alpha N_{x}$, obtained from (a)(b) a single scalar
		field, (c)(d) two scalar fields and (e)(f) more than two scalar
		fields. For comparison, the Brier score of the approximated
		committor function obtained from the uncertainty energy is
		included in all plots. The data points and the 
		shadowed zones around them show the average value and the standard
		deviation of the Brier scores obtained from different
		sampled data sets.}
	\label{fig:prediction skill F2}
\end{figure}

A primary question of the statistical analysis is to assess which predictors, among various physical fields, possess the best predictive capability, or equivalently, which choice of predictor defines a coarse-grained phase space that best captures the dynamically relevant structure of the full Navier--Stokes phase space.  For this matter we follow an evaluation procedure used
in \citet{miloshevich2023probabilistic,miloshevich2024extreme} based on
a comparison of committors trained on different predictors or
combinations of predictors albeit with a different score. Note also
that we are not concerned with the delay before the extreme event,
we just monitor the improvement of the prediction as the training
dataset size is increased. Indeed, since the statistical results depend on both the sample size and the sampling location, we train the committor functions on datasets of different sizes by varying the number of observation boxes $N_{x}$ included in the training set. For each case, the committor is trained using $\alpha N_{x}$ observation boxes, where $\alpha$ varies from $0$ to $1$. As mentioned in section \ref{sec:Data testing}, each case is simulated $N_{\mathrm{case}}$ times from the same reference field but with different randomly generated perturbations. This yields $N_{\mathrm{case}}$ Brier scores corresponding to committor functions trained from different realizations. The average
values and standard deviations of these scores are represented in the
figures as data points and shadow regions, respectively.

For both cases F1 and F2, it is observed that,
all curves in figures \ref{fig:prediction skill F1} and \ref{fig:prediction skill F2} (except the curves with all four predictors in case F1$_{128}$) reach a minimum that is about constant with $\alpha N_x$ as
$\alpha$ is increased. Note also that the shadow regions (standard deviation) become
progressively narrower with increasing $\alpha$. The more or less sustained minimum
suggests that the performance of the prediction given by the committor
function is no longer sensitive neither to the amount of input
sampling data nor their positions as $\alpha N_x$ increases. Meanwhile, the observed decrease of the error bars with increasing window size suggests that the influence of window position and inter-window correlation is progressively reduced. Equation
(\ref{eq:eception brier function}) implies the existence of a non-zero
minimum value of the Brier score, and it is probably reached by the
aforementioned trained committor functions. Among all the single predictors, for all cases $E_{\Delta}$ exhibits
	the best predictive capability with the lowest stable Brier score. This optimal Brier
score is attributed to $E_{\Delta}$ containing information from both
the reference flow field and the perturbed flow field. However, in
practice, one cannot simultaneously obtain information from both flow
fields. Therefore, it is the predictive capability of physical
quantities from only a single flow field that is most worthy of
attention. In the low Reynolds-number cases, we observe that the strain rate $\left\Vert \boldsymbol{\mathsf{S}}
\right\Vert_{F}$, although worse than uncertainty energy $E_{\Delta}$,
return the best predictions among all the predictors containing
information from only one only field in cases F1 and F2, while the
vorticity $\left\Vert \boldsymbol{\Omega} \right\Vert_{F}$ returns the worst
predictions in both cases, with the highest Brier scores in both cases F1 and F2.
However, for the higher-Reynolds-number cases, the differences among $\left\Vert [\boldsymbol{\mathsf{S}}, \boldsymbol{\Omega}]\right\Vert_{F}$, $\left\Vert\boldsymbol{\Omega} \right\Vert_{F}$, and $\left\Vert \boldsymbol{\mathsf{S}}\right\Vert_{F}$ become much less marked. In particular, in F2$_{512}$, as shown in figure \ref{fig:prediction skill singleF2_512}, $\left\Vert\boldsymbol{\Omega} \right\Vert_{F}$ gives the lowest Brier score among the three predictors. This trend is consistent with the Reynolds-number dependence of the local velocity-gradient structure and dynamics \citep{DasGirimaji2019}, and with the fact that the small-scale structure becomes more fully developed as the Reynolds number increases \citep{Elsinga2023}. In that situation, the predictive information carried by $\left\Vert [\boldsymbol{\mathsf{S}}, \boldsymbol{\Omega}]\right\Vert_{F}$, $\left\Vert\boldsymbol{\Omega} \right\Vert_{F}$, and $\left\Vert \boldsymbol{\mathsf{S}}\right\Vert_{F}$ can become more similar. At the same time, the $(Q,R)$-conditioned statistics show that strong uncertainty-production events extend over a larger fraction of the vortex-stretching quadrant at higher Reynolds numbers, as shown in figure \ref{fig:Q-F}, which makes the vorticity-based predictor comparatively more effective. In addition, the relative importance of the strain field depends not only on its magnitude but also on its alignment with the local structures associated with uncertainty-energy production \citep{ge2023production}. At higher Reynolds numbers, where strong production events occupy a broader region of the $(Q,R)$ plane, such geometric alignment effects may become more relevant, further reducing the distinction among $\left\Vert [\boldsymbol{\mathsf{S}}, \boldsymbol{\Omega}]\right\Vert_{F}$, $\left\Vert\boldsymbol{\Omega} \right\Vert_{F}$, and $\left\Vert \boldsymbol{\mathsf{S}}\right\Vert_{F}$ as predictors.

Compared to predictions based solely on the
uncertainty energy $E_{\Delta}$, In the lower-Reynolds-number cases, a modest improvement is observed for the combinations $\left\{E_{\Delta},\left\Vert [\boldsymbol{\mathsf{S}},\boldsymbol{\Omega}]\right\Vert_{F}\right\}$ and $\left\{E_{\Delta},\left\Vert \boldsymbol{\Omega}\right\Vert_{F}\right\}$, whereas $\left\{E_{\Delta},\left\Vert \boldsymbol{\mathsf{S}}\right\Vert_{F}\right\}$ shows a clearly better predictive performance in both cases F1$_{128}$ and F2$_{128}$. However, in the higher-Reynolds-number cases, the improvement becomes much less remarkable, as shown in figures \ref{fig:prediction skill doubleF1_512} and \ref{fig:prediction skill doubleF2_512}. Although $\left\{E_{\Delta},\left\Vert \boldsymbol{\mathsf{S}}\right\Vert_{F}\right\}$ still gives the best performance among the three combinations, the difference is much smaller, and all combinations involving $E_{\Delta}$ even perform worse than the predictor based on $E_{\Delta}$ alone.      For predictors formed of
combinations of vorticity $\left\Vert \boldsymbol{\Omega} \right\Vert_{F}$,
vortex-stretching $\left\Vert [\boldsymbol{\mathsf{S}},\boldsymbol{\Omega}]\right\Vert_{F}$ and strain rate
$\left\Vert \boldsymbol{\mathsf{S}} \right\Vert_{F}$, the Brier scores are
larger than the scores of the uncertainty energy $E_{\Delta}$, as shown in figures \ref{fig:prediction skill tribleF1_128}, \ref{fig:prediction skill tribleF1_512}, \ref{fig:prediction skill tribleF2_128} and \ref{fig:prediction skill tribleF2_512}.
For both lower-Reynolds-number cases, the lowest Brier score is obtained for the combination $\left\{ \left\Vert \boldsymbol{\Omega} \right\Vert_{F}, \left\Vert \boldsymbol{\mathsf{S}} \right\Vert_{F}\right\}$, as shown in figures \ref{fig:prediction skill doubleF1_128} and \ref{fig:prediction skill doubleF2_128}, whereas the combination $\left\{ \left\Vert \boldsymbol{\Omega} \right\Vert_{F}, \left\Vert [\boldsymbol{\mathsf{S}},\boldsymbol{\Omega}] \right\Vert_{F}\right\}$ yields the highest Brier score, i.e. the worst predictive performance. Notably, as in the single-predictor case, the predictive performances of the combinations built from $\left\{\left\Vert [\boldsymbol{\mathsf{S}},\boldsymbol{\Omega}] \right\Vert_{F}, \left\Vert \boldsymbol{\Omega} \right\Vert_{F}, \left\Vert \boldsymbol{\mathsf{S}} \right\Vert_{F}\right\}$ show almost no difference from one another in the higher-Reynolds-number cases, as shown in figures \ref{fig:prediction skill doubleF1_512} and \ref{fig:prediction skill doubleF2_512}. Even the full three-component combination $\left\{\left\Vert [\boldsymbol{\mathsf{S}},\boldsymbol{\Omega}] \right\Vert_{F}, \left\Vert \boldsymbol{\Omega} \right\Vert_{F}, \left\Vert \boldsymbol{\mathsf{S}} \right\Vert_{F}\right\}$ does not lead to any clear improvement, as shown in figures \ref{fig:prediction skill tribleF1_512} and \ref{fig:prediction skill tribleF2_512} . A similar contrast is observed when $E_{\Delta}$ is included: in the lower-Reynolds-number cases, combining $E_{\Delta}$ with other predictors leads to an improvement, whereas in the higher-Reynolds-number cases such combinations provide no clear benefit and may even degrade the predictive performance relative to using $E_{\Delta}$ alone, as the committor function estimated using all candidate scalars $\left\{E_{\Delta},\left\Vert[\boldsymbol{\mathsf{S}}, \boldsymbol{\Omega}]\right\Vert_{F}, \left\Vert \boldsymbol{\Omega} \right\Vert_{F}, \left\Vert\boldsymbol{\mathsf{S}} \right\Vert_{F}\right\}$ achieves the lowest Brier score in the low-Reynolds-number cases, whereas in the higher-Reynolds-number cases it gets higher Brier score than obtained with only $E_{\Delta}$.

\begin{figure}
	\centering
	\subfigure[]{
		\label{fig:committorfunctionenergyF1_128}
		\includegraphics[width=0.48\textwidth]{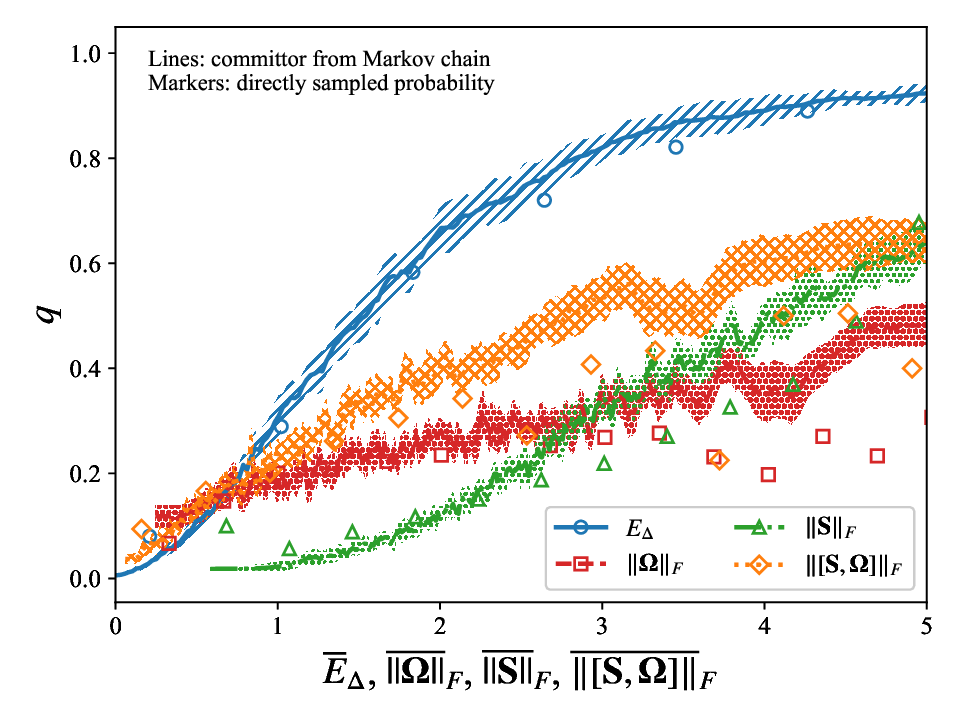}}
	\subfigure[]{
		\label{fig:committorfunctionenergyF1_512}
		\includegraphics[width=0.48\textwidth]{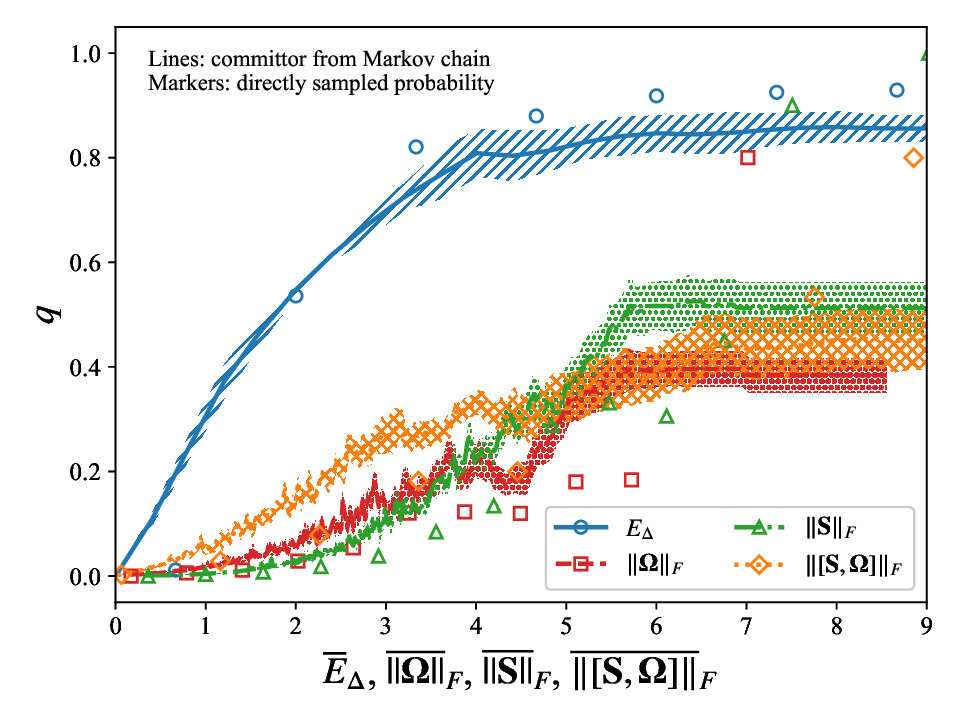}}
	\subfigure[]{
		\label{fig:committorfunctionenergyF2_128}
		\includegraphics[width=0.48\textwidth]{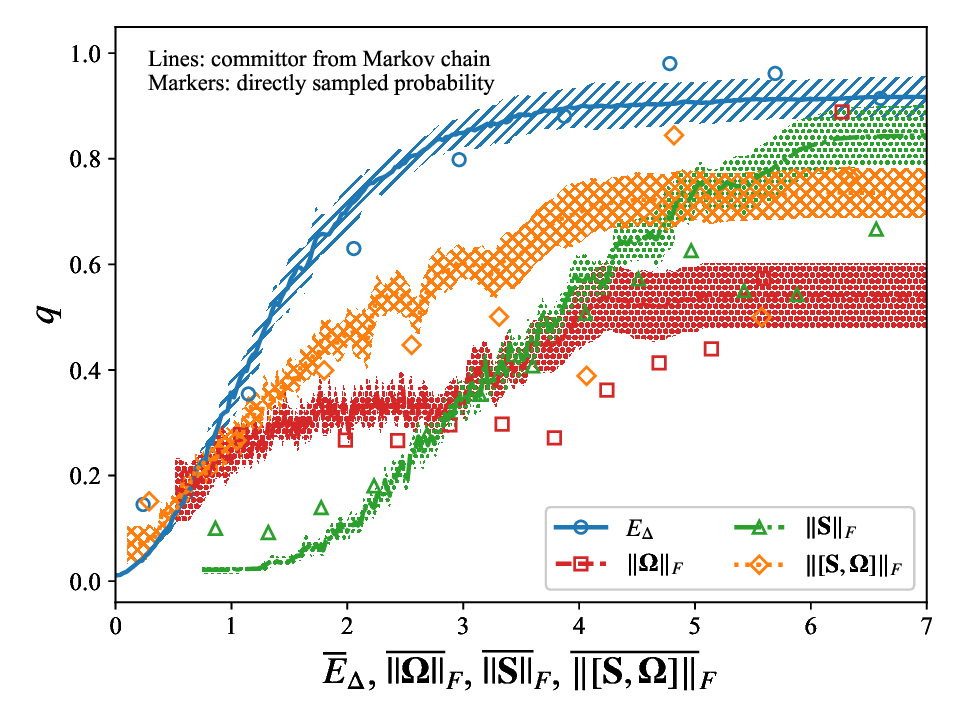}}
	\subfigure[]{
		\label{fig:committorfunctionenergyF2_512}
		\includegraphics[width=0.48\textwidth]{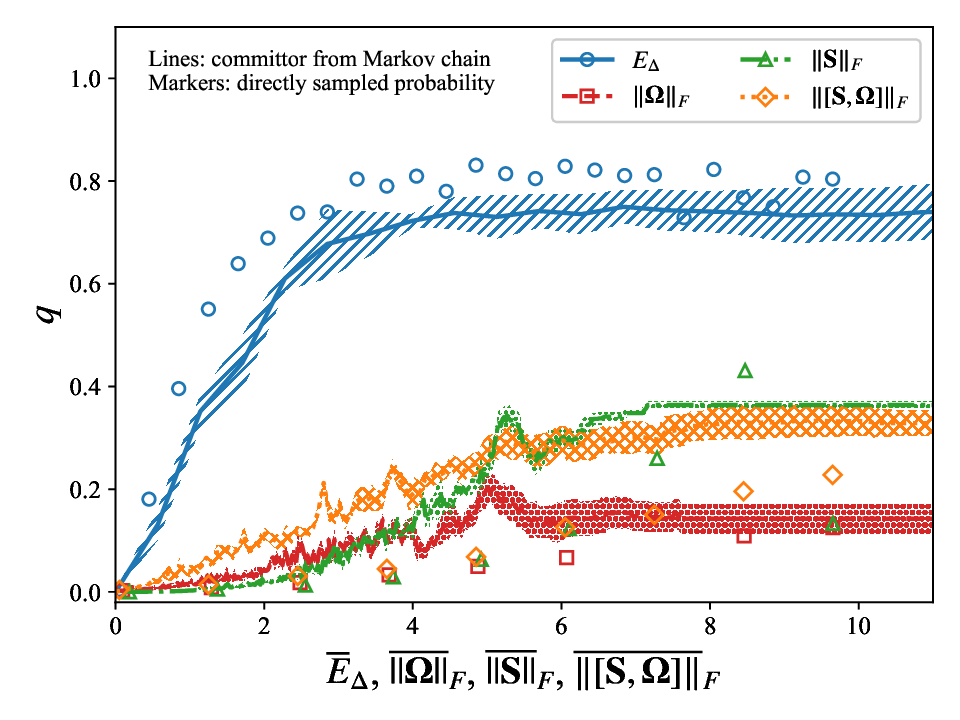}}
	\caption{For cases (a) F1$_{128}$, (b) F1$_{512}$, (c) F2$_{128}$, and (d) F2$_{512}$, committor functions estimated from different single-field predictors using the analogue Markov chain, namely $\{\overline{E_{\Delta}},\overline{\left\Vert [\boldsymbol{\mathsf{S}}, \boldsymbol{\Omega}]\right\Vert}_{F},\overline{\left\Vert \boldsymbol{\mathsf{S}} \right\Vert}_{F}, \overline{\left\Vert \boldsymbol{\Omega} \right\Vert}_{F}\}$. The lines show the averages over the $N_{\text{case}}$ realizations, and the shaded regions indicate the corresponding standard deviations. The normalization is defined as $\overline{F}\equiv F/\sigma_{F}$. The markers represent the direct empirical probabilities computed from the training dataset. }
	\label{fig:committorfunctionvalue}
\end{figure}

\subsection{Dependence of  committor functions on predictor values\label{sec:Evolution of the committor functions with the values of predictor}}

Figure~\ref{fig:committorfunctionvalue} presents the probabilities for
evolving into an extreme positive production event before a negative
production event based on committor function obtained from single
normalised field predictors. Note that this normalisation neither
improves nor degrades the estimated committor. However it helps
compare how different committors depend on different predictors
values. Indeed, a normalised variable close to zero indicates a state
close to average, while an absolute value of one or several units
indicates a state that is one or several standard deviations from
average.

It can be seen that the committor functions obtained from the same
predictor in different flow fields are qualitatively similar. For all
single predictors in $\{\overline{E}_{\Delta},\overline{\left\Vert [\boldsymbol{\mathsf{S}}, \boldsymbol{\Omega}]\right\Vert}_{F}, \overline{\left\Vert \boldsymbol{\mathsf{S}} \right\Vert}_{F}, \overline{\left\Vert \boldsymbol{\Omega} \right\Vert}_{F}\}$, the estimated probabilities increase with the input variable. Statistically, this
implies that events characterized by high values of $\overline{E_{\Delta}}$, $\overline{\left\Vert [\boldsymbol{\mathsf{S}},
	\boldsymbol{\Omega}] \right\Vert}_{F}$, 
$\overline{\left\Vert \boldsymbol{\mathsf{S}} \right\Vert}_{F}$ or $\overline{\left\Vert \boldsymbol{\Omega} \right\Vert}_{F}$ eventually
evolve into extreme production events, but with different sensibility. In all our four cases, the committor function
derived from $\overline{E_{\Delta}}$ exhibits the steepest slope, followed by $\overline{\left\Vert \boldsymbol{\mathsf{S}} \right\Vert}_{F}$, $\overline{\left\Vert [\boldsymbol{\mathsf{S}}, \boldsymbol{\Omega}]\right\Vert}_{F}$ and finally  $\overline{\left\Vert \boldsymbol{\Omega} \right\Vert}_{F}$. The rise in probability as
$\overline{\left\Vert \boldsymbol{\mathsf{S}} \right\Vert}_{F}$ increases is
consistent with the qualitative analysis based on equation
(\ref{eq:Production NS equation}). However, the increases of probability with $\overline{\left\Vert \boldsymbol{\Omega} \right\Vert}_{F}$ goes counter to our theoretical analysis in equation (\ref{eq:Production NS equation}), which suggests that high vorticity should reduce the
probability of extreme uncertainty production. The probabilities obtained directly from the training data are also shown in figure \ref{fig:committorfunctionvalue} as the markers . In all four cases, the committor curves obtained from the
	analogue Markov chain are broadly consistent with the directly sampled
	empirical probabilities, showing that the reduced Markov-chain model
	captures the main dependence of the future extreme-event probability
	on the predictor. This agreement is strongest for
	$\overline{E}_{\Delta}$, for which the markers closely follow the
	rapid increase and the subsequent plateau of the committor curve. For
	$\overline{\left\Vert \boldsymbol{\mathsf{S}} \right\Vert}_{F}$, 
	$\overline{\left\Vert [\boldsymbol{\mathsf{S}}, \boldsymbol{\Omega}] \right\Vert}_{F}$ and $\overline{\left\Vert \boldsymbol{\Omega} \right\Vert}_{F}$,
	the same overall monotonic trend is observed, although the empirical
	probabilities exhibit relatively larger deviations.

For case F1$_{128}$,
the variation of the committor function trained using
$\overline{E}_{\Delta}$ can be divided into two parts.  We find the
committor curve starts from $q=0$ at $\overline{E}_{\Delta}=0$, i.e., at the point of minimum
uncertainty energy.  The committor increases as
$\overline{E_{\Delta}}$ grows, eventually converging to the highest
probability platform of $q=0.9$ when $\overline{E_{\Delta}}=4$. Above
this value of extremely high uncertainty energy, the committor is
constant.  This means that most high uncertainty production event are
found at large values of positive standard deviation. The committor function trained using $\left\Vert [\boldsymbol{\mathsf{S}}, \boldsymbol{\Omega}]\right\Vert_{F}$ monotonically increases from 0.1 to 0.6 as $\overline{\left\Vert [\boldsymbol{\mathsf{S}}, \boldsymbol{\Omega}]\right\Vert}_{F}$ increases from 0 to 5, while for the strain rate $\left\Vert \boldsymbol{\mathsf{S}} \right\Vert_{F}$, the committor function also increases from 0 to 0.6 over the range $\overline{\left\Vert \boldsymbol{\mathsf{S}} \right\Vert}_{F}\in[0,5]$. It should be noted, however, that the effective onset for the strain-rate predictor is at $\overline{\left\Vert \boldsymbol{\mathsf{S}} \right\Vert}_{F}\approx1$, below which the committor remains nearly zero. By contrast, the committor obtained from $\left\Vert \boldsymbol{\Omega} \right\Vert_{F}$ is the flattest among the three, increasing only from about 0.1 to 0.4 as $\overline{\left\Vert \boldsymbol{\Omega} \right\Vert}_{F}$ varies from 0 to 5.  The
change of slope of the committor function indicates that the
probability of evolving into extreme uncertainty production is most
sensitive to the uncertainty energy and subsequently to the strain
rate. For case F2$_{128}$ the slopes of the committor
function are steeper than for case F1$_{128}$. For case F2$_{128}$, the committor
function trained using $E_{\Delta}$ increases from 0 at
$\overline{E}_{\Delta}=0$, to $q=0.9$ for
$\overline{E_{\Delta}}>3$. The committor function trained using
$\left\Vert [\boldsymbol{\mathsf{S}} , \boldsymbol{\Omega}]\right\Vert_{F}$ monotonically increases from 0 to
0.7 as $\overline{\left\Vert [\boldsymbol{\mathsf{S}} , \boldsymbol{\Omega}] \right\Vert}_{F}$ varies from 0
to 4, while for the strain rate $\left\Vert \boldsymbol{\mathsf{S}}
\right\Vert_{F}$, the committor function increases from 0 at
$\overline{\left\Vert \boldsymbol{\mathsf{S}} \right\Vert}_{F}=1$ to $q=0.85$
for $\overline{\left\Vert \boldsymbol{\mathsf{S}} \right\Vert}_{F}>6$. All in
all, the committor functions in case F2$_{128}$ appear more sensitive than in
case F1$_{128}$.

For the higher-Reynolds-number cases, it can be observed that the slopes of the committor functions become systematically weaker in both F1$_{512}$ and F2$_{512}$. Although the four predictors retain the same qualitative ordering as in the lower-Reynolds-number cases, namely a monotonic increase of the committor and the same hierarchy of plateau levels, the plateau probabilities are all reduced compared with their lower-Reynolds-number counterparts. This reduction is particularly obvious in case F2$_{512}$. Part of this trend can be attributed to the lower probability of extreme events, especially those in set $\mathcal{B}$, as reported in Table~\ref{tab:main parameters of statistics}. However, this effect alone does not explain why the magnitude of the reduction differs across predictors and across cases. One possible reason is that, at higher Reynolds numbers, the small-scale velocity-gradient structure becomes more fully developed and more strongly coupled, so that the future evolution is less controlled by any single local velocity-gradient predictor taken in isolation. More importantly, uncertainty growth is influenced not only by the magnitude of the strain rate, but also by how the strain-rate field is aligned with the local uncertainty. At higher Reynolds numbers, such alignment effects may become increasingly important as the velocity-gradient dynamics become more strongly coupled. This may further weaken the predictive skill of predictors based on a single scalar measure, and thus contribute to the lower plateau values of the committor functions in F1$_{512}$ and F2$_{512}$.

\begin{figure}
	\centering
	\subfigure[]{
		\label{fig:committorfunctionStrain-VorF1_128}
		\includegraphics[width=0.48\textwidth]{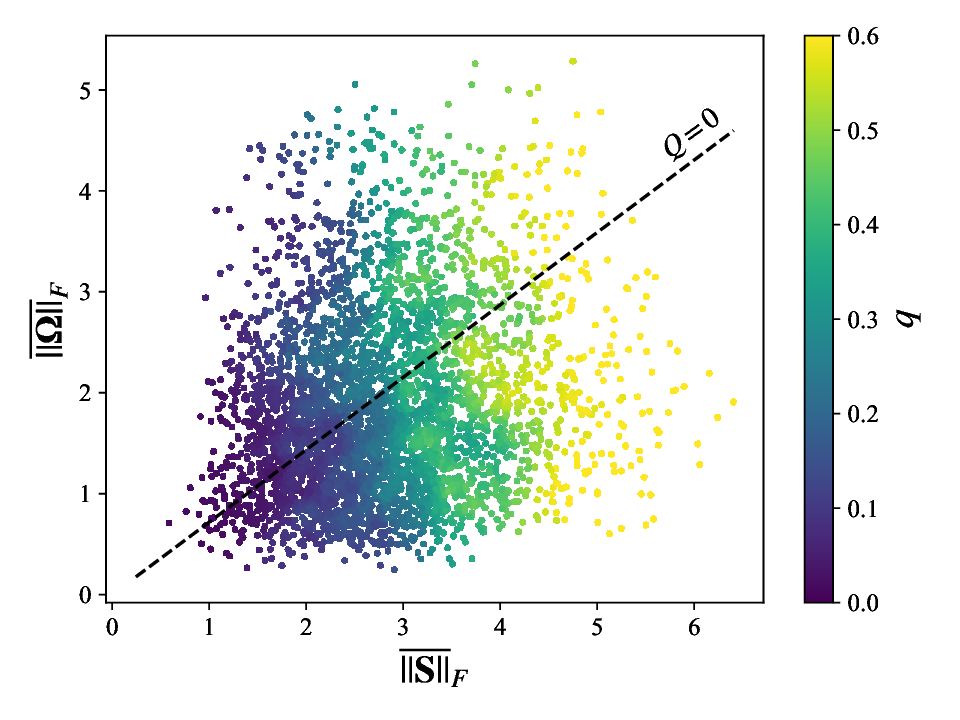}}
	\subfigure[]{
	\label{fig:committorfunctionStrain-VorF1_512}
	\includegraphics[width=0.48\textwidth]{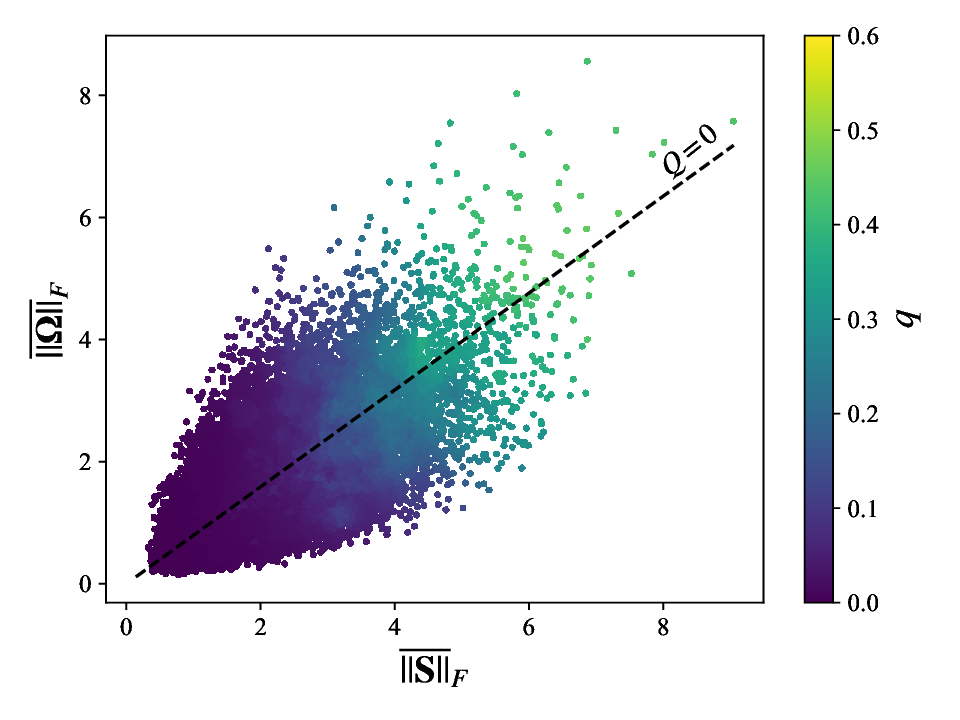}}
	\subfigure[]{
		\label{fig:committorfunctionStrain-StrainVorF1_128}
		\includegraphics[width=0.48\textwidth]{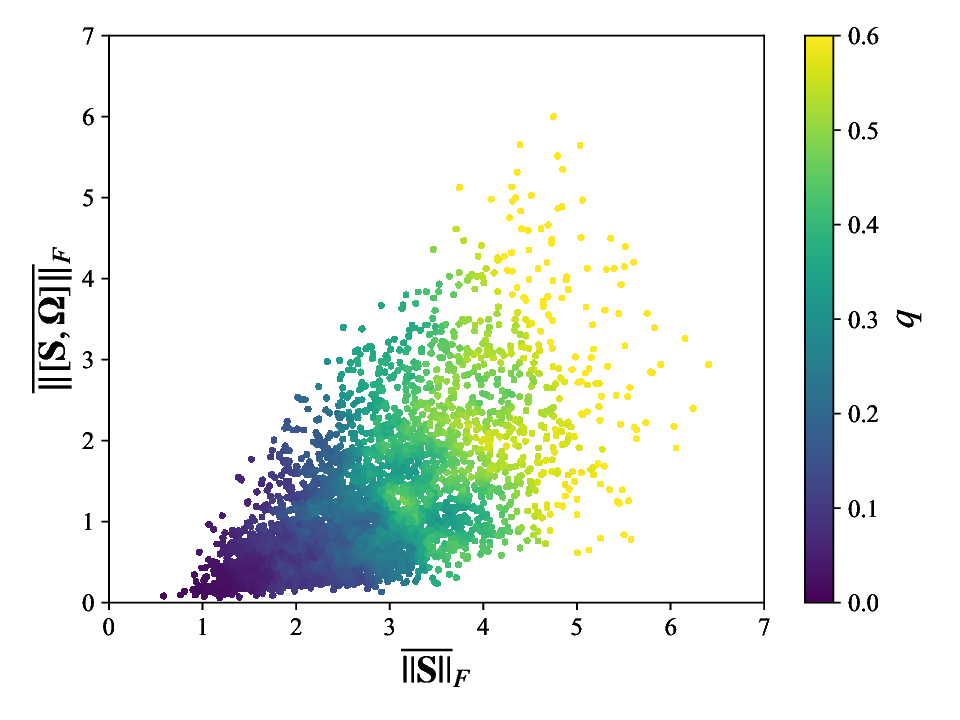}}
	\subfigure[]{
	\label{fig:committorfunctionStrain-StrainVorF1_512}
	\includegraphics[width=0.48\textwidth]{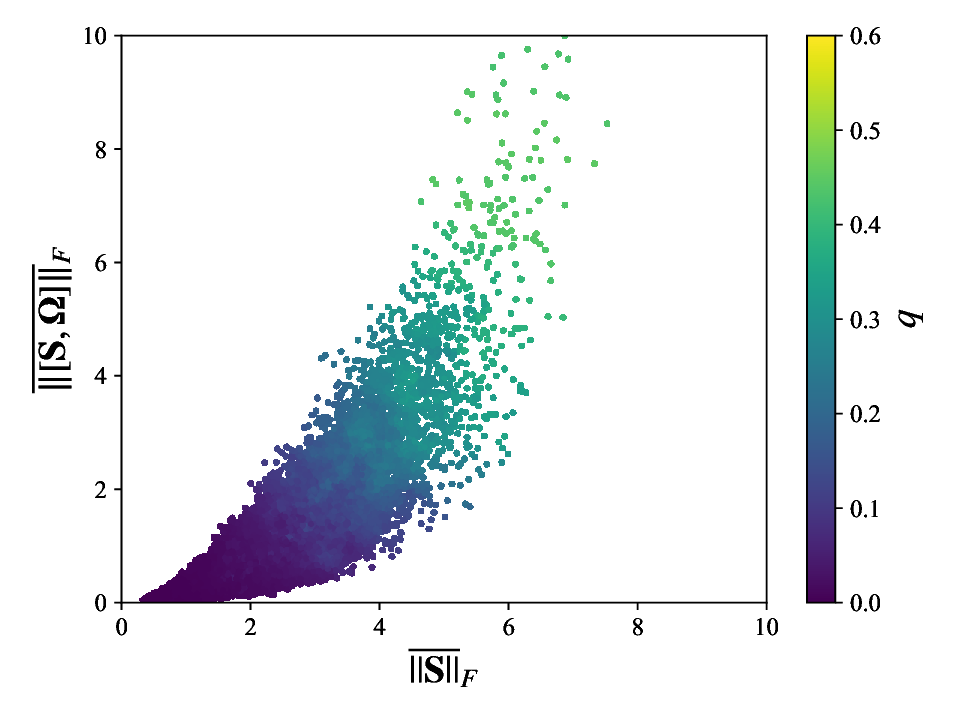}}
	\subfigure[]{
		\label{fig:committorfunctionVor-StrainVorF1_128}
		\includegraphics[width=0.48\textwidth]{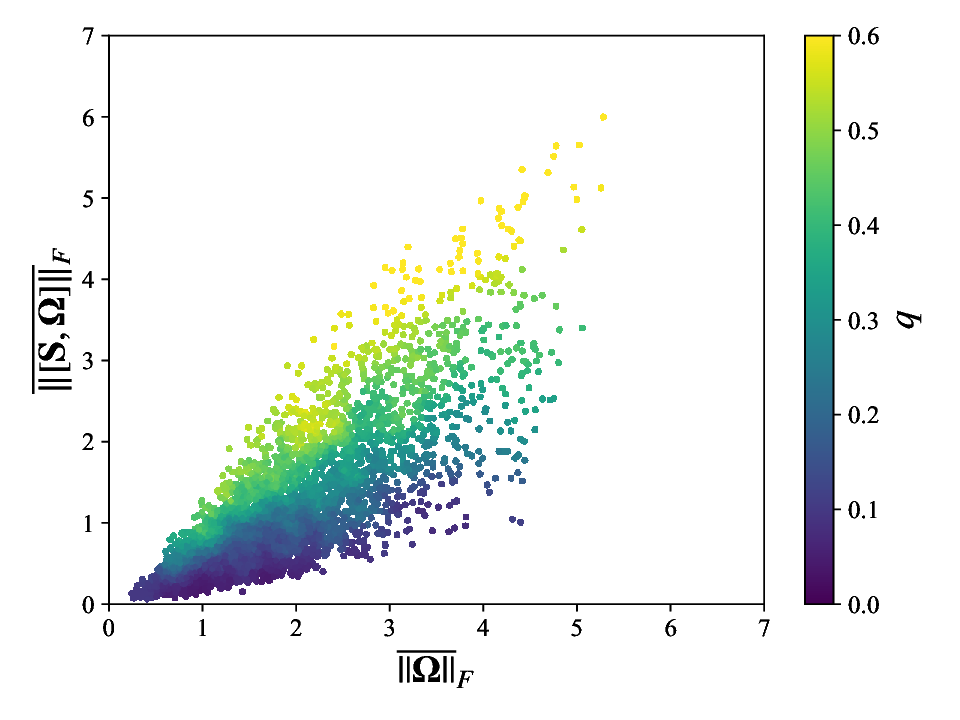}}
	\subfigure[]{
	\label{fig:committorfunctionVor-StrainVorF1_512}
	\includegraphics[width=0.48\textwidth]{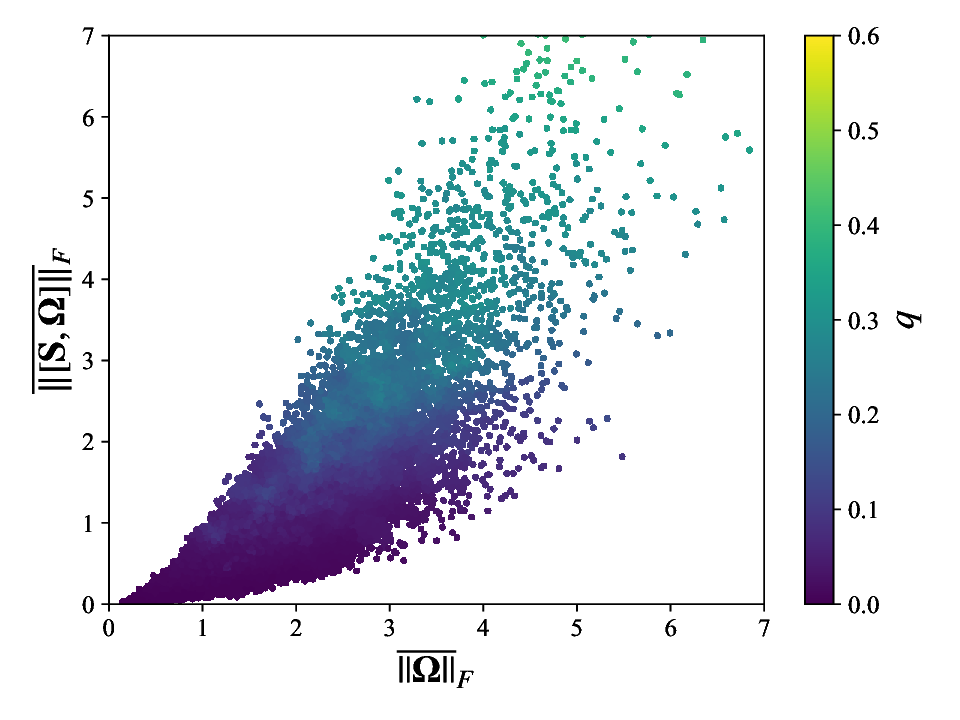}}
	\caption{For (a)(c)(e) case F1$_{128}$ in left panel and (b)(d)(f) case F1$_{512}$ in right panel, the committor function estimated with different predictors:
		(a)(b) normalised strain rate $\overline{\left\Vert \boldsymbol{\mathsf{S}} \right\Vert}_{F}$ - normalsed vorticity $\overline{\left\Vert \boldsymbol{\Omega} \right\Vert}_{F}$. The dashed line indicates $Q=0$ and separates the strain-dominated region ($Q<0$) from the rotation-dominated region ($Q>0$);
		(c)(d) normalised vortex stretching $\overline{\left\Vert [\boldsymbol{\mathsf{S}}, \boldsymbol{\Omega}]\right\Vert}_{F}$
		- normalised strain rate $\overline{\left\Vert \boldsymbol{\mathsf{S}} \right\Vert}_{F}$ and
		(e)(f) normalised vorticity $\overline{\left\Vert \boldsymbol{\Omega} \right\Vert}_{F}$ - normalised vortex-straining
		$\overline{\left\Vert [\boldsymbol{\mathsf{S}},\boldsymbol{\Omega}]\right\Vert}_{F}$.
		The normalisation is defined as $\overline{F}\equiv (F)/\sigma_{F}$.
		We use all the testing boxes (see section \ref{sec:Data testing}) at initial time as input.}
	\label{fig:committorfunction2valueF1}
\end{figure}

\begin{figure}
	\centering
	\subfigure[]{
		\label{fig:committorfunctionStrain-VorF2_128}
		\includegraphics[width=0.48\textwidth]{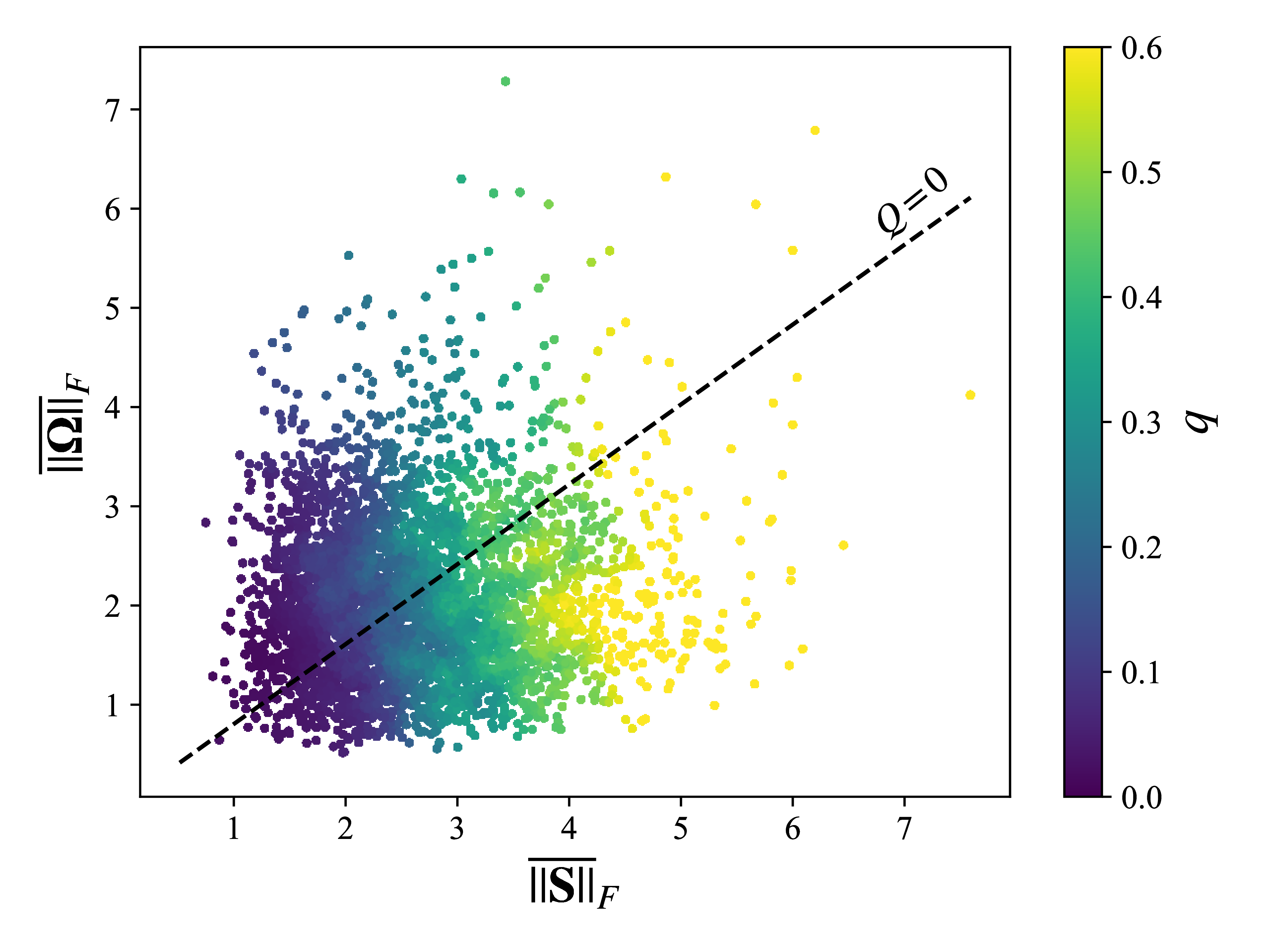}}
	\subfigure[]{
	\label{fig:committorfunctionStrain-VorF2_512}
	\includegraphics[width=0.48\textwidth]{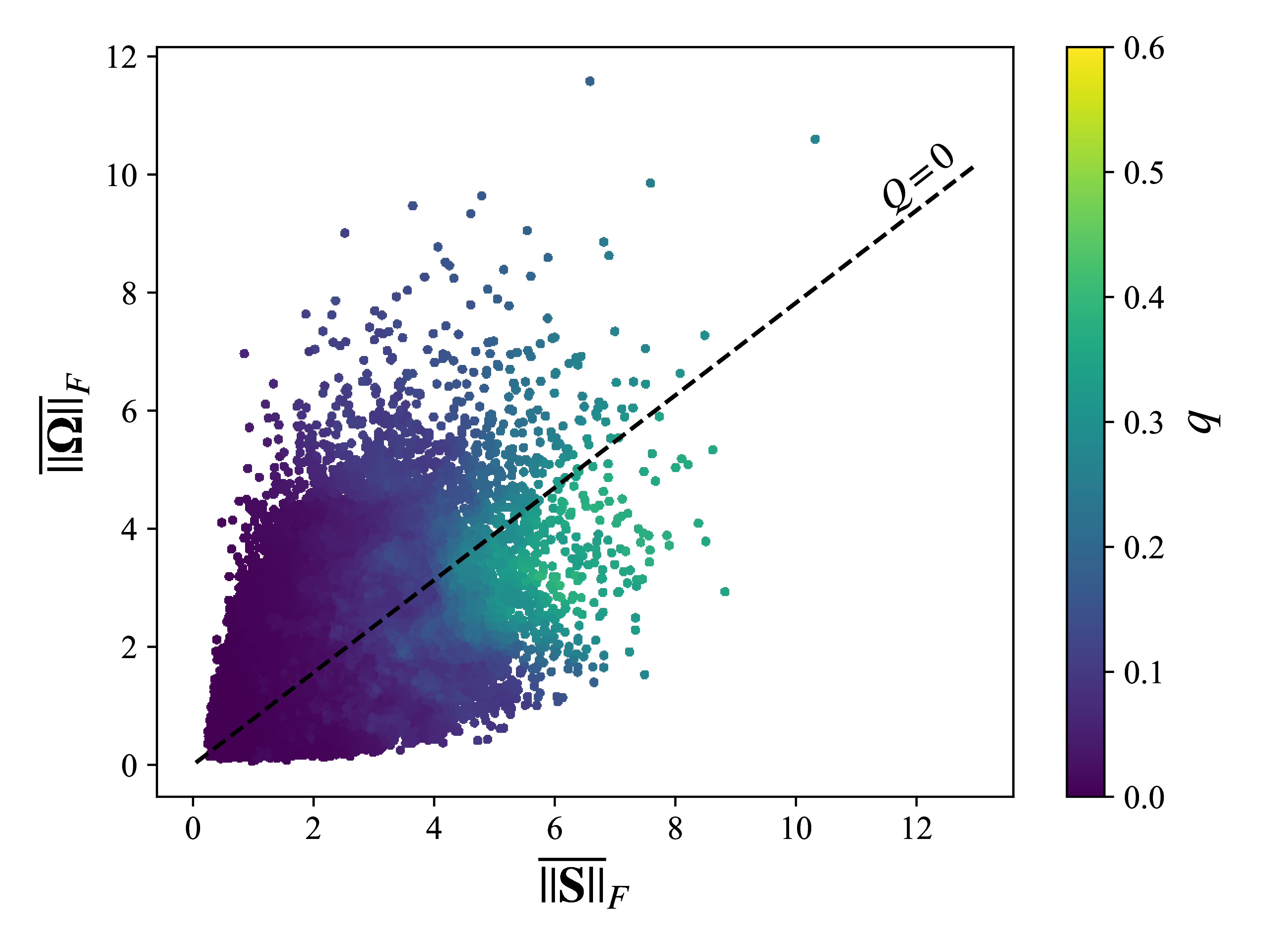}}
	\subfigure[]{
		\label{fig:committorfunctionStrain-StrainVorF2_128}
		\includegraphics[width=0.48\textwidth]{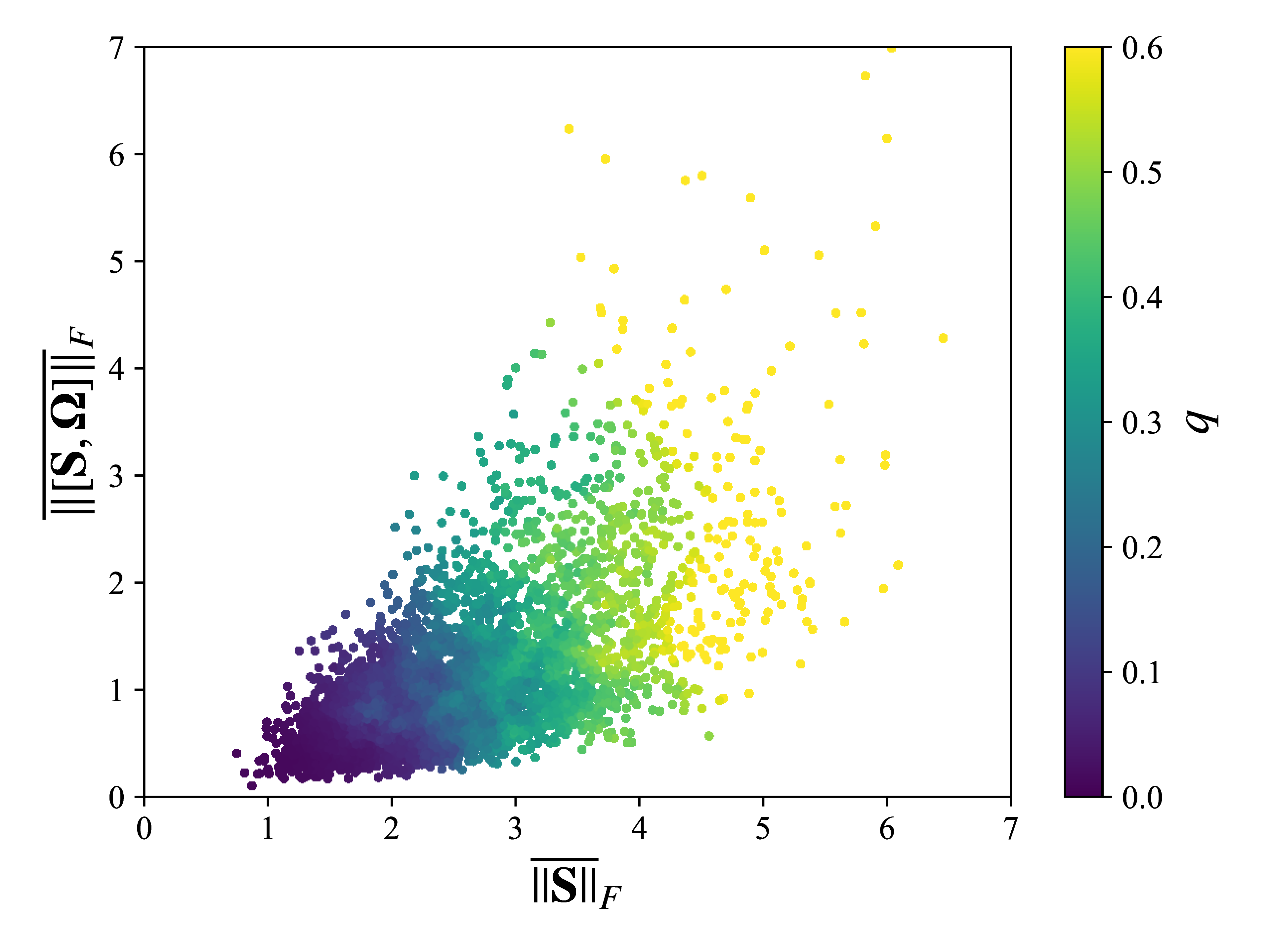}}
	\subfigure[]{
	\label{fig:committorfunctionStrain-StrainVorF2_512}
	\includegraphics[width=0.48\textwidth]{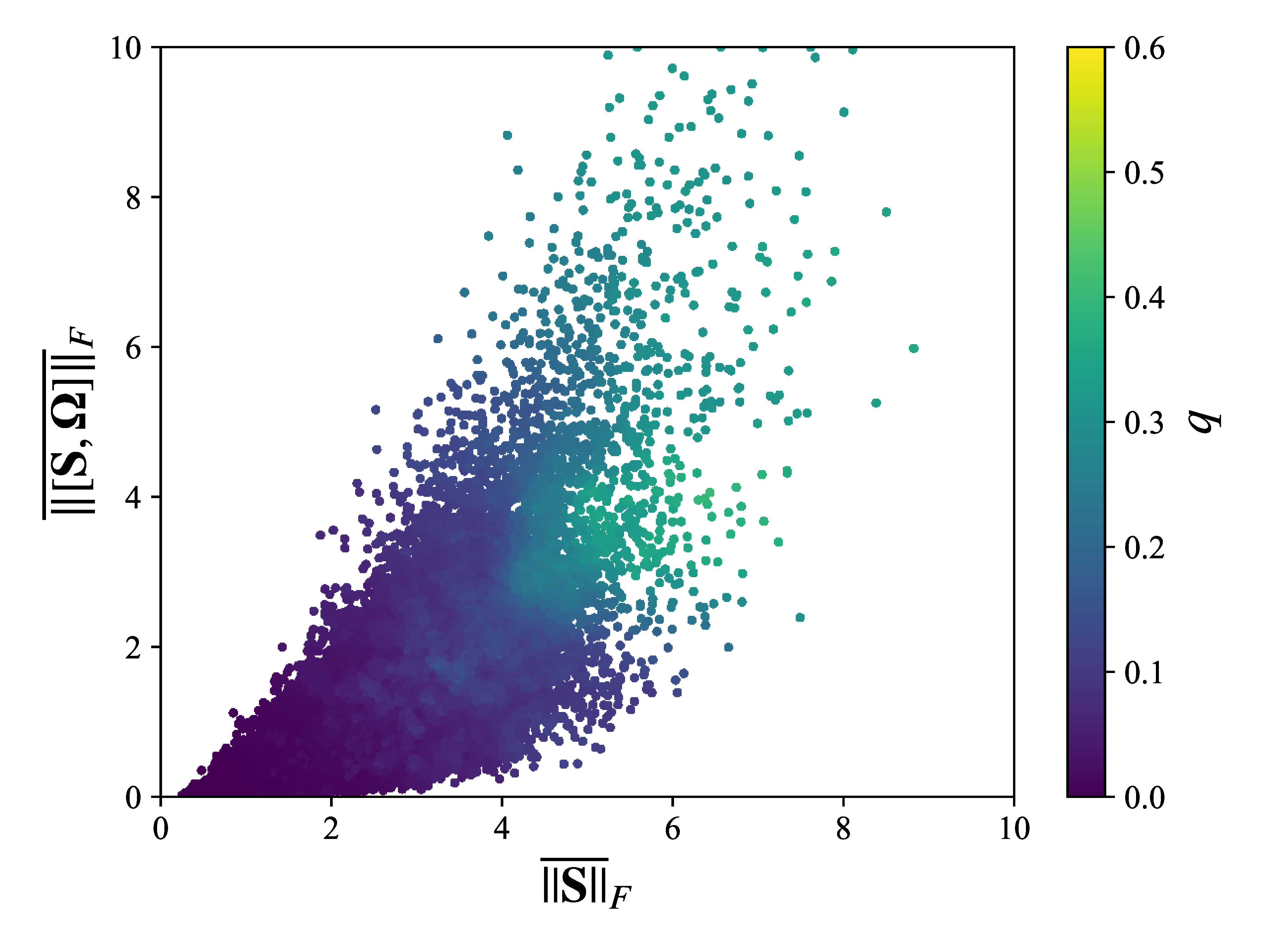}}
	\subfigure[]{
		\label{fig:committorfunctionVor-StrainVorF2_128}
		\includegraphics[width=0.48\textwidth]{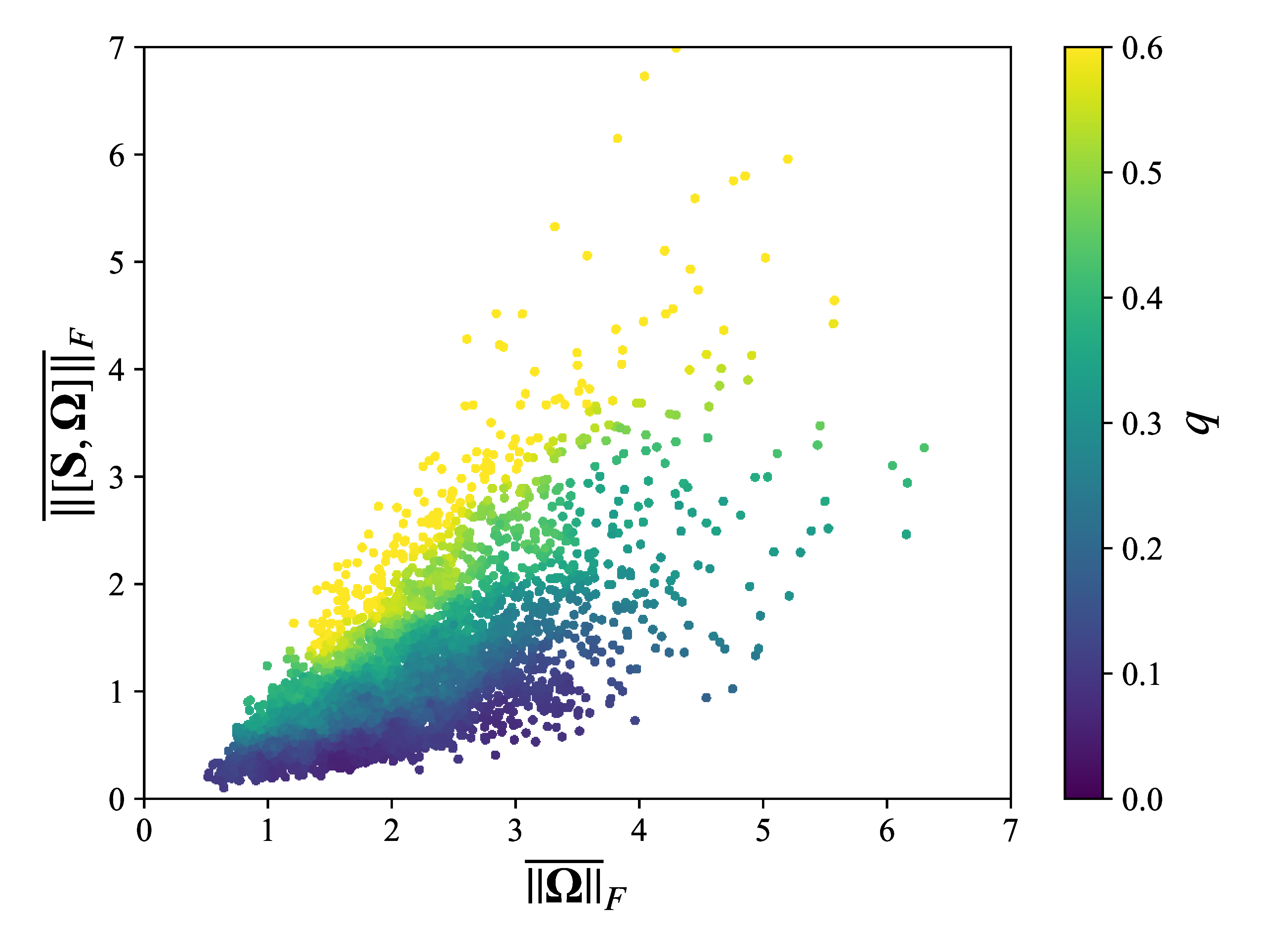}}
	\subfigure[]{
	\label{fig:committorfunctionVor-StrainVorF2_512}
	\includegraphics[width=0.48\textwidth]{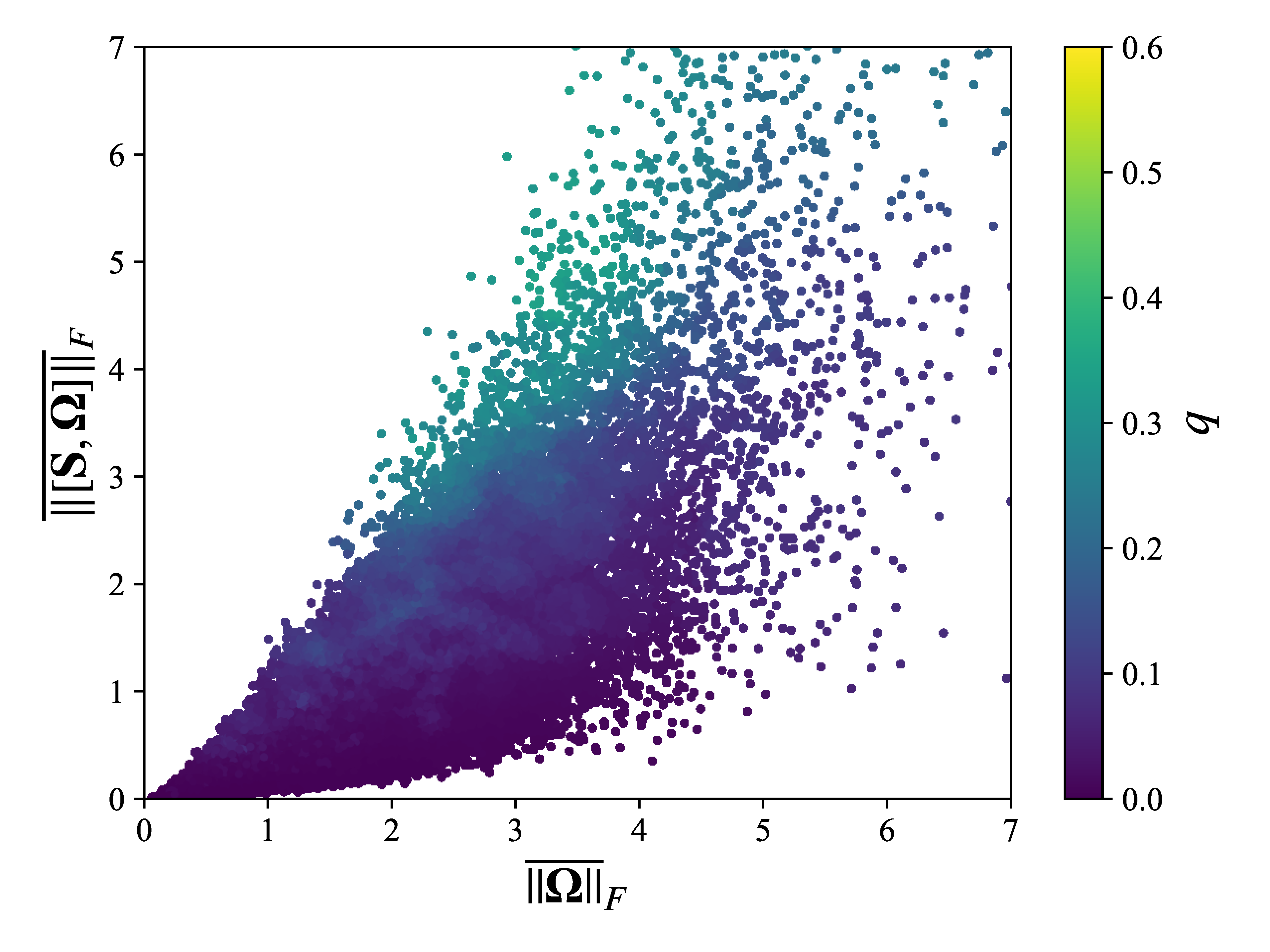}}
	\caption{For (a)(c)(e) case F2$_{128}$ and (b)(d)(f) case F2$_{512}$, the committor function estimated with
		different predictors: (a)(b) normalised strain rate $\overline{\left\Vert \boldsymbol{\mathsf{S}} \right\Vert}_{F}$ - normalsed vorticity $\overline{\left\Vert \boldsymbol{\Omega} \right\Vert}_{F}$.  The dashed line indicates $Q=0$ and separates the strain-dominated region ($Q<0$) from the rotation-dominated region ($Q>0$);
		(c)(d) normalised vortex stretching $\overline{\left\Vert [\boldsymbol{\mathsf{S}}, \boldsymbol{\Omega}]\right\Vert}_{F}$
		- normalised strain rate $\overline{\left\Vert \boldsymbol{\mathsf{S}} \right\Vert}_{F}$ and
		(e)(f) normalised vorticity $\overline{\left\Vert \boldsymbol{\Omega} \right\Vert}_{F}$ - normalised vortex-straining
		$\overline{\left\Vert [\boldsymbol{\mathsf{S}},\boldsymbol{\Omega}]\right\Vert}_{F}$.
		The normalisation is defined as $\overline{F}\equiv (F)/\sigma_{F}$.
		We use all the testing boxes (see section \ref{sec:Data testing}) at initial time as input.}
	\label{fig:committorfunction2valueF2}
\end{figure}

Figures \ref{fig:committorfunction2valueF1} and
\ref{fig:committorfunction2valueF2} show two-dimensional plots of
committor functions trained using two normalized scalar variables as
predictors. Each point in the plot corresponds to values of the used
predictors and the colour of the point indicates the value of the
committor. It can be seen that the distributions of
$\left\{\overline{\left\Vert[\boldsymbol{\mathsf{S}},\boldsymbol{\Omega}]\right\Vert}_{F},
\overline{\left\Vert \boldsymbol{\Omega}\right\Vert}_{F}\right\}$ and
$\left\{\overline{\left\Vert[\boldsymbol{\mathsf{S}},\boldsymbol{\Omega}]\right\Vert}_{F},
\overline{\left\Vert\boldsymbol{\mathsf{S}}\right\Vert}_{F}\right\}$ clearly
trend along the diagonal, suggesting a correlation
between the fluctuations of
$[\boldsymbol{\mathsf{S}},\boldsymbol{\Omega}]$ with
those of $\boldsymbol{\Omega}$ and $\boldsymbol{\mathsf{S}}$ (fluctuations above and
below their respective averages). This agrees with our aforementioned
observation that the inclusion of $[\boldsymbol{\mathsf{S}},\boldsymbol{\Omega}]$ in predictor combinations does not
enhance prediction performance. Notably, in all four cases, the statistical outcomes
derived from the committor function for the set
$\left\{\overline{\left\Vert \boldsymbol{\Omega} \right\Vert}_{F},
\overline{\left\Vert \boldsymbol{\mathsf{S}} \right\Vert}_{F}\right\}$
demonstrate that the probability of extreme events occurring in the
future exhibits an increasing trend from the lower left to the upper
right along the diagonal, suggesting that high $\overline{\left\Vert
	\boldsymbol{\Omega} \right\Vert}_{F}+\overline{\left\Vert \boldsymbol{\mathsf{S}}
	\right\Vert}_{F}$ values
are more likely to trigger future extreme uncertainty production.
Furthermore, it can be seen that the trained committor functions are not symmetric with respect to the dashed line corresponding to $Q=0$. This line separates the strain-dominated region, where $Q<0$, from the rotation-dominated region, where $Q>0$. For the same value of
$\overline{\left\Vert \boldsymbol{\Omega} \right\Vert}_{F}
+
\overline{\left\Vert \boldsymbol{\mathsf{S}} \right\Vert}_{F}$,
the probability of evolving towards an extreme uncertainty-production event is significantly higher on the strain-dominated side, i.e. in the region below the dashed line, as shown in figures \ref{fig:committorfunctionStrain-VorF1_128}, \ref{fig:committorfunctionStrain-VorF1_512}, \ref{fig:committorfunctionStrain-VorF2_128} and \ref{fig:committorfunctionStrain-VorF2_512}.  This
indicates that appearance of extreme uncertainty production events are
more likely when the strain rate is above average than when vorticity
is above average.
Similar observations can be made for the other figures, where the probability
increases with increasing $\overline{\left\Vert \boldsymbol{\mathsf{S}}
	\right\Vert}_{F}$ and decreasing $\overline{\left\Vert \boldsymbol{\Omega}
	\right\Vert}_{F}$ for a given value of
$\overline{\left\Vert[\boldsymbol{\mathsf{S}},\boldsymbol{\Omega}]\right\Vert}_{F}$.
Such results indicate that extreme high strain rate is the most
important factor for extreme positive uncertainty production.

\begin{figure}
	\centering
	\subfigure[]{
		\label{fig:committorfunctionEveryPointStrain-VorF1_128}
		\includegraphics[width=0.48\textwidth]{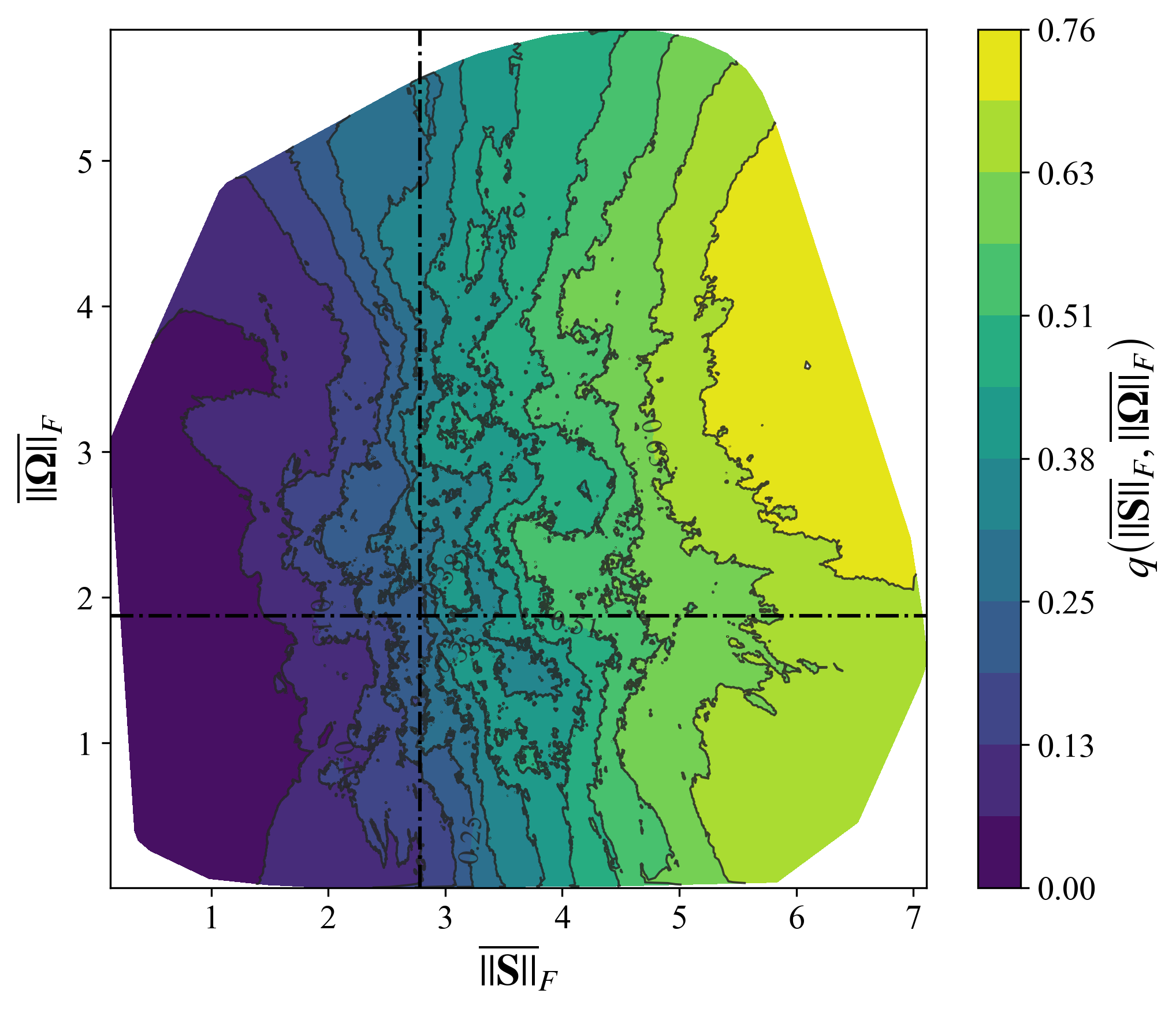}}
	\subfigure[]{
	\label{fig:committorfunctionEveryPointStrain-VorF1_512}
	\includegraphics[width=0.48\textwidth]{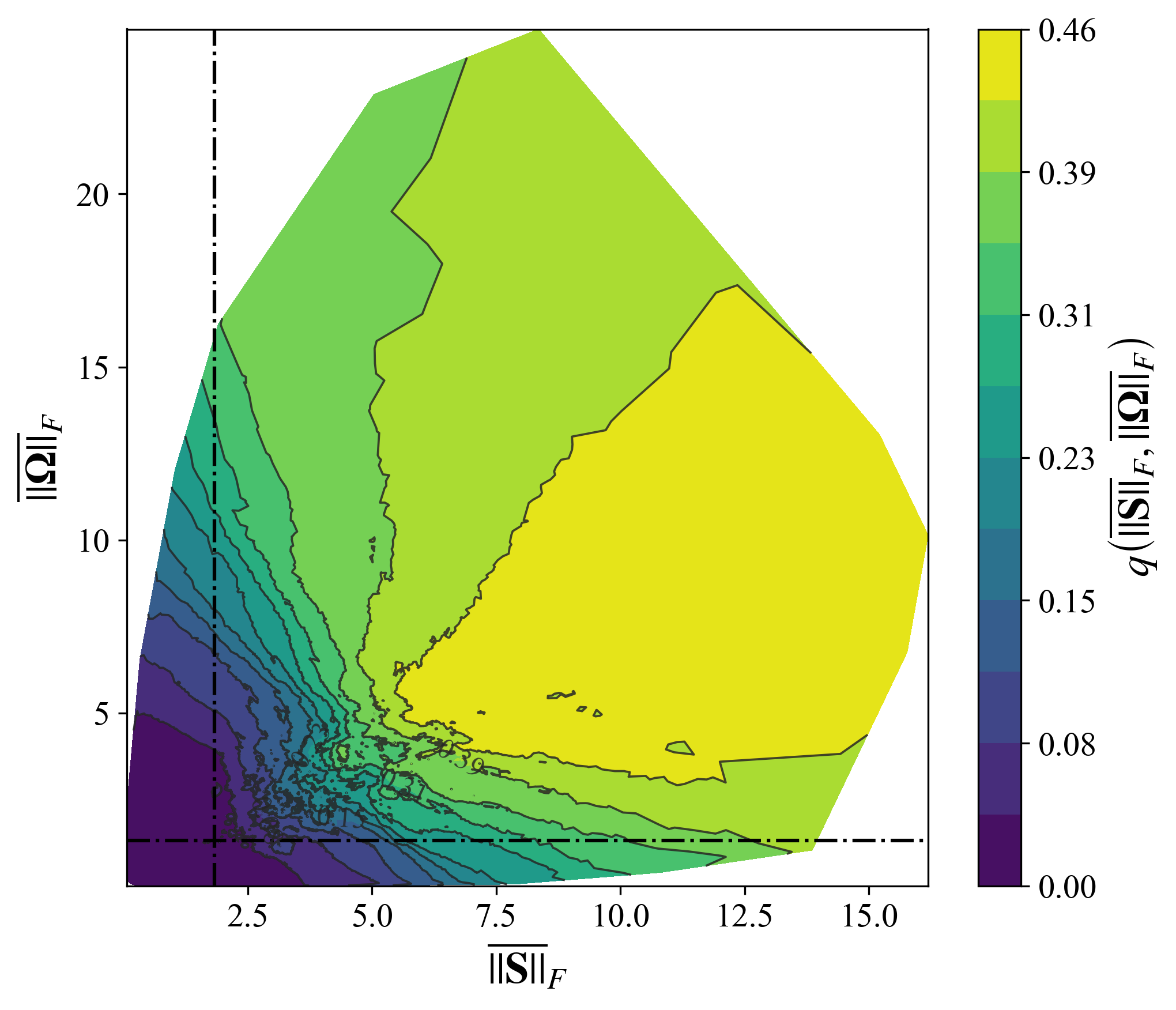}}
	\subfigure[]{
		\label{fig:committorfunctionEveryPointStrain-VorF2_128}
		\includegraphics[width=0.48\textwidth]{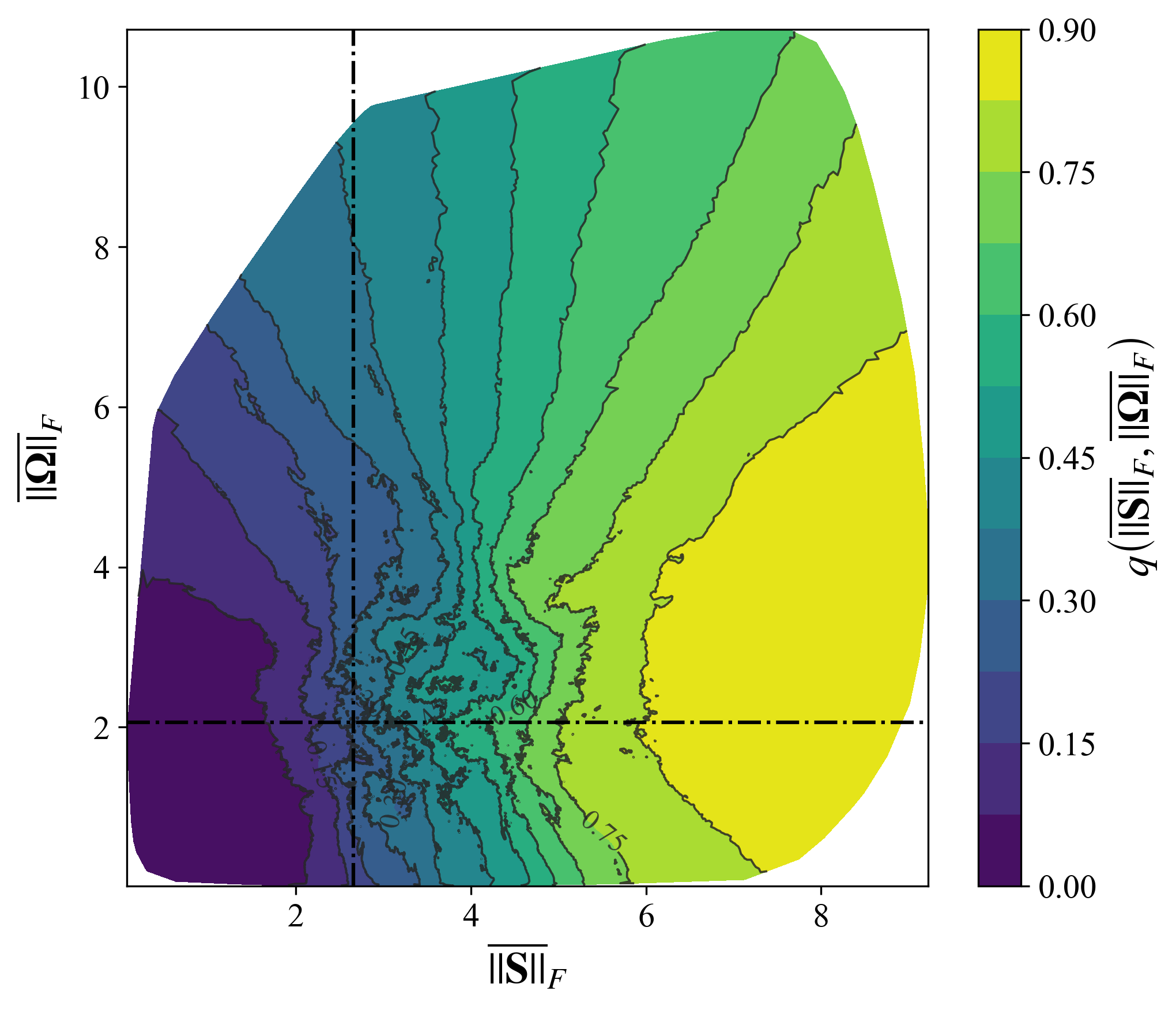}}
	\subfigure[]{
	\label{fig:committorfunctionEveryPointStrain-VorF2_512}
	\includegraphics[width=0.48\textwidth]{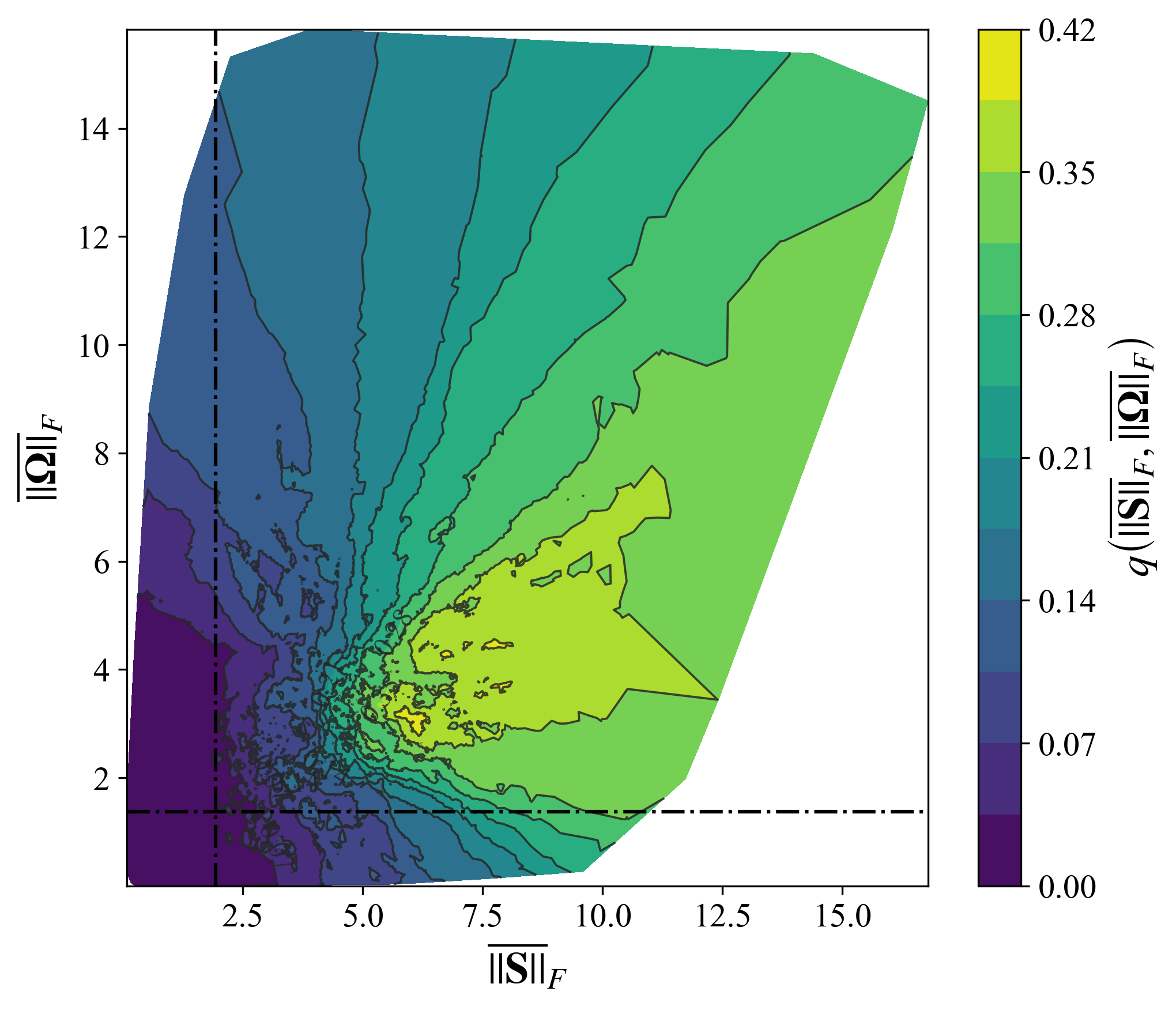}}
	\caption{For (a) F1$_{128}$, (b) F1$_{512}$, (c) F2$_{128}$ and (d) F2$_{512}$, the probability (in
		colours) based on the estimated committor function, using the input $\left\{\left\Vert \boldsymbol{\Omega}
		\right\Vert_{F}, \left\Vert \boldsymbol{\mathsf{S}}
		\right\Vert_{F}\right\}$ at every grid point of the DNS at
		its corresponding starting time of sampling. The vertical and horizontal dash-dotted lines denote the spatial average of $\left\Vert \boldsymbol{\mathsf{S}}
		\right\Vert_{F}$ and $\left\Vert \boldsymbol{\Omega}
		\right\Vert_{F}$ respectively.}
	\label{fig:committorfunctionEveryPointStrain-Vor}
\end{figure}

To further clarify this point, we estimate the committor function at
any state of turbulence existing in the phase space.
Therefore, instead of displaying the committor for predictors averaged
over testing boxes (as described in section \ref{sec:Data testing}),
we compute this probability as a function of
$\left\{\overline{\left\Vert \boldsymbol{\Omega} \right\Vert}_{F},
\overline{\left\Vert \boldsymbol{\mathsf{S}} \right\Vert}_{F}\right\}$ at
every grid point of the DNS (i.e. $N^3$ data points) at the starting time,
for all four cases. For each of these input values, we obtain
the probability by using the nearest neighbor method in section
\ref{sec:Computing the committor function from the analogue Markov
	chain} and the committor functions.  Figure
\ref{fig:committorfunctionEveryPointStrain-Vor} presents colour levels
of our four interpolated committors.
For all four cases, The committor functions reveal a complex region where both strain rate and vorticity are slightly larger than
their spatial averaged value. In this region, the estimated committor function
varies rapidly without a simple trend. Not only does the Navier-Stokes
system have a strong deterministic sensitivity to initial conditions
because of its chaoticity, the probability of evolution towards either
an extreme uncertainty production event or a much more common
uncertainty reducing event may also be very sensitive to the initial
condition. Situations where both strain rate and vorticity are close
to their spatial averaged value may display what has been termed
probabilistic unpredictability in stochastic systems
\citep{lucente2022committor} or it may be that more predictor
information is required, for example in terms of the pressure field
and two-point correlations rather than just one-point quantities. Note that the surface region of probabilistic unpredictability is qualitatively similar in all four cases, despite the differences in forcing and Reynolds number. For the
region away from the complex region, where the strain rate or vorticity is
close to the two tails of their PDF, it is observed that the contours
of probability go through the complex region.  In this region, more definitive
probability predictions can be made. Similar to figures
\ref{fig:committorfunctionStrain-VorF1_128}, \ref{fig:committorfunctionStrain-VorF1_512},
\ref{fig:committorfunctionStrain-VorF2_128} and \ref{fig:committorfunctionStrain-VorF2_512}, the region with high velocity
gradients, particularly on the side of high strain rates, is more
likely to evolve into extreme uncertainty production events, while low
probabilities are mainly concentrated on the high vorticity side with
smaller strain rates.

In this section we have investigated the future evolution of
uncertainty production using the statistical committor function.
Among all the candidate predictors suggested by the theoretical analysis, the strain rate is found to be the quantity most strongly related, in statistical terms, to future extreme positive uncertainty production in the lower-Reynolds-number cases, as indicated by its lowest Brier score. For the higher-Reynolds-number cases, however, this statistical advantage becomes much less obvious, as all three candidate predictors give very similar Brier scores.  Furthermore, the
non-uniform distribution of the probability of extreme events
is notable in figure \ref{fig:committorfunctionEveryPointStrain-Vor} and
corresponds to the observation that extreme uncertainty production
probablities are localised in high convergence regions (section
\ref{sec:The relation of extreme uncertainty production events and
	flow topology}). It is worth stressing that even probablities
may not be predictable
in regions of the flow field where strain rate and vorticity are near
their flow field average. In such regions, the committor function is
very dependent on the input predictor values, making it potentially
impossible to provide stable probabilistic forecasts.

\section{Conclusion\label{sec:Conclusion}}															
We have explored how to predict and locate the extreme uncertainty
production events, which are rare in turbulence but play a dominant
role in the early exponential growth of very small initial
uncertainties. We have investigated the evolution towards extreme
uncertainty production events from three perspectives. 

We have derived from the Navier-Stokes equations the
evolution equation for $P_{\Delta}$ and in the present work we focus primarily on the local inertial contribution to $P_{\Delta}$. We have seen that the
time-evolution of $P_{\Delta}$ is influenced by the eigenvalues of the
reference field’s vortex-deformation rate, rotation rate and strain
rate tensors, as well as the distribution of uncertainty energy
alignment with the eigenvectors of these tensors. Equation
(\ref{eq:Production NS equation}) suggests that vortex-stretching and
strain (both compressing and stretching) increase uncertainty
production, whereas vortex-compression and vorticity decrease
uncertainty production. However, since these three tensors influence
each other, the cumulative impact of these mechanisms on the evolution
of uncertainty production is unclear from equation (\ref{eq:Production
NS equation}). For instance, while vortex stretching
increases uncertainty production, it concurrently enhances vorticity,
which in turn reduces uncertainty production.  Therefore, the sign of the individual local contributions alone does not determine which mechanism dominates the occurrence of extreme events, which motivates the statistical analyses developed below.

Given that the vorticity $\left\Vert \boldsymbol{\Omega}
\right\Vert_{F}$ and the strain rate $\left\Vert \boldsymbol{\mathsf{S}}
\right\Vert_{F}$ are widely used to define flow topology, we decided
to investigate the uncertainty production from a flow topology
perspective.
We therefore used the $Q-R$ diagram for our analysis
and observed that extreme uncertainty production events mainly
concentrate on zones characterized by strain self-amplification and
zones characterised by vortex stretching with relatively large values of $Q$
and $R$, with the latter becoming more prominent at higher Reynolds numbers, whereas the uncertainty-production term in the vortex-compression quadrant ($Q>0$, $R>0$) is dominated by negative values for all forcing types and Reynolds-number cases.

Finally, we considered probabilistic forecasts and used the committor
function to investigate the overall impact of vortex deformation,
vorticity and strain rate on the time evolution of $P_{\Delta}$. In
the present work the committor function has been computed for events
that may occur anywhere in the domain and at any time, rather than
restricting it to a pre-selected region and/or temporal window as in
some previous works \citep{lucente2022committor}. While this increases
the complexity of the problem, it also enhances the robustness and
applicability of the results to a broader class of transitions. Our
statistical results show that, in lower-Reynolds-number cases, the strain rate has the closest
relationship with future extreme uncertainty production events given its lowest Brier score,
whereas the vorticity has weak predictive power.  However, in the higher-Reynolds-number cases, the differences in predictive performance among the three velocity-gradient-based quantities become much less obvious, which may be related to the alignement between uncertainty and the strain field, and the stronger coupling of the local velocity-gradient dynamics at higher Reynolds numbers. Although the
probability of extreme uncertainty production predicted in terms of
strain rate increases with strain rate, consistent with direct
inspection of the evolution equation of $P_{\Delta}$, probability of extreme uncertainty production predicted in terms of vorticity appears to also increase with vorticity, an effect that runs counter to the trend suggested by a direct inspection of the evolution equation of $P_{\Delta}$. When using both $\left\Vert
\boldsymbol{\Omega} \right\Vert_{F}$, $\left\Vert \boldsymbol{\mathsf{S}}
\right\Vert_{F}$ as predictors, it is observed that there exists a
region, corresponding to strain rate and vorticity near their spatial
averages over the entire flow field where even probabilities are
effectively unpredictable. However, away from these average values,
significantly higher probabilities of extreme uncertainty production
are more closely associated with the strain rate than vorticity in all type of forcing and Reynolds number,
suggesting that, compared to high vorticity, high strain rate is more
important for the occurrence of future extreme uncertainty production
events.

A remarkable observation is that the Brier scores obtained using the
	committor function still differ between the large-scale forcings F1
	and F2. Since a similar difference remains visible even in the
	higher-Reynolds-number cases, it is unlikely that this effect is due
	simply to a direct influence of the forcing on the smallest scales.
	We thus speculate that the difference is mediated by
	terms that are not explicitly analysed in the present work, in
	particular the pressure-related contributions. Because pressure
	contains nonlocal information about the global flow organization, it
	may implicitly retain a dependence on the forcing configuration and
	thus affect the predictability of extreme uncertainty-production
	events. However, this interpretation remains speculative and requires
	further investigation.


The insights gained from analyzing uncertainty production in
turbulence through both theoretical and probabilistic approaches
highlight the crucial role of the strain rate in predicting extreme
$P_{\Delta}$ events.
Future work should focus on refining our understanding of the
interactions between vortex deformation, strain rate, and vorticity to
better capture the mechanisms driving extreme uncertainty
production. Furthermore, the roles of other effects, such as those
arising from the pressure field, external forcing and viscous
diffusion, must also be addressed in the future.

\section*{Acknowledgments}
Jin Ge acknowledges financial support from the China Scholarship
Council. We are grateful for the access to the computing resources
supported by the Zeus supercomputers (Mésocentre de Calcul
Scientifique Intensif de l'Université de Lille) and
Lantuxinsuan (Chengdu) Technology Co., Ltd.. We thank the anonymous referees for very valuable comments that help us improve this paper very significantly.

\section*{Funding}
This research received no specific grant from any funding agency, commercial or not-for-profit sectors. 

\section*{Declaration of interests}
The authors report no conflict of interest.

\bibliographystyle{jfm}



\end{document}